\documentclass[a4paper,10pt]{article}
\usepackage[numbers]{natbib}
\setcitestyle{sort, square, comma}
\usepackage[utf8]{inputenc}
\usepackage{amsmath,amssymb}
\usepackage{authblk}
\usepackage{float}
\usepackage{array}
\usepackage{mathrsfs}
\usepackage{setspace}
\usepackage{subfig}
\usepackage{cancel}
\usepackage{graphicx}
\usepackage{txfonts}
\usepackage{graphicx}
\usepackage{anysize}
\usepackage{hyperref}
\usepackage{textcomp}
\usepackage{caption}
\usepackage{float}
\usepackage{epsfig}
\usepackage[normalem]{ulem}
\title{A Cure for numerical shock instability in HLLC Riemann solver using antidiffusion control}
\author[1]{Sangeeth Simon}
\author[2]{J. C. Mandal \thanks{Corresponding author: mandal@iitb.ac.in}}

\affil[1,2]{Department of Aerospace Engineering, Indian Institute of Technology Bombay, Mumbai-400076}

\date{}

\begin{document}

\maketitle

\begin{abstract}
Various forms of numerical shock instabilities are known to plague many contact and shear preserving approximate Riemann solvers, including the popular
Harten-Lax-van Leer with Contact (HLLC) scheme, during high speed flow simulations. 
In this paper we propose a simple and inexpensive novel strategy to prevent the HLLC scheme from developing such spurious solutions without compromising on its linear wave resolution
ability.
The cure is primarily based on a reinterpretation of the HLLC scheme as a combination of its well-known diffusive counterpart, the HLL scheme, 
and an antidiffusive term responsible for its accuracy on linear wavefields.
In our study, a linear analysis of this alternate form indicates that shock instability in the HLLC scheme could be triggered 
due to the unwanted activation of the antidiffusive terms of its mass and interface-normal flux components on interfaces that are not aligned with the shock front. This inadvertent
activation results in weakening of the favourable dissipation provided by its inherent HLL scheme and causes unphysical mass flux variations along the shock front.
To mitigate this, we propose a modified HLLC scheme that employs a simple differentiable pressure based multidimensional shock sensor to achieve smooth control of 
these critical antidiffusive terms near shocks.
Using a linear perturbation analysis and a matrix based stability analysis, we establish that the resulting scheme, called HLLC-ADC (Anti-Diffusion Control),
is shock stable over 
a wide range of freestream Mach numbers. 
Results from standard numerical test cases demonstrate that the HLLC-ADC scheme is indeed free from the
most common manifestations of shock instability including the Carbuncle phenomenon without significant loss of accuracy on shear dominated viscous flows.

\end{abstract}

\section{Introduction}

Approximate Riemann solvers are considered to be a simple, cost effective and practically employable choice for the simulation of high speed gasdynamical flows.
Some famous numerical schemes from this category include  Roe scheme \cite{roe1981}, HLL scheme\cite{harten1983}, HLLC scheme \cite{toro1994}, HLLEM scheme \cite{einfeldt1991} 
and
AUSM class of schemes \cite{liou1993} etc. Among these, the HLL-family of schemes have gathered interest because of their accuracy, mathematical simplicity, 
inherent entropy satisfying ability, positivity, lack of demand for knowledge of complete eigenstructure of flux Jacobians and their easy extensibility to various 
hyperbolic systems of governing equations. While most of the schemes from this family are well designed to satisfactorily handle shocks and expansion fans,
the ability to resolve the linearly degenerate wavefields distinguishes them from each other. Thus, while the HLLC scheme, the HLLEM scheme, HLLE+ scheme \cite{park2003}, 
HLLEMS \cite{xie2017} etc are complete-wave Riemann solvers capable of resolving the contact and shear waves of the Euler system, HLLE scheme \cite{einfeldt1988}, 
HLLS scheme and HLLCM scheme \cite{shen2016} and HLL-CPS scheme \cite{mandal2011} etc are incomplete-wave Riemann solvers that omit one or both of these linear waves 
in their construction.
Naturally then, the complete-wave counterparts are in demand for computing practically relevant flows that involve 
shear dominated phenomenon, mixing flows, flame fronts, material interfaces etc \cite{vanleer1987}.

Unfortunately, the wide employablility of contact-shear preserving approximate Riemann solver, including the complete-wave schemes from the HLL family, are 
hampered due to the presence of various forms of numerical shock instability that afflict them. 
The occurrence of the instability are also known to be 
highly sensitive to parameters like grid aspect ratio\cite{henderson2007}, inflow Mach number\cite{dumbser2004}, 
numerical shock structure \cite{xie2017,dumbser2004,chauvat2005} and order of accuracy of the solution \cite{gressier2000}. 
The Carbuncle phenomenon \cite{peery1988}, the moving shock instability\cite{quirk1994}, 
the standing shock instability\cite{sanders1998}, 
the kinked Mach stems\cite{gressier2000} etc are some common manifestations of the instability. 
A general understanding is that the instability is essentially a multidimensional phenomena occurring due to the inability of these contact-shear preserving schemes
in providing adequate crossflow dissipation to dampen numerical perturbations in physical quantities near the shock front \cite{sanders1998,shen2014,pandolfi2001}. 
It is increasingly suspected that antidiffusive terms in their numerical viscosity, that are responsible for accuracy on the linear waves, could be a major 
factor in triggering the instability \cite{park2003,shen2014,pandolfi2001}. Some authors show that merely controlling the antidiffusive terms corresponding to 
the shear wave, wherever they could be explicitly identified (for eg. in Roe scheme and HLLEM scheme), could guarantee shock stability \cite{xie2017,farhad2006} although, 
such observations could benefit from further confirmation. 
However, injudicious control of the antidiffusive terms with an aim to cure shock instability 
could compromise the overall accuracy of the scheme on viscous problems especially by smearing the contact and shear wave itself. 
In this regard, Gressier et al \cite{gressier2000} conjectures that a numerical scheme cannot be both shock stable and contact-shear preserving. 
This contradictory requirement in terms of dissipation characteristics of a robust and accurate
Riemann solver have opened up opportunities for research.

Almost all of the cures proposed towards improving the robustness of schemes against shock instability are based upon
improving the numerical dissipation provided by them in the vicinity of shocks. 
For example, several authors have demonstrated that Roe scheme can be cured by providing supplementary numerical dissipation 
through suitable modification of the 
eigenvalues of its dissipation matrix \cite{peery1988,sanders1998,pandolfi2001,lin1995}. Another interesting approach has been to develop rotated Riemann 
solvers \cite{ren2003,nishikawa2008,zhang2016,huang2011}. In such schemes the extra numerical dissipation is achieved
either by adaptively employing a contact-shear dissipative scheme across a shock or through the rotation mechanism itself.
Among various cures proposed heretofore, the idea of hybridizing complementary Riemann solvers, pioneered by Quirk \cite{quirk1994}, seems
to have gathered much attention. Following this idea several other  
hybrid Riemann solvers like Roe-HLLE \cite{quirk1994,nishikawa2008,wu2010,dongfang2016}, HLLC-HLL \cite{shen2016,huang2011,kim2009,kim2010}
and AUSM based schemes \cite{zhang2017,kim1998,wada1997} have come into existence. Each of these hybrid fluxes combine a contact-shear preserving scheme with a complementary 
contact-shear diffusive one through a carefully designed switching sensor that 
automatically engages the appropriate scheme depending on the flow conditions encountered. Thus, while the diffusive scheme is engaged near shock waves to suppress the
instability, the accurate scheme is used to achieve resolution of linear-wave phenomena.
Some hybrid fluxes \cite{quirk1994} chose to engage its diffusive component everywhere in the vicinity of the shock to deal with the instability while inadvertently
smearing the discontinuity itself in the process. Experiments show that on the contrary, it may be desirable to engage the dissipative 
fluxes only along the shock front and not necessarily across it to deal with the instability \cite{xie2017,shen2014}. Some hybrid fluxes 
\cite{quirk1994,wu2010,kim1998,zoltak1998} that require complete evaluation of both flux components may suffer the disadvantage of being computationally expensive 
to implement. In such cases, careful choice of schemes that share complementary dissipation characteristics which allows for a unified hybrid framework may prove
beneficial \cite{shen2016,dongfang2016,zhang2017}. 

Another strategy to reduce cost of hybrid fluxes could be to identify and hybridize only certain flux components that may be critical to the 
instability phenomenon. However several opinions exist regarding which of the flux components markedly influence the instability behavior of a scheme.
Liou\cite{liou2000} identifies the massflux discretization alone to be critical to shock stability. He recommends using a scheme that does not introduce pressure terms
into massflux discretization to ensure shock stability.
Wu et al\cite{wu2010} hybridizes both the mass and interface-normal flux component to achieve shock stability.
Shen et al\cite{shen2014} on the other hand recommends hybridizing only the interface-normal flux component on interfaces that 
are orthogonal to the shock front to avoid shock instability while Zhang et al\cite{zhang2017} employs hybridization only on the momentum components. 
Clearly, more studies are necessary not only to rightly identify the 
critical flux components but also to explain how their discretizations trigger instability.  
Apart from the type of Riemann solvers used to constitute a hybrid scheme, the 
choice of the switching sensor also plays a vital role in the overall effectiveness of the framework. This is because abrupt switching between
fluxes are known to hamper convergence \cite{zhang2017}.

In this paper a new strategy for curing shock instabilities in the HLLC scheme is proposed. 
Firstly, we rewrite the HLLC scheme into an alternative form wherein its inherent diffusive HLL component and the antidiffusive 
component responsible for its accuracy on contact and shear waves can be clearly distinguished. 
Secondly, we perform some linear analyses and numerical experiments on this alternative HLLC form to identify the most critical flux components that affect shock 
instability behavior of the HLLC scheme. Through a linear scale analysis of a numerically perturbed shock, we probe the role of the antidiffusive terms in these critical flux components in triggering
the instability. Motivated by this, we formulate a shock stable version of the HLLC scheme called 
HLLC-ADC (Anti Diffusion Control) which uses a smooth multidimensional pressure ratio based shock sensor to achieve the requisite antidissipation control on these
critical flux components. This scheme can be thought of as a hybrid scheme that takes advantage of the proposed alternative HLLC form to switch efficiently 
between the HLLC and the HLLE schemes without necessitating an explicit computation of each flux seperately. We employ this hybridization only on the 
critical flux components identified earlier. Also, since the HLLC and the HLLE scheme are both derived from the HLL-family of
schemes and inherit identical properties in terms of shock capturing, they are excellent choice as complementary schemes to constitute an hybrid flux.
These properties make the proposed scheme computationaly cheaper and easier to implement.
To establish the robustness of the scheme against shock instabilities we perform a linear perturbation analysis \cite{quirk1994,pandolfi2001} of it 
to study its damping characteristics followed by a matrix based stability analysis \cite{dumbser2004} over a wide range of Mach numbers. 
Lastly, a carefully chosen suite of numerical test problems are used to demonstrate the efficacy of the proposed scheme.

The outline of the paper is as follows. In Sections \ref{sec:governingequations} and \ref{sec:FVM_framework} we briefly outline the governing equation and 
the Finite Volume framework used in the present work.  In Section \ref{sec:reviewofHLLCandHLL} we review the HLL and the HLLC Riemann solvers.
In Section \ref{sec:linearanalysisofhllcandhll} we first introduce the alternate form of the HLLC scheme that forms the basis of our proposed scheme and employ it to 
uncover the critical flux components that affects the shock instability characteristics of the HLLC scheme using a linear perturbation analysis and a matrix based
stability analysis.
In Section \ref{sec:orderofmagnitudeanalysis} we analyze the dissipation characteristics of these critical flux components of the HLLC scheme 
in comparison to that of the HLLE scheme  to identify the role of antidiffusion terms in 
proliferation of instabilities. In Section \ref{sec:formulation} we give details for constructing a shock stable version of the HLLC scheme called the HLLC-ADC scheme and 
use the linear perturbation method to investigate its perturbation damping characteristics. In Section \ref{sec:matrixanalysis} we perform a matrix based stability analysis of the
HLLC-ADC scheme. In Section \ref{sec:numericaltests} we present the results for various numerical test problems. Some concluding remarks are given in Section \ref{sec:conclusions}.

\section{Governing equations}
\label{sec:governingequations}
The governing equations for two dimensional inviscid compressible flow can be expressed in their conservative form as, 
\begin{align}
 \frac{\partial \mathbf{\acute{U}}}{\partial t} + \frac{\partial \mathbf{\acute{F}(U)}}{\partial x} + \frac{\partial \mathbf{\acute{G}(U)}}{\partial y} = 0
 \label{equ:EE-differentialform}
\end{align}
where $\mathbf{\acute{U}}, \mathbf{\acute{F}(U)} , \mathbf{\acute{G}(U)}$ are the vector of conserved variables and x and y directional fluxes respectively. These are 
given by, 

\begin{align}
 \mathbf{\acute{U}} = \left [ \begin{array}{c}
\rho \\
\rho u\\
\rho v\\
\rho E
\end{array}\right ],
\mathbf{\acute{F}(U)} = \left [ \begin{array}{c}
\rho u \\
\rho u^2 + p\\
\rho u v\\
(\rho E + p)u
\end{array}\right ],
\mathbf{\acute{G}(U)}= \left [ \begin{array}{c}
\rho v \\
\rho u v\\
\rho v^2 + p\\
(\rho E + p)v
\end{array}\right ]
\end{align}
In the above expressions, $\rho, u, v, p$ and $E$ stands for density, x-velocity, y-velocity, pressure and specific total energy. The system of 
equations are closed through the equation of state,
\begin{align}
 p = (\gamma-1)\left(\rho E - \frac{1}{2}\rho(u^2 + v^2)\right)
\end{align}
where $\gamma$ is the ratio of specific heats. Present work assumes a calorifically perfect gas with $\gamma =1.4$. A particularly useful form of the 
Eq.(\ref{equ:EE-differentialform}) is the integral form given by,

\begin{align}
 \frac{\partial}{\partial t} \int_{\varOmega} \mathbf{\acute{U}} dxdy + \oint_{d\varOmega} [(\mathbf{\acute{F},\acute{G}}).\mathbf{n}] dl = 0 
 \label{eqn:EE-integralform}
\end{align}
where $\varOmega$ denotes a control volume over which Eq.(\ref{eqn:EE-integralform}) describes a Finite Volume balance of the conserved
quantities, $dx$ and $dy$
denotes the x and y dimensions of the control volume respectively, $d\varOmega$ denotes the boundary 
surface of the control volume and $dl$ denotes an infinitesimally small element on $d\varOmega$. $\mathbf{n}$ is the outward pointing unit normal vector to the surface $d\varOmega$. 

\section{Finite volume discretization}
\label{sec:FVM_framework}
\begin{figure}[H]
	    \centering
	    \includegraphics[scale=0.3]{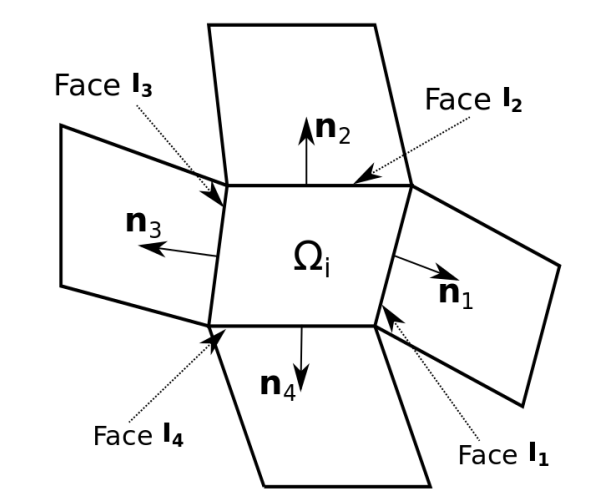}
	    \caption{Typical control volume $i$ with its associated interfaces $I_k$ and respective normal vectors $\mathbf{n}_k$.}
	    \label{fig:finitevolume}
	    \end{figure}

In this paper we seek a Finite Volume numerical solution of Eq.(\ref{eqn:EE-integralform}) by discretizing the equation on a computational 
mesh consisting of structured quadrilateral cells as shown in Fig.(\ref{fig:finitevolume}). For a typical cell $i$ belonging to this mesh, a 
semi-discretized version
of such a solution can be written as,

\begin{align}
 \frac{d\mathbf{U}_i}{dt} = -\frac{1}{|\varOmega|_i} \sum_{k=1}^{4} [({\mathbf{\acute{F},\acute{G}}})_k.\mathbf{n}_k] \Delta s_k
 \label{eqn:EE-partialdiscretized}
\end{align}
where ${\mathbf{U}_i}$ is an appropriate cell averaged conserved state vector, $({\mathbf{\acute{F},\acute{G}}})_k$ denotes the flux vector at the 
mid point of each interface $I_k$ while $\mathbf{n}_k$ and $\Delta s_k$ denotes the unit normal vector  and the length of each $I_k$ interface 
respectively. These are shown in Fig.(\ref{fig:finitevolume}). The interface flux $({\mathbf{\acute{F},\acute{G}}})_k.\mathbf{n}_k$ can be obtained by various methods. 
One of the most popular class of methods 
to compute this are the approximate Riemann solvers. A conventional two state approximate Riemann solver uses the 
rotational invariance property of Euler equations to express the term $({\mathbf{\acute{F},\acute{G}}})_k.\mathbf{n}_k$ as, 

\begin{align}
 \frac{d\mathbf{U}_i}{dt} = -\frac{1}{|\varOmega|_i} \sum_{k,m=1}^{4} [\mathbf{T}^{-1}_{k} \mathbf{F} (\mathbf{U_L},\mathbf{U_R})]\Delta s_k
\label{eqn:EE-rotationalinvariance}
 \end{align}
 where $ \mathbf{U_L} = \mathbf{T}_{k}(\mathbf{U}_i), \mathbf{U_R}=\mathbf{T}_{k}(\mathbf{U}_m)$ indicates the initial conditions of a local
Riemann problem across $k^{th}$ interface constituted by cells $i$ and $m$. The matrices $\mathbf{T}_{k}$ and $\mathbf{T}_{k}^{-1}$ are rotation matrices at the $k^{th}$ interface given by,

\begin{align}
 \mathbf{T}_k= \left [ \begin{array}{r}
1\\
0\\
0\\
0
\end{array}
\begin{array}{r}
0\\
n_{xk}\\
-n_{yk}\\
0
\end{array}
\begin{array}{r}
0\\
n_{yk}\\
n_{xk}\\
0 
\end{array}
\begin{array}{r}
0\\
0\\
0\\
1 
\end{array}\right ],\mathbf{T}^{-1}_k= \left [ \begin{array}{r}
1\\
0\\
0\\
0
\end{array}
\begin{array}{r}
0\\
n_{xk}\\
n_{yk}\\
0
\end{array}
\begin{array}{r}
0\\
-n_{yk}\\
n_{xk}\\
0 
\end{array}
\begin{array}{r}
0\\
0\\
0\\
1 
\end{array}\right ] 
\end{align}
where $n_{xk},n_{yk}$ denote the components of the normal vector $\mathbf{n}$.

\noindent In the next section we briefly describe two approximate Riemann solvers named the HLLE scheme and the HLLC scheme that can be used to estimate the flux 
$\mathbf{F} (\mathbf{U_L},\mathbf{U_R})$
at a given interface. To avoid extra notations, henceforth $\mathbf{F}$ simply represent the local Riemann flux at any interface with outward pointing normal $\mathbf{n}$.

\section{Recap of the HLLE and the HLLC schemes}
\label{sec:reviewofHLLCandHLL}
Amongst various approximate Riemann solvers that exist in the literature, the HLLE and the HLLC schemes have gained popularity because of their simplicity and accuracy. 
In this work, we intend to closely study the numerical dissipation features of these schemes and their relation to the shock instability characteristics exhibited by them. 
To this end, we present a brief review of the HLLE and the HLLC schemes below.

\subsection{The HLLE scheme}
The original HLL scheme was devised by Harten, Lax and van Leer \cite{hll1983}. It assumes a wave structure consisting of two waves that seperates three constant states.
Using the integral form of the conservation laws on this wave structure, the average HLL state enclosed by these waves can be written as,

\begin{align}
  \mathbf{U}_{*}^{HLL}  &= \frac{S_R \mathbf{U}_R - S_L \mathbf{U}_L + \mathbf{F}_L - \mathbf{F}_R }{S_R - S_L}
  \label{eqn:HLLstate}
\end{align}
The corresponding HLL Riemann flux can then be written as, 

\begin{align}
    \mathbf{F}_{HLL}= \frac{S_R \mathbf{F}_L - S_L \mathbf{F}_R + S_LS_R (\mathbf{U}_R - \mathbf{U}_L) }{S_R - S_L}
\label{eqn:hllformualtion}
\end{align}
Here $\mathbf{F}_L = \mathbf{F(\mathbf{U}_L)}$
and $\mathbf{F}_R = \mathbf{F(\mathbf{U}_R)}$ are the exact local fluxes at either side of the interface.
$S_L$ and $S_R$ are numerical approximations to the speeds of the left most and right most running characteristics that emerge as the solution of 
the Riemann problem at an interface.
It has been shown that under appropriate choice of wavespeeds $S_L$ and $S_R$, the HLL scheme is both positivity preserving and entropy
satisfying \cite{einfeldt1988}. This choice of wavespeeds are given as, 

\begin{align}
\nonumber
 S_L = min(0,u_{nL}-a_L, \tilde{u}_n-\tilde{a})\\
 S_R = max(0,u_{nR}+a_R, \tilde{u}_n +\tilde{a})
 \label{eqn:HLLEwavespeedestimate}
\end{align}
where $u_{nL,R}$ are the normal velocities across an interface, $a_{L,R}$ are the respective sonic speeds and $\tilde{u}_n,\tilde{a}$ are the standard Roe averaged 
quantities at the interface \cite{roe1981}. Using these wavespeed estimates, the HLL scheme is also known as the 
HLLE scheme \cite{einfeldt1988}.

Although quite accurate in resolution of nonlinear waves, a major drawback of the HLLE scheme is its inability to 
exactly resolve the contact and shear waves. The loss of accuracy on these waves occur because of the assumption of constant average state between the two wave
structure. Since exact contact ability is a prerequisite for accurate resolution of viscous 
phenomenon like boundary layer flows \cite{vanleer1987,toro_vaz2012}, the HLLE solver is not a popular choice for simulating flows with such features. 
However, it has been observed that the HLLE scheme is free from various forms of numerical shock instabilities \cite{pandolfi2001}. Gressier et al \cite{gressier2000} have 
conjectured that upwind schemes that possess exact contact ability cannot be free from shock instabilities. Hence by corollary, the shock stable behaviour of the HLLE scheme 
is attributed to its dissipative nature on contact and shear waves. Later in this paper, we provide insights into how the numerical dissipative characteristics of the 
HLLE scheme results in its shock stable behaviour.

\subsection{The HLLC scheme}

The inability of HLLE solver to resolve contact and shear waves was mitigated through the development of the HLLC Riemann solver ( C for Contact) by Toro et al \cite{toro1994}. 
This improvement was achieved by adding a third wave called a contact wave with speed $S_M$ to the pre-existing two wave HLL structure. 
Using the integral form of the conservation laws, closed form expressions for the conserved quantities in the states across the additional wave $S_M$ can be derived as, 
\begin{align}
    \mathbf{U}_{*L/R}^{HLLC} &= \rho_{L/R} \left(\frac{S_{L/R}-u_{nL/R}}{S_{L/R}-S_M}\right) \left( \begin{array}{c}
                                                                              1\\
                                                                              S_M\\
                                                                              u_{tL/R}\\
                                                                              \frac{(\rho E)_{L/R}}{\rho_{L/R}} + (S_M-u_{nL/R})(S_M + \frac{p_{L/R}}{\rho_{L/R}(S_{L/R}-u_{nL/R})})                                                        
                                                                              \end{array} \right)\\
    \label{eqn:HLLCstate}
\end{align}
where, apart from the already defined variables, $u_{tL/R}$ denote the tangential velocities across an interface.
Based on this the HLLC interface flux $\mathbf{F}_{HLLC}$ can be written as, 

\begin{align}
    \mathbf{F}_{HLLC}= 
    \begin{cases}
    \mathbf{F}_L + S_L(\mathbf{U}_{*L}^{HLLC} - \mathbf{U}_L),& \text{if } S_L\leq 0 \leq S_M\\
    \mathbf{F}_R + S_R(\mathbf{U}_{*R}^{HLLC} - \mathbf{U}_R),& \text{if } S_M \leq 0 \leq S_R\\
    \end{cases}
\label{eqn:hllcformulation}
\end{align}
In the above expressions, $S_L$ and $S_R$ can be obtained using Eq.(\ref{eqn:HLLEwavespeedestimate}). Batten et al \cite{batten1997} provides a
closed form expression for $S_M$ as,
\begin{align}
 \label{eqn:HLLCmiddlewaveestimate}
  S_M = \frac{p_R - p_L + \rho_Lu_{nL}(S_L - u_{nL}) - \rho_Ru_{nR}(S_R - u_{nR})}{\rho_L(S_L - u_{nL}) -\rho_R(S_R - u_{nR})}
\end{align}
The HLLC scheme is one of the simplest known Riemann solver to be able to resolve both linearly degenerate and genuinely nonlinear wavefields accurately.
Like the HLLE scheme, it is also an entropy satisfying and positively conservative scheme under appropriate wave speed selection \cite{toro1994,batten1997}.
However, albeit all the attractive features the scheme
posesses, the HLLC scheme is known to be highly susceptible to numerical shock instabilities. As observed by some authors \cite{shen2016,gressier2000,pandolfi2001}
this undesirable behaviour can be directly attributed to its accuracy on contact and shear waves. Later on in this paper, we will compare the numerical dissipation of the HLLC scheme
to that of the HLLE scheme in order to identify the terms within the HLLC scheme that makes it vulnerable to shock instabilities. 

Before we proceed to develop a shock stable HLLC scheme, it is imperative to understand how the HLLC scheme tend 
to develop shock 
unstable solutions while the HLLE scheme is capable of producing shock stable solution. 
In the following sections we utilize two popular linear perturbation analysis methods to study the shock instability characteristics of these schemes. 

\section{Shock instability characteristics of the HLLE and the HLLC schemes}
\label{sec:linearanalysisofhllcandhll}

As mentioned in the last section, the HLLC scheme is accurate on contact and shear waves but is plagued by shock instabilities
while the HLLE scheme is dissipative on these waves and possess strong robustness towards shock instabilities. 
A common strategy to harness the advantages of both these schemes is to hybridize these fluxes using a switch between them that engages the appropriate scheme
based on the flow conditions encountered \cite{huang2011,kim2009}. For example, in the vicinity of a strong normal shock, the dissipative HLLE scheme is engaged to 
keep the overall scheme shock stable while in the vicinity of weaker shocks, contacts and shear wave phenomenon, the HLLC scheme is used for its accuracy. 
Unfortunately, as pointed out by some authors \cite{shen2016,zhang2017}, 
an explicit combination of these schemes is not only expensive to implement but may also
result in convergence issues due to abrupt switching. 
Hence, an alternative strategy to combine these schemes without resorting to explicit hybridization is necessary.
In this regard, we propose to express the HLLC scheme as, 
\begin{align}
    \mathbf{F}_{HLLC}= 
\begin{cases}
    \mathbf{F}_{HLL} + S_L(\mathbf{U}_{*L}^{HLLC} - \mathbf{U}_{*}^{HLL}),& \text{if } S_L\leq 0 \leq S_M\\
    \mathbf{F}_{HLL} + S_R(\mathbf{U}_{*R}^{HLLC} - \mathbf{U}_{*}^{HLL}),& \text{if } S_M \leq 0 \leq S_R\\
\end{cases}
\label{eqn:hllc-modifiedform}
\end{align}
Note that expressing the HLLC scheme in this form allows for precise identification of its inherent diffusive HLL component and the antidiffusive component (represented by
terms $S_{L/R}(\mathbf{U}_{*L/R}^{HLLC} - \mathbf{U}_{*}^{HLL})$)
that helps restore contact and shear ability to this embedded HLL component. It is important to note that explicit identification of the numerical dissipation corresponding 
to the contact and the shear waves can be distinguished in case of Roe scheme and HLLEM scheme whereas, it is not possible in case of the HLLC scheme. In the present 
framework represented in Eq.(\ref{eqn:hllc-modifiedform}) we have managed to isolate the numerical dissipation terms corresponding to these linearly degenerate wavefields. 
Hence this provides an alternative framework to quickly switch between the highly accurate HLLC scheme and the dissipative HLLE scheme. 
This can be achieved by introducing a switching parameter $\omega$ as,

\begin{align}
    \mathbf{F}_{HLLE/HLLC}= 
\begin{cases}
    \mathbf{F}_{HLL} + \omega S_L(\mathbf{U}_{*L}^{HLLC} - \mathbf{U}_{*}^{HLL}),& \text{if } S_L\leq 0 \leq S_M\\
    \mathbf{F}_{HLL} + \omega S_R(\mathbf{U}_{*R}^{HLLC} - \mathbf{U}_{*}^{HLL}),& \text{if } S_M \leq 0 \leq S_R\\
\end{cases}
\label{eqn:hllc-modifiedformwithomega}
\end{align}
Notice that the $\omega$ is employed only on the antidiffusive component. When $\omega=0$, the contribution from the antidiffusive component is removed and 
we obtain the HLLE scheme. On the other hand when $\omega=1$, we recover the whole HLLC scheme. The aim of this paper is to utilize the combined framework presented 
above to develop an accurate and robust HLLC scheme that is shock stable.
In the next section, we utilize this combined framework to investigate shock instability characteristics
of the HLLE and the HLLC schemes using two popular linear analysis tools. 

\subsubsection{ Evolution of a saw-tooth initial profile by the HLLE and HLLC schemes}
\label{sec:quirkanalysis_hllandhllc}
To understand the instability characteristics of the HLLE and the HLLC schemes, it is instructive to study how these schemes evolve an initially perturbed profile.
This technique was pioneered by Quirk \cite{quirk1994} and discussed by others \cite{gressier2000,pandolfi2001}.
Although the technique deals only with perturbations superimposed on a steady mean flow without shocks and does not consider the effects of grid or boundary conditions, 
it neverthless provides a useful linearized temporal evolution model for these perturbations and their mutual interactions.
Based on these evolution equations, Quirk conjectured that \textit{schemes in which pressure perturbations feed into density perturbations are prone to produce shock instabilities} \cite{quirk1994}.  
The analysis is set up as follows. A steady mean flow with normalised 
state values of $\rho_0=1$, $u_0\neq0$, $v_0=0$ and $p_0=1$ is chosen as the base flow. 
Since shock instability is noticed to occur primarily along the length of the shock front, an imbalance in the fluxes in a direction parallel to the shock front is assumed to reveal its 
occurrence. Hence the analysis restricts the governing equations in Eq.(\ref{equ:EE-differentialform}) to include only the y-directional $\mathbf{\acute G}$ 
fluxes while the x-directional $\mathbf{\acute F}$ fluxes are assumed to balance out each other. The reduced
form of governing equation for the analysis is,
  \begin{align}
  \frac{\partial \mathbf{\acute U}}{\partial t} + \frac{\partial \mathbf{\acute {G}(U)}}{\partial y} \approx 0
  \label{equ:EE-reduceddifferentialform}
  \end{align}
On a typical y-directional stencil chosen anywhere in the computational domain, the cells are marked as \textquotedblleft even \textquotedblright and \textquotedblleft odd \textquotedblright \cite{pandolfi2001}.
The perturbations are introduced as a saw tooth profile in density, 
x-velocity and pressure variables. For a typical cell 'j', these are initialized as,
\begin{align}
\begin{cases}
    \rho_j=\rho_0+\hat{\rho}, \;\;p_j=p_0+\hat{p},\;\;u_j=u_0+\hat{u},\;\;v_j=0 ,& \text{if } j\;\; is\;\; even\\
    \rho_j=\rho_0-\hat{\rho}, \;\;p_j=p_0-\hat{p},\;\;u_j=u_0-\hat{u},\;\;v_j=0 ,& \text{if } j\;\; is\;\; odd
\end{cases}
\label{eqn:sawtoothinitialconditions}
\end{align}
Such profiles are typical of most shock unstable solutions. The crux of the analysis is to develop equations that describe the 
temporal evolution of the saw tooth perturbations described in Eq.(\ref{eqn:sawtoothinitialconditions}).  
Note that we avoid introducing perturbations $\hat{v}$ in y-directional velocity in the present analysis
because it has been reported that effect of this component is to dampen out perturbations in other components eventually stabilizing the scheme \cite{shen2014,pandolfi2001}. 
Had it been included, then isolating the effect of the numerical scheme on the evolution equations would have been difficult.

\begin{figure}[H]
	    \setcounter{subfigure}{0}
	    \subfloat[ $\mathbf{HLLE}$ $\boldsymbol{\left(\hat{\rho} = 0.01, \hat{u} = 0, \hat{p} = 0\right)} $ ]{\label{fig:HLLE_density_perturbation}\includegraphics[scale=0.24]{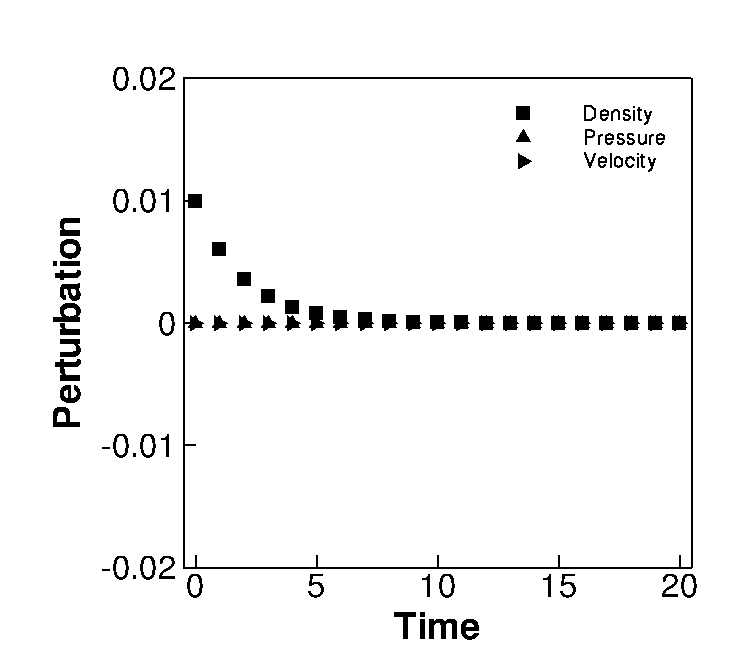}}
            \qquad
            \qquad
            \subfloat[$\mathbf{HLLC}$ $ \boldsymbol{\left(\hat{\rho} = 0.01, \hat{u} = 0, \hat{p} = 0\right) } $]{\label{fig:HLLC_density_perturbation}\includegraphics[scale=0.25]{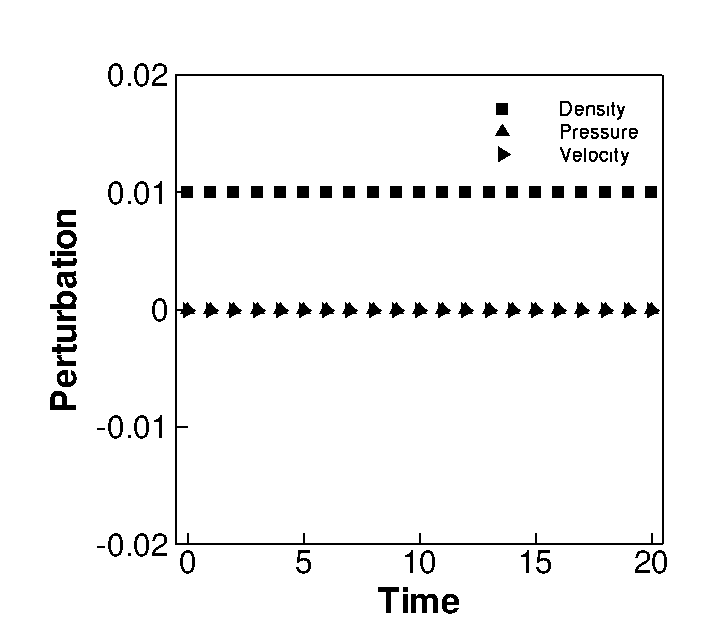}}\\
            \subfloat[$\mathbf{HLLE}$ $\boldsymbol{\left( \hat{\rho} = 0, \hat{u} = 0.01, \hat{p} = 0 \right) }$]{\label{fig:HLLE_velocity_perturbation}\includegraphics[scale=0.24]{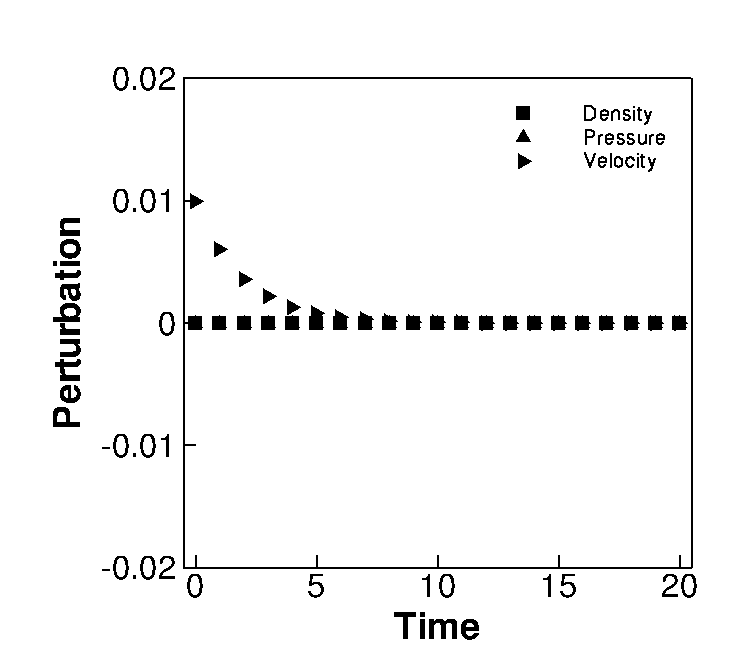}}
            \qquad
            \qquad
            \subfloat[$\mathbf{HLLC}$  $\boldsymbol{\left(\hat{\rho} = 0, \hat{u} = 0.01, \hat{p} = 0\right) }$]{\label{fig:HLLC_velocity_perturbation}\includegraphics[scale=0.25]{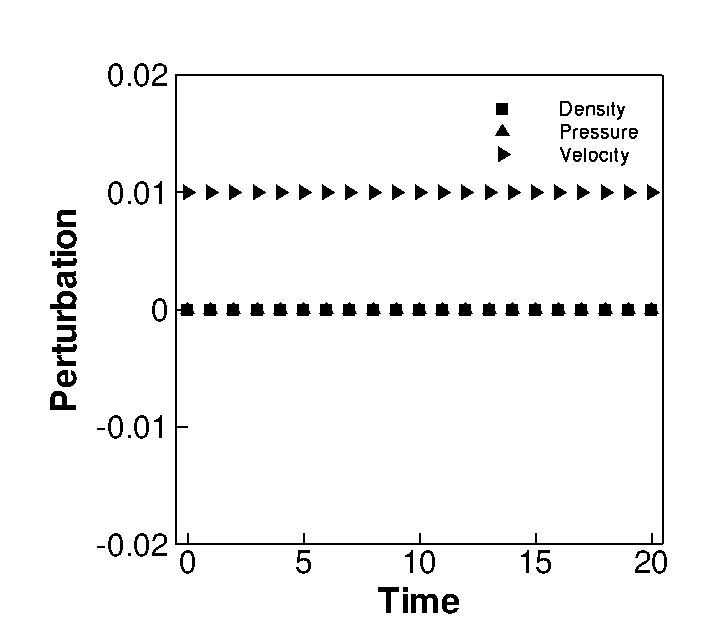}}\\
            \subfloat[$\mathbf{HLLE}$ $\boldsymbol{ \left(\hat{\rho} = 0, \hat{u} = 0, \hat{p} = 0.01\right) }$]{\label{fig:HLLE_pressure_perturbation}\includegraphics[scale=0.24]{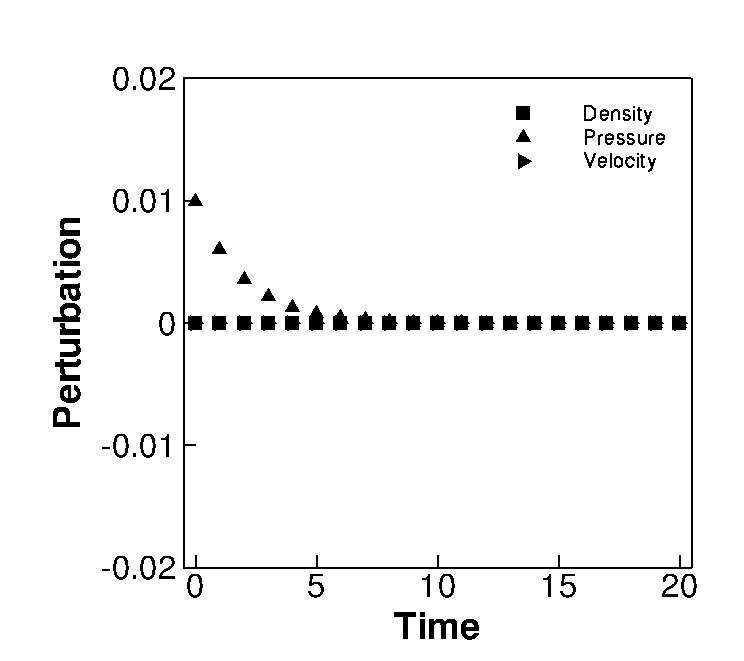}} 
	    \qquad
            \qquad
            \subfloat[$\mathbf{HLLC}$ $\boldsymbol{ \left(\hat{\rho} = 0, \hat{u} = 0, \hat{p} = 0.01\right) } $]{\label{fig:HLLC_pressure_perturbation}\includegraphics[scale=0.25]{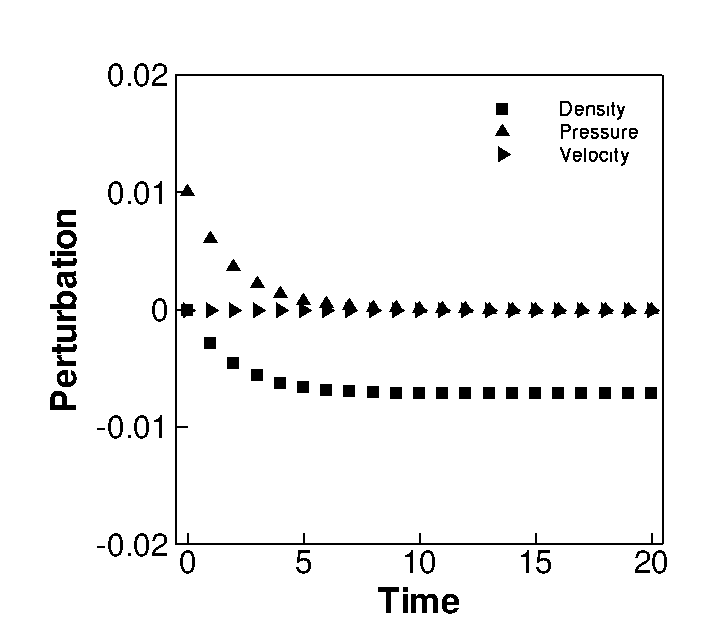}}
	    \caption{Comparison of evolution of density, x-velocity and pressure perturbations in the HLLE and the HLLC schemes.}
	    \label{fig:perturbationstudies_hllc_hll}
	    \end{figure}

\noindent For the HLLE scheme (which corresponds to setting $\omega=0$ in Eq.(\ref{eqn:hllc-modifiedformwithomega})), these equations are,
\begin{align}
 \nonumber
\hat{\rho}^{n+1}&= \hat{\rho}^n (1-2\lambda)\\\nonumber
\hat{u}^{n+1} &= \hat{u}^n (1-2\lambda)\\  
\hat{p}^{n+1}&= \hat{p}^n(1-2 \lambda)
 \label{eqn:quirk_analysis_hll}
\end{align}
Corresponding evolution equations for the HLLC scheme (setting $\omega=1$ in Eq.(\ref{eqn:hllc-modifiedformwithomega})) are,  
\begin{align}
 \nonumber
\hat{\rho}^{n+1}&= \hat{\rho}^n - \frac{2 \lambda}{\gamma}\hat{p}^n\\\nonumber
\hat{u}^{n+1} &= \hat{u}^n \\ 
\hat{p}^{n+1}&= \hat{p}^n(1-2 \lambda)
\label{eqn:quirk_analysis_hllc}
\end{align}		      
where $\lambda = \sqrt{\gamma}\frac{\Delta t}{\Delta y}$ denotes a linearized CFL value. The eigenvalues of the coefficient matrices that describe these system of equations
represent
the amplification factors of the respective perturbation quantities. While amplification factors for density, x-velocity and pressure in case of the HLLE scheme are 
$(1-2\lambda,1-2\lambda,1-2\lambda)$ respectively, those for the HLLC scheme are $(1,1,1-2\lambda)$. A von Neumann type stability bound on $\lambda$ can be obtained from these as, 
\begin{align}
  0<\lambda<1 
  \label{eqn:stability_hllc}
\end{align}
For $0<\lambda<1$, it is seen from Eq.(\ref{eqn:quirk_analysis_hll}) that any initial perturbation would be effectively damped by the HLLE scheme. Also notice that 
the HLLE scheme does not allow any mutual interactions between the perturbations.
The behaviour of these
evolution equations are plotted on the left panel of Fig.(\ref{fig:perturbationstudies_hllc_hll}). Each plot in these figures
indicate the evolution of perturbation in either density, x-velocity or pressure (indicated by $\hat{\rho}, \hat{u}, \hat{p}$ respectively) for a 
given initial perturbation in one of those quantities (while setting the other initial perturbations to be 0). These plots correspond to a value of $\lambda = 0.2$.

On the other hand, in case of the HLLC scheme, the evolution equations in Eq.(\ref{eqn:quirk_analysis_hllc}) describe a different behaviour. 
It is seen that the HLLC scheme is unable to damp the
perturbations in $\rho$ and $u$ while those in $p$ are attenuated for $0<\lambda<1$. However, the HLLC scheme also causes any finite $\hat{p}$ to be  
repeatedly fed into $\hat{\rho}$ which results in the existence of a residual $\hat{\rho}$ that is growing with time. These behaviours can be observed in the right 
panel of Fig.(\ref{fig:perturbationstudies_hllc_hll}) which also corresponds to a value of $\lambda = 0.2$.
These undamped perturbations in $\rho$ and $u$ may eventually amplify leading to unphysical mass flux 
jumps across the shock wave. Such jumps could then force a distortion of the shock profile causing the typical 'bulge' in shock structure and eventually result in shock 
unstable solutions \cite{shen2014}. Since the HLLC scheme is known to produce shock instabilites while the HLLE scheme is free from them, 
it can be inferred from these observations that Quirk's conjecture is perfectly valid for these schemes.
Gressier et al \cite{gressier2000} have observed that schemes with decaying amplification factors, termed as 
\textit{strictly stable} schemes  are known to be shock instability free while those with atleast one non-decaying amplifcation factor, 
termed as \textit{marginally stable} schemes are known to produce shock instabilities. 
By due consideration the HLLE scheme is a \textit{strictly stable} scheme while the HLLC scheme is clearly an example of a \textit{marginally stable} scheme.

\subsubsection{Numerical stability of an isolated two dimensional steady thin shock computed by the HLLE and the HLLC schemes}
\label{sec:matrixanalysis_hllandhllc}
A matrix based linear analysis proposed by Dumbser et al \cite{dumbser2004} is a useful tool to characterize the instability behaviour of a given flux function
computing the simple case of an  isolated two dimensional steady shock. The advantage of this analysis as compared to the one presented in Sec.(\ref{sec:quirkanalysis_hllandhllc})
is that it allows us to 
incorporate the features of a numerical shock (through Rankine-Hugoniot conditions), the effects of grid configuration and the boundary conditions.
The analysis aims to study the nature of the temporal evolution of random perturbations introduced into the initial conditions of the flowfield as 
computed by the flux function of interest. Using appropriate linearization of the fluxes, evolution of these errors in 
conserved quantities of all cells can be developed from Eq.(\ref{eqn:EE-rotationalinvariance}) (without neglecting the contribution from the $\mathbf{\acute F}$ flux as done
in Sec.\ref{sec:quirkanalysis_hllandhllc}). These can be arranged as a system of linear ODE as,
  \begin{align}
  \frac{d}{d t} \left(\begin{array}{r}
			\delta \mathbf{U}_1\\
			\vdots\\
			\delta \mathbf{U}_q
		      \end{array}\right )= S \left(\begin{array}{r}
						    \delta \mathbf{U}_1\\
						    \vdots\\
						    \delta \mathbf{U}_q
						      \end{array}\right)
  \end{align}
where $q=N\times M$ denotes the total number of cells in the computational mesh consisting of $N$ rows and $M$ coloums.   
Here $S$ is the stability matrix, unique for each flux function under a prescribed initial and boundary conditions for a specified grid. 
$S$ is comprised of flux Jacobians which 
represents the interactions of the errors in cells with those in its neigbours. The eigenvalues of this matrix represents the growth rate of various error modes 
of the system. According to this
analysis, a flux function is termed unstable if the stability matrix $S$ has atleast one eigenvalue $\theta$ with positive real part that indicates exponential growth 
in perturbations. 
Alternatively, the inequality,
  \begin{align}
  max(Re(\theta(S)))< 0
  \end{align}
must hold for a numerical scheme to be strictly shock stable under the given conditions. 

\begin{figure}[H]
	    \centering
	    \subfloat[HLLE]{\label{fig:hlle_matrixanalysis}\includegraphics[scale=0.25]{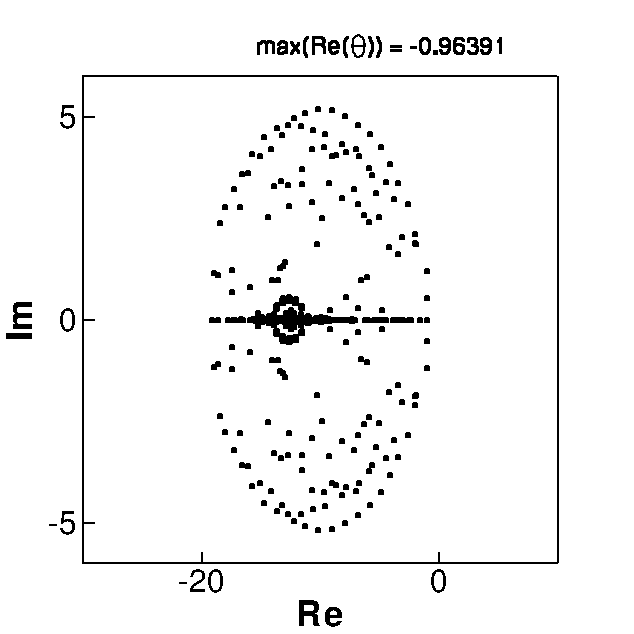}}
            \subfloat[HLLC]{\label{fig:hllc_matrixanalysis}\includegraphics[scale=0.25]{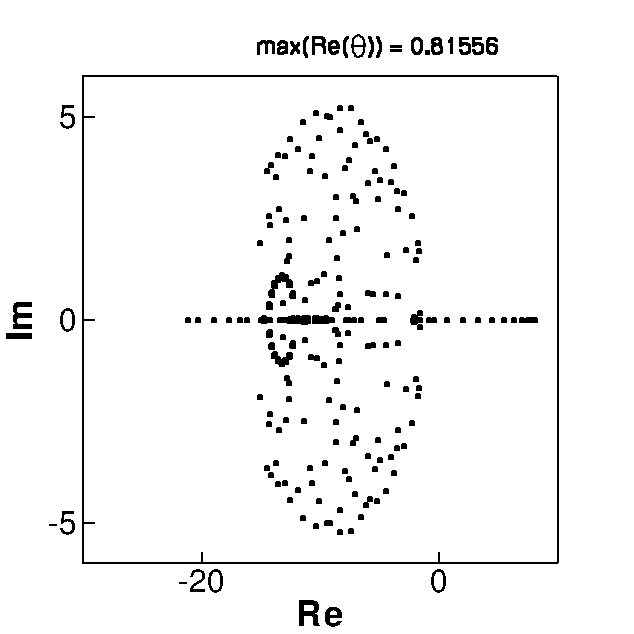}}
	   \caption{ Eigenvalue spectrum for the HLLE and HLLC scheme showing the distribution of eigenvalues in the complex plane. The 
	   spectrum correspond to a $M=7$ two dimensional isolated steady thin shock located on a $11$ by $11$ grid computed using these schemes. The $max(Re(\theta))$ which indicates
	   maximum error growth rate of the entire configuration is indicated on the top of each Figure.}
	   \label{fig:matrixanalysis_hllandhllc}
	    \end{figure}

For the present study a $M=7$ two dimensional thin stationary shock is chosen to be 
located between the 5th and 6th cells of a regular Cartesian grid comprising of 11 $\times$ 11 
cells on a 1.0$\times$
1.0 domain. The initial conditions on the supersonic side of the shock are chosen as $(\rho,u,v,p)_L=(1.0,1.0,0.0,0.01457)$ while the Rankine-Hugoniot 
conditions are used to determine the corresponding conditions on the subsonic side. The top and the bottom boundaries are maintained periodic to each other.
Random numerical perturbations of the order of $10^{-7}$ are introduced into the entire flow field to initiate the instabilities.
Since the HLLC scheme exactly preserves a thin shock, 
the initial conditions can be exactly specified \cite{dumbser2004}. The eigenvalues of this problem is obtained using the eigenvalue 
function \textit{eig()} from \textit{linalg} package in python version 2.7.6. 
The Jacobians in the $S$ matrix are computed numerically by using a centered approximation as, 
\begin{align}
 \frac{\partial F_i}{\partial U_j}\approx \frac{F_i(U_j+\Delta U_j)-F_i(U_j-\Delta U_j)}{2\Delta U_j} \/\/\ (i,j=1...4)
 \label{numerical_jacobian}
\end{align}
where the perturbation quantity $\Delta U_j$ is taken to be $10^{-6}$ \cite{dumbser2004}. The result of this analysis performed on the HLLE and the HLLC schemes are given 
in Fig.(\ref{fig:matrixanalysis_hllandhllc}). The plots show the distribution of the eigenvalues of the matrix $S$ on a complex plane, 
constituted for this problem computed using these schemes.

\begin{figure}[H]
	    \centering
	    \subfloat[HLLE]{\label{fig:hlle_standingshock}\includegraphics[scale=0.20]{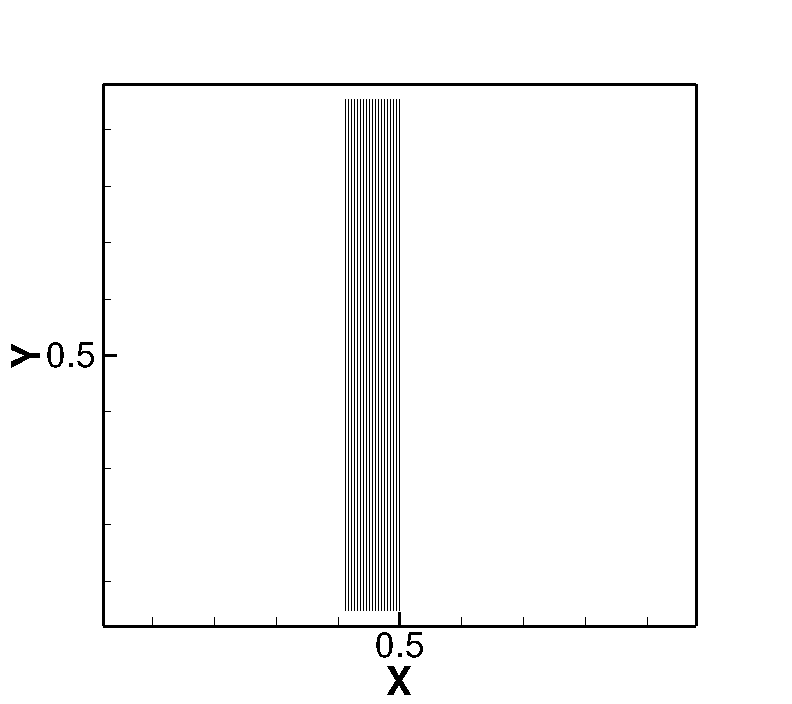}}
            \subfloat[HLLC]{\label{fig:hllc_standingshock}\includegraphics[scale=0.20]{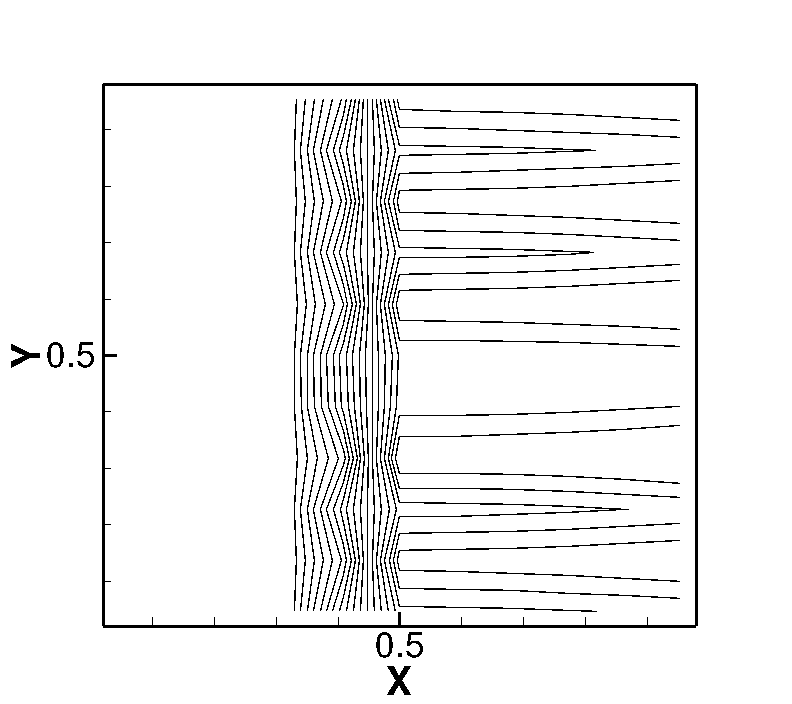}}
	   \caption{ Results at t=20 for a $M=7$ two dimensional isolated steady thin shock located on a $11$ by $11$ grid computed using the HLLE and the HLLC schemes. The HLLE
	   scheme is seen to be able to compute a stable shock profile while perturbations are visible in the profile computed by the HLLC scheme.}
	   \label{fig:standingshock_hllandhllc}
	    \end{figure}

\noindent It can be seen that the maximum linear error growth rate ($max(Re(\theta))$) for the HLLE scheme is -0.96391 which corresponds to an exponential 
decaying of initial perturbations in the flow field while that
of the HLLC scheme is +0.81556 indicating an exponential growth of these perturbations. The exponentially growing perturbations could hence result in an shock unstable 
solution as mentioned in Sec.(\ref{sec:quirkanalysis_hllandhllc}). Results from actual numerical experiments shown in Fig.(\ref{fig:standingshock_hllandhllc}) that depicts
thirty isodensity contours spanning values from 1.4 to 5.44 confirm these linear predictions.

In the next section we narrow down our investigation into the relationship between numerical dissipation in each flux component and the propensity to develop 
shock instability. 
This study helps us identify the most critical flux component that contributes to the phenomenon of shock instability and eventually direct us towards uncovering 
the cause of these spurious solutions.

\subsection{Numerical discretization of flux components and their effect on shock instability }
\label{sec:effectoftansversemomentumdissipation}

In this section, we seek to understand the influence of numerical discretization of flux components on shock instability. Identifying the flux components
that are crucial to shock instability is important in two respects. Firstly it helps us understand the origin of the instability better and secondly it helps us
apply any proposed cures only on these components which in effect leads to cheaper implementation of such fixes.
We aim to perform this study by using the contact-shear preserving HLLC scheme and its dissipative HLLE counterpart selectively on various flux components. Since 
they provide non-equal quantity of numerical dissipation we presume that their use on a critical flux component could result in dissimilar instability characteristics.
In this regard we choose to use the modified form of the HLLC scheme presented in Eq.(\ref{eqn:hllc-modifiedformwithomega}) because of its facility to 
quickly switch between these schemes by simply retaining or withdrawing the antidiffusive component using the factor $\omega$. 

\begin{figure}[H]
	    \centering
	    \includegraphics[scale=0.35]{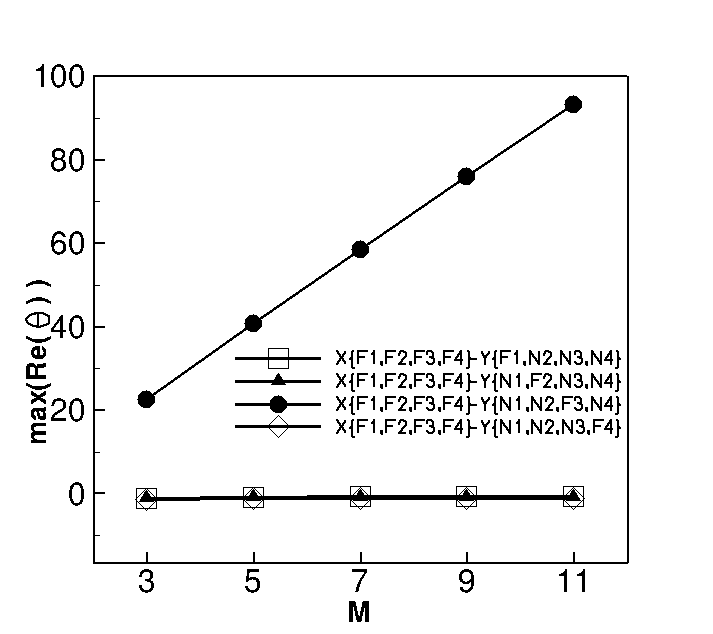}
	    \caption{Plot showing the effect of antidiffusive terms on each of the y-directional flux component for the case of a two dimensional isolated steady thin shock. 
	    $max(Re(\theta))$ is plotted for various Mach numbers.}
	    \label{fig:effectoftransverseantidissipation_hllc_hll_hybrid}
	    \end{figure}

We choose the $M=7$ isolated two dimensional steady thin shock problem on $11$ by $11$ grid discussed in Sec.(\ref{sec:matrixanalysis_hllandhllc}). 
The following conventions are adopted for this experiment. Flux component discretization of a numerical scheme is denoted 
by $X\{1,2,3,4\}-Y\{1,2,3,4\}$ where X and Y 
are the x and y-directional, interface normal, Riemann fluxes and $1,2,3,4$ denotes each of the flux components: mass, x-momentum, y-momentum and energy.
To denote the presence of antidiffusion terms or the absence of it in a particular case of the experiment, the prefix $F$ or $N$ is assigned to the numbers 
representing the flux components. 
For any flux component number, the prefix $F$ represents the presence of full antidiffusive terms ($\omega=1$) and $N$ represents the lack of it ($\omega=0$). 
For example, the notation $X\{F1,F2,F3,F4\}-Y\{N1,F2,N3,N4\}$ denotes a scheme configuration where the antidiffusive terms are applied on all flux components 
in the x-direction and
the interface-normal momentum flux component in the y-direction while the remaining flux components in this direction are discretized using the HLLE scheme.
Previous research \cite{sanders1998,shen2014} indicates that numerical dissipation along interfaces that are normal to the shock front (y-directional in this case) or not 
aligned with it, have the most significant effect on 
the instability as compared to those on interfaces that lie parallely to it. Hence we restrict our experiments only to the flux components along y-direction.
Fig.(\ref{fig:effectoftransverseantidissipation_hllc_hll_hybrid}) shows the variation of $max(Re(\theta))$ with inlet Mach number for four configurations: each case
corresponds to a flux configuration where the antidiffusive term is only retained in one of the flux component in y-direction. 

\begin{figure}[H]
	    \centering
	    \includegraphics[scale=0.20]{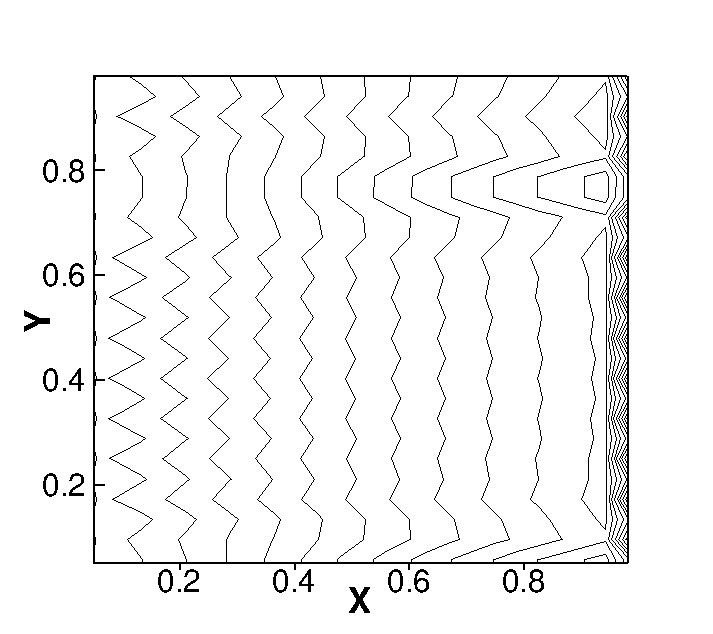}
	    \caption{Results at t=50 computed using the configuration $X\{F1,F2,F3,F4\}-Y\{N1,N2,F3,N4\}$ for a $M=7$ two dimensional isolated steady thin shock located on 
	    a grid with $26$ by $26$ cells. }
	    \label{fig:X(1F2F3F4F)_Y(1N2N3F4N)_standingshock_finergrid}
	    \end{figure}

\noindent It is interesting to note from Fig.(\ref{fig:effectoftransverseantidissipation_hllc_hll_hybrid}) that among the four cases considered it is only the configuration 
$X\{F1,F2,F3,F4\}-Y\{N1,N2,F3,N4\}$ wherein the antidiffusive term is retained in the interface-normal momentum flux in the y-direction that leads to instability. This component corresponds
to $\rho u v$ flux in the global flux vector $\mathbf{\acute{G}(U)}$ and deals with convection of the quantity $\rho u$ parallel to the shock front with a velocity $v$. 
Infact the instability in this case linearly scales with inlet Mach number confirming the criticality of this flux component. Similar observation has been made in \cite{shen2014}. 
To confirm these predictions, we resort to numerical experiments. We simulate this problem on a finer grid (a grid with $26$ by $26$ cells wherein the initial thin shock is 
now located on the interface between cells $12$ and $13$) for a configuration $X\{F1,F2,F3,F4\}-Y\{N1,N2,F3,N4\}$. The corresponding isodensity contours are shown in 
Fig.(\ref{fig:X(1F2F3F4F)_Y(1N2N3F4N)_standingshock_finergrid}). It is evident from this that using the full HLLC scheme on the interface-normal momentum component in the 
y-direction results in complete break up of the shock profile on this problem. 

\begin{figure}[H]
	    \centering
	    \subfloat[]{\label{fig:X(1F2F3F4F)_Y(1F2F3N4F)_standingshock_finergrid}\includegraphics[scale=0.20]{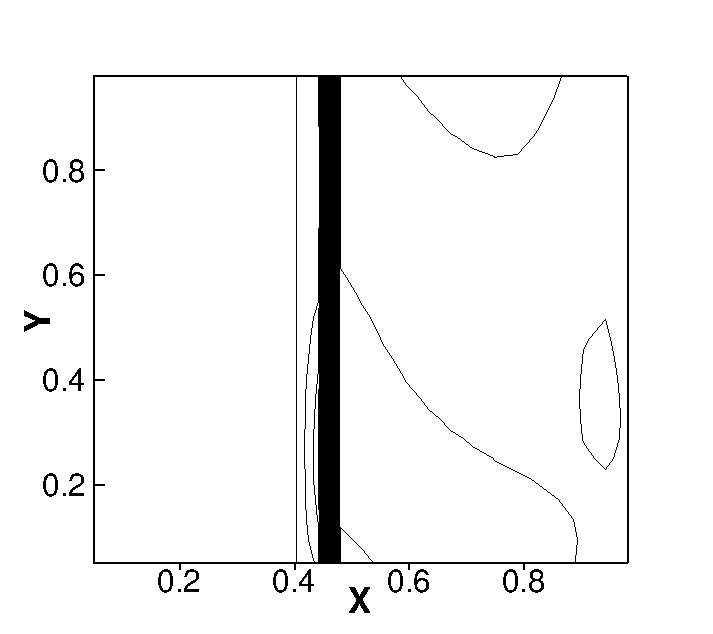}}
	    \subfloat[]{\label{fig:X(1F2F3F4F)_Y(1F2F3N4F)_movingshock}\includegraphics[scale=0.20]{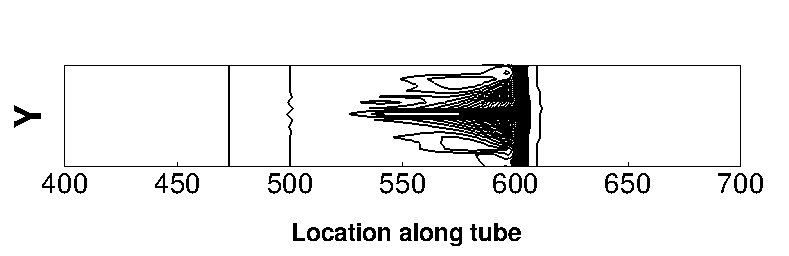}}
	   \caption{ Result for computation using configuration $X\{F1,F2,F3,F4\}-Y\{F1,F2,N3,F4\}$ (a) Results at t=50 for a $M=7$ two dimensional isolated steady 
	   thin shock located on the $26$ by $26$ grid (b) Results at t=100 for a $M=6$ moving shock problem on 800 by 20 grid. It is evident that withdrawing the antidiffusive terms
	   from the interface-normal momentum component in the y-direction alone does not guarentee shock stability.}
	   \label{fig:X(1F2F3F4F)_Y(1F2F3N4F)_instabilities}
	    \end{figure}

We also confirmed the insignificance of numerical discretization of x-directional flux components on instability by performing the above analysis and associated test case
with the x-directional fluxes now computed using the HLLE scheme. Once again (results not shown here), it was noticed that only the configuration 
$X\{N1,N2,N3,N4\}-Y\{N1,N2,F3,N4\}$ resulted in instability that scaled linearly with Mach number.	
The experiments above leads us to believe that a easy cure for shock instability can be constructed by simply withdrawing the antidiffusion terms in the numerical 
discretization of the interface-normal momentum flux in transverse direction of a shock front. To test this hypothesis we use the configuration $X\{F1,F2,F3,F4\}-Y\{F1,F2,N3,F4\}$ to compute
the $M=7$ steady thin shock problem on the $26$ by $26$ finer grid. Note that in this new configuration the HLLC scheme is used to compute 
all the x-directional flux components in addition to mass, interface-tangential momentum and energy flux components of the y-directional flux while the HLLE scheme is used to compute the 
interface-normal momentum component in this direction. The result for this experiment is shown in Fig.(\ref{fig:X(1F2F3F4F)_Y(1F2F3N4F)_standingshock_finergrid}). It is noticed from this 
result that the proposed strategy is able to successfully prevent the instability associated with this problem. To test whether such a strategy will be 
robust on other forms of instability,
we use this configuration to compute a $M=6$ moving shock problem on 800 by 20 perturbed grid reported in \cite{quirk1994}. The results for this case at t=100 is shown in 
Fig.(\ref{fig:X(1F2F3F4F)_Y(1F2F3N4F)_movingshock}). It is understood from this result that withdrawing the antidiffusive terms from the interface-normal momentum equation in the
y-direction alone may not be enough to ensure a shock stable result. This could also mean that not all forms of instabilities are triggered in the same 
fashion and that a more comprehensive understanding about their triggering mechansim is needed. We present such an understanding in Sec.(\ref{sec:orderofmagnitudeanalysis}).
	     
\begin{figure}[H]
	    \centering
	    \includegraphics[scale=0.35]{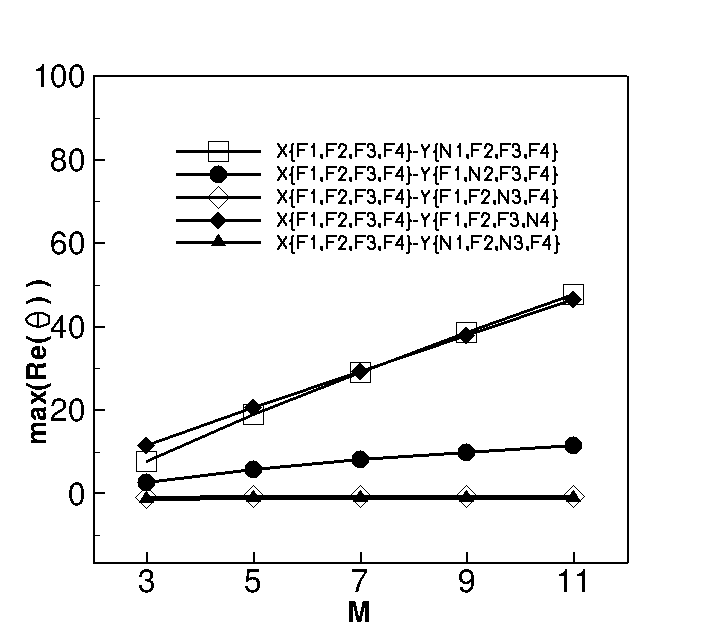}
	    \caption{Plot showing the effect of withdrawing antidiffusive terms on each of the y-directional flux component for the case of a two dimensional isolated 
	    steady thin shock. 
	    $max(Re(\theta))$ is plotted for various Mach numbers.}
	    \label{fig:effectofhllontransversefluxes_hllc_hll_hybrid}
	    \end{figure}

Since dissipative treatment of the interface-normal momentum component in the y-direction is clearly not enough, it is imperative to understand which other flux components must be
treated to ensure shock stability. To achieve this, we use the matrix stability analysis of two dimensional isolated steady thin shock problem on $11$ by $11$ grid. 
However, this time
instead of selectively adding antidiffusive terms as in experiments above, we choose to remove them from each component and study its effect on stability. 
Fig.(\ref{fig:effectofhllontransversefluxes_hllc_hll_hybrid}) shows the variation of $max(Re(\theta))$ with inlet Mach number for these configurations.

\begin{figure}[H]
	    \centering
	    \subfloat[]{\label{fig:X(1F2F3F4F)_Y(1N2F3N4F)_standingshock_finergrid}\includegraphics[scale=0.20]{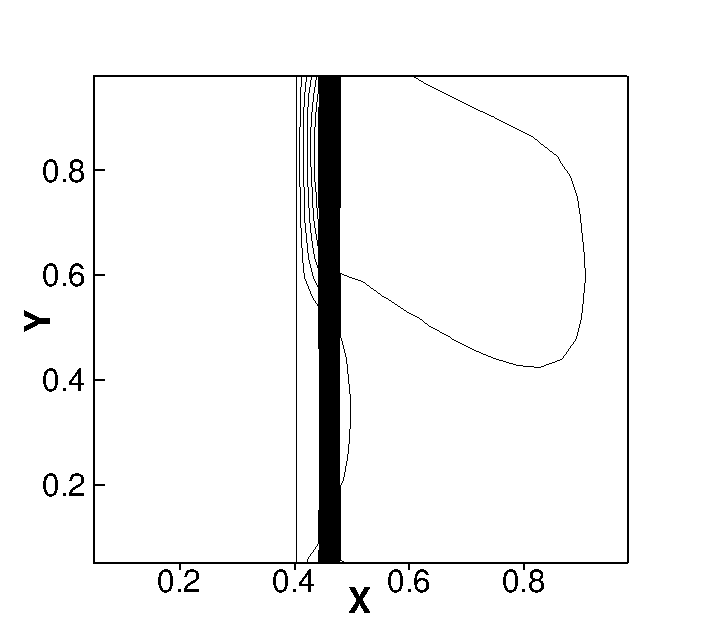}}
            \subfloat[]{\label{fig:X(1F2F3F4F)_Y(1N2F3N4F)_movingshock}\includegraphics[scale=0.20]{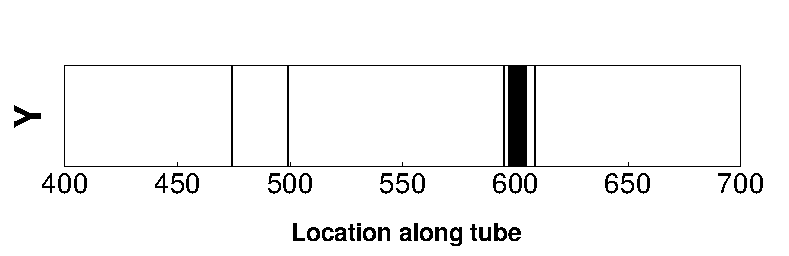}}
	   \caption{ Result for computation using configuration $X\{F1,F2,F3,F4\}-Y\{N1,F2,N3,F4\}$ (a) Results at t=50 for a $M=7$ two dimensional isolated steady 
	   thin shock located on a $26$ by $26$ grid (b) Results at t=100 for a $M=6$ moving shock problem on 800 by 20 grid. It is seen that this configuration suffices to 
	   ensure shock stability on both problems.}
	   \label{fig:X(1F2F3F4F)_Y(1N2F3N4F)_instabilities}
	    \end{figure}

Fig.(\ref{fig:effectofhllontransversefluxes_hllc_hll_hybrid}) reveals some interesting facts. Firstly, withdrawing of the antidiffusive terms from the mass flux component, 
which is achieved through $X\{F1,F2,F3,F4\}-Y\{N1,F2,F3,F4\}$ configuration, alone does not guarantee shock stability on this problem. However in comparison to the 
extent of instability (measured by $max(Re(\theta))$ at each Mach number) observed in Fig.(\ref{fig:effectoftransverseantidissipation_hllc_hll_hybrid}) when using 
configuration $X\{F1,F2,F3,F4\}-Y\{N1,N2,F3,N4\}$, 
dissipative treatment of the mass flux component reduces the growth rate of unstable modes at each Mach number.
Additionally, note that $X\{F1,F2,F3,F4\}-Y\{N1,F2,F3,F4\}$ configuration essentially deploys the HLLE scheme on the mass flux. 
It has been claimed in \cite{liou2000} that since the HLLE scheme does not have a 
pressure term in its mass flux discretization, it would be shock stable. Clearly then, the result shown here demonstrates that such a 
conclusion need not be always true and needs further refinement. 
Secondly, it is noticed that increasing the dissipation in the interface-normal momentum component in the y-direction does aid in stability on this problem. 
This was already confirmed by the result in 
Fig.(\ref{fig:X(1F2F3F4F)_Y(1F2F3N4F)_standingshock_finergrid}). Importantly, it is seen that an enhanced complete theoretical stability on all Mach numbers considered
can be guaranteed by the configuration $X\{F1,F2,F3,F4\}-Y\{N1,F2,N3,F4\}$
wherein the antidiffusive terms are withdrawn from both the mass and the interface-normal momentum component in the y-direction. Results for numerical experiments confirming this 
prediction is shown in Fig.(\ref{fig:X(1F2F3F4F)_Y(1N2F3N4F)_instabilities}) where it is seen that this configuration is capable of ensuring shock stability on both problems.
Lastly, it is observed that treatment of interface-tangential momentum component and energy component in the y-direction does not assure stability on this problem. 
We confirmed these with numerical test as well (not shown here). 
Thus in conclusion, it can be said that the most critical flux components that affect the shock instability behavior of the HLLC scheme are the mass flux and
the interface-normal momentum flux component on interfaces that are not aligned with the shock front.

In the next section we analyze the numerical dissipation of the full HLLC scheme in comparison to that of the full HLLE scheme, in the vicinity of a normal 
shock subjected to numerical perturbations, to explain how numerical discretization of the mass flux and interface-normal momentum flux components on the transverse 
interfaces triggers shock instability.

\section{Link between numerical discretization of transverse directional mass and interface-normal momentum flux component and shock instability}
\label{sec:orderofmagnitudeanalysis}

\begin{figure}[H]
	    \centering
	    \subfloat[]{\label{fig:initialshockstructure}\includegraphics[scale=0.2]{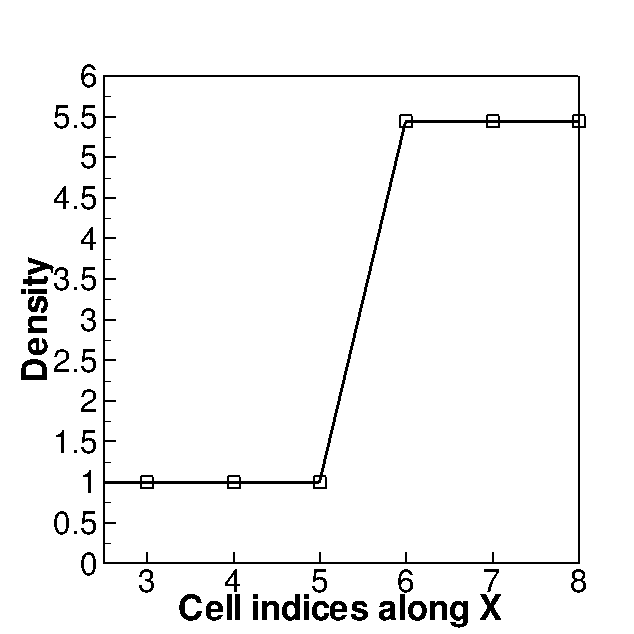}}
            \subfloat[]{\label{fig:finalshockstructure_hlle}\includegraphics[scale=0.2]{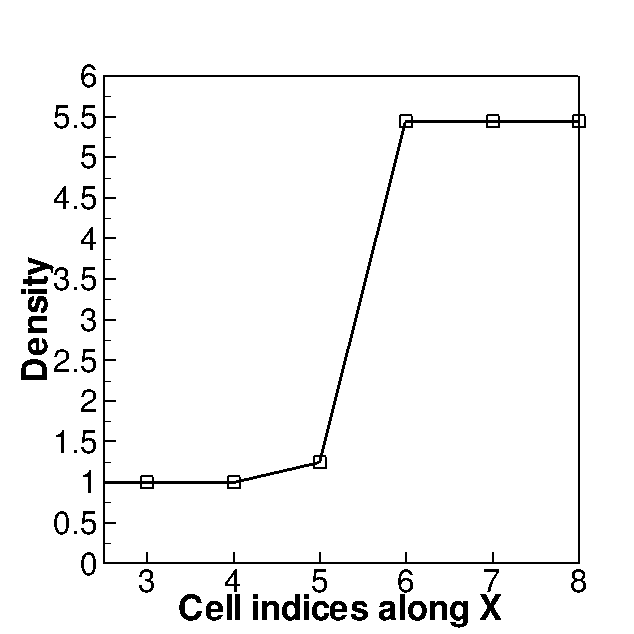}}
            \subfloat[]{\label{fig:finalshockstructure_hllc}\includegraphics[scale=0.2]{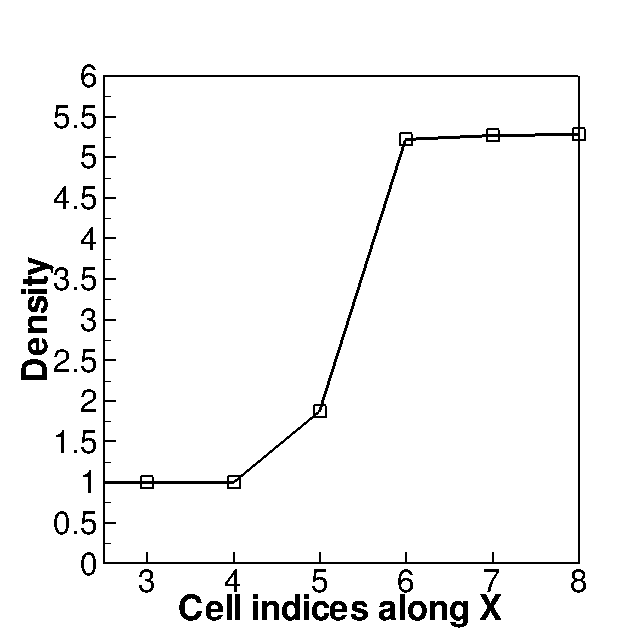}}
	   \caption{ (a) Initial shock structure (b) Computed shock structure by the HLLE scheme (c) Computed shock structure by the HLLC scheme. Results shown at t=20.}
	   \label{fig:initialandfinalshockstuctureforperturbationandhllceverywhere}
	    \end{figure}

\begin{figure}[H]
	    \centering
	    \includegraphics[scale=0.5]{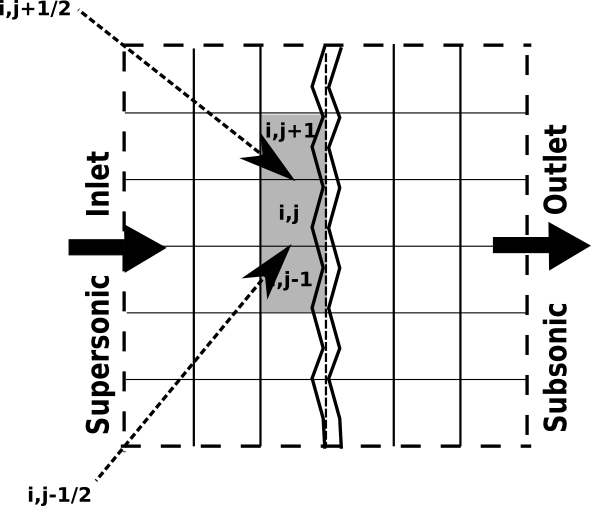}
	    \caption{Schematic showing the stencil chosen for performing the dissipation analysis on the HLLE and the HLLC schemes.}
	    \label{fig:stencilfororderanalysis}
	    \end{figure}

It was concluded in Sec.(\ref{sec:effectoftansversemomentumdissipation}) that shock instability behaviour is largely influenced by the type of numerical discretization 
used for mass flux component and interface-normal momentum flux component that come into play on interfaces that are normal to the shock front. 
In this section we 
provide an explanation for this observation by analyzing the numerical dissipation characteristics of the full HLLE and HLLC schemes in the vicinity of a normal shock.
Some researchers have indicated that the source of shock instability may lie in the numerical shock region of the captured shock \cite{xie2017,dumbser2004,chauvat2005}. 
Hence we focus on perfoming our analysis of these schemes in the set of cells that comprise the numerical shock region. In our experience we noticed that the initial
thin shock that we considered in Sec.(\ref{sec:matrixanalysis_hllandhllc}) would later, during actual numerical computations, spread over to include atleast 
one intermediate state in case of the HLLE and the HLLC scheme. Typically, the shock spreads upstream and causes 
the coloum of cells lying just ahead of the initial shock to constitute this intermediate state as shown in Fig.(\ref{fig:initialandfinalshockstuctureforperturbationandhllceverywhere}).
Interestingly, the numerical shock structure shown in Fig.(\ref{fig:finalshockstructure_hllc}) corresponds to a shock unstable solution computed by the HLLC scheme for an
initial condition wherein random perturbations were introduced only in the single coloum of cells lying upstream of the shock. This underlines the importance of 
the coloum of cells lying just ahead of the shock which has been noted earlier [cf.\cite{dumbser2004}].
Now, consider a y-directional stencil comprising of three candidate cells namely $(i,j), (i,j+1)$ and $(i,j-1)$ located within the upstream coloum of cells 
of an isolated strong normal shock that exists in a steady supersonic flow as shown in Fig.(\ref{fig:stencilfororderanalysis}). 

Being in the vicinity of a normal shock, it is safe to assume that the flow happens only in the positive x direction on these cells. 
Hence, in these cells, the following assumptions can be made:\newline
\begin{align}
 \nonumber
  \rho,u,p \neq 0 \\ 
  v =0 \ \ 
  \label{eqn:assumptionsforupstreamcellsnearshock}
\end{align}

\noindent Our particular concern is the evolution of the conserved quantities $\rho$ and $\rho u$ in cell ($i,j$) due to fluxes that cross the interfaces ($i,j\pm1/2$). 
While the fluxes $(\rho v)_{i,j\pm1/2}$ affect the evolution of quantity $\rho$, the fluxes  $(\rho u v) _{i,j\pm1/2}$ affect the quantity $\rho u$. 
The saw-tooth like perturbations in primitive and conserved variables that are characteristic of a shock unstable solution can be thought to occur due to imbalances in these 
fluxes only \cite{quirk1994}. 
An evolution equation for these quantities can be written as, 

\begin{align}
  \{\rho\}_{i,j}^{n+1} \approx  \{\rho\}_{i,j}^{n} - \frac{\Delta t}{\Delta y} \left [ \{\rho v\}_{i,j+1/2} - \{\rho v\}_{i,j-1/2} \right ]\\
 \{\rho u\}_{i,j}^{n+1} \approx  \{\rho u\}_{i,j}^{n} - \frac{\Delta t}{\Delta y} \left [ \{\rho u v\}_{i,j+1/2} - \{\rho u v\}_{i,j-1/2} \right ]
 \label{eqn:evolutionofxmomentum}
\end{align}
Notice that since there is no primary flow in the y-direction, subsonic fluxes corresponding to either the HLLE or the HLLC schemes will be engaged on these transverse cell interfaces. In the absence of any 
perturbations, these fluxes would only transmit information regarding pressure waves across these interfaces. However if numerical perturbations does exist, then unphysical 
fluxes would occur through these interfaces. The objective then is to see how the HLLE and HLLC schemes differ in their treatment of these unphysical fluxes.  

\subsection{Dissipation analysis of the HLLE scheme}
\label{sec:dissipationanalysisofhllescheme}
\subsubsection{Mass flux}
\label{sec:dissipationanalysisofhllescheme_massflux}
A HLLE subsonic discretization for $\{\rho v\}_{i,j+1/2}$ and $\{\rho v\}_{i,j-1/2}$ can be written as,

\begin{align}
 \nonumber
 \{\rho v\}_{j+1/2} = \frac{1}{2} \left[\{\rho v\}_j + \{\rho v\}_{j+1}  \right] + \frac{S_R + S_L}{2(S_R-S_L)} \left[\{\rho v\}_j - \{\rho v\}_{j+1}  \right] 
 -\frac{S_LS_R}{S_R-S_L} \left[ \{\rho \}_{j} - \{\rho \}_{j+1} \right]
\end{align}

\begin{align}
 \nonumber
 \{\rho v\}_{j-1/2} = \frac{1}{2} \left[\{\rho v\}_{j-1} + \{\rho v\}_{j}  \right] + \frac{S_R + S_L}{2(S_R-S_L)} \left[\{\rho v\}_{j-1} - \{\rho v\}_{j}  \right] 
 -\frac{S_LS_R}{S_R-S_L} \left[ \{\rho \}_{j-1} - \{\rho \}_{j} \right]
\end{align}
where we have dropped the $i$ from the interface index for convenience. The flux difference is then, 

\begin{align}
 \nonumber 
  \{\rho v\}_{j+1/2} - \{\rho v\}_{j-1/2} = \frac{1}{2} \left[\{\rho v\}_{j+1} - \{\rho v\}_{j-1}  \right] + \frac{S_R + S_L}{2(S_R-S_L)} \left[\{\rho v\}_{j} - \{\rho v\}_{j+1} 
  -\left( \{\rho v\}_{j-1} - \{\rho u v\}_{j} \right)  \right]\\ \nonumber
  -\frac{S_LS_R}{S_R-S_L} \left[ \{\rho \}_{j}-\{\rho \}_{j+1} - \left( \{\rho u \}_{j-1} - \{\rho \}_{j} \right) \right]
\end{align}
Since our interest lies in contrasting the numerical dissipation behaviours of the HLLE and the HLLC schemes, it suffices to only consider the dissipation component of the 
flux difference which will be henceforth denoted as ${\Delta D}^{\{\rho v\}_{j+1/2} - \{\rho v\}_{j-1/2}}_{HLLE}$. Using the notation $\Delta(.) = (.)_R - (.)_L$, 
the dissipation component of the above flux difference can be written as, 
\begin{align}
  {\Delta D}^{\{\rho v\}_{j+1/2} - \{\rho v\}_{j-1/2}}_{HLLE} = - \frac{S_R + S_L}{2(S_R-S_L)} \left[ \Delta \{\rho v\}_{j+1/2} - \Delta \{\rho v\}_{j-1/2}  \right]
  +\frac{S_LS_R}{S_R-S_L} \left[ \Delta \{\rho \}_{j+1/2} - \Delta \{\rho \}_{j-1/2} \right]
  \label{eqn:dissipationofmassfluxdifferences_hll}
\end{align}
To introduce the effect of small numerical perturbations that are thought to eventually result in shock unstable solutions, consider the existence of a 
random numerical perturbations of the order 
$\delta$ (eg. due to round of errors) in the primitive variables that exists during the course of computation in the stencil considered. 
Then, the following approximations are assumed to hold true on the stencil,

\begin{align}
 \Delta \rho, \Delta u, \Delta v, \Delta p, \Delta (\rho u) \sim \mathcal{O}({\delta}) 
 \label{eqn:assumptionsforupstreamcellswithperturbations}
 \end{align}
The term $\Delta \{\rho v\}_{j\pm1/2}$ can be expanded in terms of the mass flux as $ \tilde{\rho} \Delta v + \tilde{v} \Delta(\rho) $ where $\tilde{(.)}$ represent Roe averaged 
 quantities. Under these assumptions $\Delta \{\rho v\}_{j\pm1/2}$ is $\mathcal{O}(\delta)$. Assuming $S_L=(\tilde{v}-\tilde{a}) \sim \mathcal{O}({\tilde{a}})$ and 
 $S_R =(\tilde{v}+\tilde{a}) \sim \mathcal{O}({\tilde{a}})$, 
 using Eq.(\ref{eqn:assumptionsforupstreamcellsnearshock}) 
 and Eq.(\ref{eqn:assumptionsforupstreamcellswithperturbations}), Eq.(\ref{eqn:dissipationofmassfluxdifferences_hll}) can be simplified as, 

 \begin{align}
  \Delta D^{\{\rho v\}_{j+1/2} - \{\rho v\}_{j-1/2}}_{HLLE} = - \frac{\cancelto{\mathcal{O}({\delta})}{\tilde{v}}}{2\tilde{a}} \left[ \cancelto{\mathcal{O}({\delta})}{\Delta \{\rho v\}_{j+1/2}} -\cancelto{\mathcal{O}({\delta})}{ \Delta \{\rho v\}_{j-1/2}}  \right]
  \ \ \ \ -\frac{\tilde{a}}{2} \left[ \cancelto{\mathcal{O}({\delta})}{\Delta \{\rho \}_{j+1/2}} - \cancelto{\mathcal{O}({\delta})}{\Delta \{\rho \}_{j-1/2}} \right]
  \label{eqn:dissipationofmassfluxdifferences_hll_simplified1}
\end{align}
Effectively, 
\begin{align} 
  \Delta D^{\{\rho v\}_{j+1/2} - \{\rho v\}_{j-1/2}}_{HLLE} \sim \mathcal{O}(\delta)  
  \label{eqn:dissipationofmassfluxdifferences_hll_simplified2}
\end{align}
Thus in the presence of perturbations of $\mathcal{O}(\delta)$ in flow quantities, the HLLE scheme infuses a net dissipation of the same order into the mass flux 
discretization. Refering to Eq.(\ref{eqn:quirk_analysis_hll}) we see that this
amount of dissipation is sufficient to cause damping of perturbations ($\hat{\rho}$) in density variable. 

\subsubsection{Momentum flux}
\label{sec:dissipationanalysisofhllescheme_momentumflux}
A HLLE subsonic discretization for $\{\rho u v\}_{j+1/2}$ and $\{\rho u v\}_{j-1/2}$ can be written as, 

\begin{align}
 \nonumber
 \{\rho u v\}_{j+1/2} = \frac{1}{2} \left[\{\rho u v\}_j + \{\rho u v\}_{j+1}  \right] + \frac{S_R + S_L}{2(S_R-S_L)} \left[\{\rho u v\}_j - \{\rho u v\}_{j+1}  \right] 
 -\frac{S_LS_R}{S_R-S_L} \left[ \{\rho u \}_{j} - \{\rho u \}_{j+1} \right]
\end{align}

\begin{align}
 \nonumber
 \{\rho u v\}_{j-1/2} = \frac{1}{2} \left[\{\rho u v\}_{j-1} + \{\rho u v\}_{j}  \right] + \frac{S_R + S_L}{2(S_R-S_L)} \left[\{\rho u v\}_{j-1} - \{\rho u v\}_{j}  \right] 
 -\frac{S_LS_R}{S_R-S_L} \left[ \{\rho u \}_{j-1} - \{\rho u \}_{j} \right]
\end{align}

\noindent The dissipation component of the flux difference in this case denoted as $\Delta D^{\{\rho u v\}_{j+1/2} - \{\rho u v\}_{j-1/2}}_{HLLE}$ will be, 
\begin{align}
  \Delta D^{\{\rho u v\}_{j+1/2} - \{\rho u v\}_{j-1/2}}_{HLLE} = - \frac{S_R + S_L}{2(S_R-S_L)} \left[ \Delta \{\rho u v\}_{j+1/2} - \Delta \{\rho u v\}_{j-1/2}  \right]
  +\frac{S_LS_R}{S_R-S_L} \left[ \Delta \{\rho u \}_{j+1/2} - \Delta \{\rho u \}_{j-1/2} \right]
  \label{eqn:dissipationofmomentumfluxdifferences_hll}
\end{align}
 The term $\Delta \{\rho u v\}_{j\pm1/2}$ can be expanded in terms of the mass flux as $ \tilde{\rho} \tilde{u} \Delta v + \tilde{v} \Delta(\rho u) $ and under the 
 assumptions made in Eq.(\ref{eqn:assumptionsforupstreamcellswithperturbations}) 
 is $\mathcal{O}(\delta)$. Using Eq.(\ref{eqn:assumptionsforupstreamcellsnearshock}) and Eq.(\ref{eqn:assumptionsforupstreamcellswithperturbations}), 
 Eq.(\ref{eqn:dissipationofmomentumfluxdifferences_hll}) can be simplified as, 
\begin{align}
  \Delta D^{\{\rho u v\}_{j+1/2} - \{\rho u v\}_{j-1/2}}_{HLLE} = - \frac{ \cancelto{\mathcal{O}({\delta})}{\tilde{v}}}{2\tilde{a}} \left[ \cancelto{\mathcal{O}({\delta})}{ \Bigl({\tilde{\rho} \tilde{u} \Delta v + \tilde{v} \Delta(\rho u)}\Bigr)_{j+1/2}} -\cancelto{\mathcal{O}({\delta})}{ \Bigl({\tilde{\rho} \tilde{u} \Delta v + \tilde{v} \Delta(\rho u)}\Bigr)_{j-1/2}}  \right]
  \ \ \ -\frac{\tilde{a}}{2} \left[ \cancelto{\mathcal{O}({\delta})}{\Delta \{\rho u \}_{j+1/2}} - \cancelto{\mathcal{O}({\delta})}{\Delta \{\rho u \}_{j-1/2}} \right]
  \label{eqn:dissipationofmoemtumfluxdifferences_hll_simplified1}
\end{align}
Which indicates, 
\begin{align}
  \Delta D^{\{\rho u v\}_{j+1/2} - \{\rho u v\}_{j-1/2}}_{HLLE} \sim \mathcal{O}(\delta)  
  \label{eqn:dissipationofmomentumfluxdifferences_hll_simplified2}
\end{align}
Thus in the presence of perturbations of $\mathcal{O}(\delta)$ in flow quantities, the HLLE scheme infuses a net dissipation of the same order into the discretization of 
x-momentum flux in the y-direction. Once again refering to Eq.(\ref{eqn:quirk_analysis_hll})
we see that this quantity of dissipation provided by the HLLE scheme in the transverse direction is enough to suppress any perturbations ($\hat{u}$) that may arise in the x-velocity.

\subsection{Dissipation analysis of the HLLC scheme}
\label{sec:dissipationanalysisofhllcscheme}

\subsubsection{Mass flux}
\label{sec:dissipationanalysisofhllcscheme_massflux}
Consider the subsonic HLLC flux form given in Eq.(\ref{eqn:hllc-modifiedform}) where the scheme is rewritten as a combination of diffusive HLL flux and an antidiffusive 
term responsible for restoring contact and shear wave ability. The analysis that follows is performed assuming that interfaces ($i,j\pm1/2$) will have a wave structure
such that $S_L \leq0\leq S_R$. Then, the corresponding HLLC fluxes for these interfaces can be expressed as, 

\begin{align}
 \{\rho v\}_{j+1/2} =  \{ \rho v \}_{j+1/2}^{HLL} + S_L \left[ \{\rho \}_j \left( \frac{S_L-\{v\}_j}{S_L - \{S_M\}_{j+1/2}} \right) - 
 \frac{S_R \{\rho \}_{j+1} -S_L\{\rho \}_{j} + \{\rho v \}_{j} - \{\rho v\}_{j+1} }{S_R - S_L} \right]
\end{align}

\begin{align}
 \{\rho v\}_{j-1/2} =  \{ \rho v \}_{j-1/2}^{HLL} + S_L \left[ \{\rho  \}_{j-1} \left( \frac{S_L-\{v\}_{j-1}}{S_L - \{S_M\}_{j-1/2}} \right) - 
 \frac{S_R \{\rho \}_{j} -S_L\{\rho \}_{j-1} + \{\rho v \}_{j-1} - \{\rho v\}_{j} }{S_R - S_L} \right]
\end{align}
with $\{S_M\}_{j+1/2}$ and $\{S_M\}_{j-1/2}$ defined as, 

\begin{align}
\label{eqn:middlewaveatj+1/2}
 \{S_M\}_{j+1/2} = \frac{ \{p\}_{j+1} - \{p\}_j + \{\rho v\}_{j}\left(S_L - \{v\}_j\right) - \{\rho v \}_{j+1}\left(S_R - \{v\}_{j+1}\right) }{\{\rho\}_j\left(S_L - \{v\}_j\right) - \{\rho\}_{j+1} \left(S_R - \{v\}_{j+1}\right) }
\end{align}

\begin{align}
\label{eqn:middlewaveatj-1/2}
 \{S_M\}_{j-1/2} = \frac{ \{p\}_{j} - \{p\}_{j-1} + \{\rho v\}_{j-1}\left(S_L - \{v\}_{j-1}\right) - \{\rho v \}_{j}\left(S_R - \{v\}_{j}\right) }{\{\rho\}_{j-1}\left(S_L - \{v\}_{j-1}\right) - \{\rho\}_{j} \left(S_R - \{v\}_{j}\right) }
\end{align}

\noindent If we consider only the dissipation term in the flux difference it would turn out to be,

\begin{align}
 \Delta D^{\{\rho v\}_{j+1/2} - \{\rho v\}_{j-1/2}}_{HLLC} = \Delta D^{\{\rho v\}_{j+1/2} - \{\rho v\}_{j-1/2}}_{HLL} + S_L \Biggl[ \{\rho \}_j \left( \frac{S_L-\{v\}_j}{S_L - \{S_M\}_{j+1/2}} \right)  -  \{\rho \}_{j-1} \left( \frac{S_L-\{v\}_{j-1}}{S_L - \{S_M\}_{j-1/2}} \right) \nonumber \\  
  + \frac{S_R \left( -\{\Delta \rho \}_{j+1/2}\right) + S_L \left( \{\Delta \rho \}_{j-1/2} \right) - \{\Delta \rho v \}_{j-1/2} + \{\Delta \rho v\}_{j+1/2} }{S_R -S_L}  \Biggr]
  \label{eqn:dissipationofmassfluxdifferences_hllc}
\end{align}

\noindent where the term $\Delta D^{\{\rho v\}_{j+1/2} - \{\rho v\}_{j-1/2}}_{HLL}$ denotes the HLL-type diffusion component that is inherent in the HLLC scheme. 
The order of magnitude of this term has been already estimated in Sec.(\ref{sec:dissipationanalysisofhllescheme_massflux}) and turns out to be $\mathcal{O}(\delta)$. 
The remaining terms
arise from the the antidiffusion component of the HLLC flux (identified in Eq.(\ref{eqn:hllc-modifiedform})) and will be referred to
as $\Delta A^{\{\rho v\}_{j+1/2} - \{\rho v\}_{j-1/2}}_{HLLC}$. To obtain the overall order of magnitude of $\Delta D^{\{\rho v\}_{j+1/2} - \{\rho v\}_{j-1/2}}_{HLLC}$, 
an estimate for $\Delta A^{\{\rho v\}_{j+1/2} - \{\rho v\}_{j-1/2}}_{HLLC}$ has to be obtained. To do this, firstly we need to  
estimate the order of magnitudes of $\{S_M\}_{j\pm1/2}$. Consider the case of $\{S_M\}_{j+1/2}$. 
Using assumptions stated in Eq.(\ref{eqn:assumptionsforupstreamcellsnearshock}) and (\ref{eqn:assumptionsforupstreamcellswithperturbations}) $\{S_M\}_{j+1/2}$
can be simplified as,

\begin{align}
 \{S_M\}_{j+1/2} = \frac{  \{\Delta p\}_{j+1/2} }{ -\hat{a} \left( \{\rho\}_{j} + \{\rho\}_{j+1} \right) }
\end{align}
Now since pressure perturbation $\{\Delta p\}_{j+1/2} \sim \mathcal{O}({\delta})$, it can be said that $\{S_M\}_{j+1/2} \sim \mathcal{O}(\delta)$. Similar observation can be made about $\{S_M\}_{j-1/2}$ too.
Using these estimates for $\{S_M\}_{j\pm1/2}$, assuming $S_L=(\tilde{v}-\tilde{a}) \sim \mathcal{O}({\tilde{a}})$ and $S_R =(\tilde{v}+\tilde{a}) \sim \mathcal{O}({\tilde{a}})$ and under assumptions in Eq.(\ref{eqn:assumptionsforupstreamcellsnearshock}), 
the terms arising from the antidiffusive component can be further simplified as,

\begin{align}
  \Delta A^{\{\rho v\}_{j+1/2} - \{\rho v\}_{j-1/2}}_{HLLC} =  (\tilde{v}-\tilde{a}) \Biggl[ \cancelto{\mathcal{O}(\delta)} {\{\Delta \rho \}_{j-1/2}} \ \ \ \ \ \ \ \ \ \ \ \cancelto{\mathcal{O}(1)}{ \left( \frac{(\tilde{v}-\tilde{a})}{(\tilde{v}-\tilde{a}) - \mathcal{O}(\delta)} \right)}  \nonumber \\  
  + \frac{-(\tilde{v}+\tilde{a}) \cancelto{\mathcal{O}(\delta)}{\{\Delta \rho \}_{j+1/2}} +(\tilde{v}-\tilde{a})\cancelto{\mathcal{O}(\delta)}{\{\Delta \rho \}_{j-1/2}}  - \cancelto{\mathcal{O}(\delta)}{\{\Delta \rho v \}_{j-1/2}} + \cancelto{\mathcal{O}(\delta)}{\{\Delta \rho v\}_{j+1/2}} }{2\tilde{a}}  \Biggr]
  \label{eqn:dissipationofmassfluxdifferences_hllc_simplified1}
\end{align}
Indicating that,
\begin{align}
 \Delta A^{\{\rho v\}_{j+1/2} - \{\rho v\}_{j-1/2}}_{HLLC} \sim \mathcal{O} (\delta)
 \label{eqn:orderofantidissipationterms_hllcmassflux}
\end{align}
It is interesting to observe that in the presence of perturbations of $\mathcal{O}(\delta)$ in flow quantities, the antidiffusive terms in the mass flux discretization
of the HLLC scheme are activated along with their inherent HLL-type diffusive terms. Moreover, these terms are of the same order of the perturbations themselves.

\subsubsection{Momentum flux}
\label{sec:dissipationanalysisofhllcscheme_momentumflux}
A HLLC flux for the x-momentum flux components $\{\rho u v\}_{j\pm1/2}$ in y-direction can be written as, 

\begin{align}
 \{\rho u v\}_{j+1/2} =  \{ \rho u v \}_{j+1/2}^{HLL} + S_L \left[ \{\rho u \}_j \left( \frac{S_L-\{v\}_j}{S_L - \{S_M\}_{j+1/2}} \right) - 
 \frac{S_R \{\rho u\}_{j+1} -S_L\{\rho u\}_{j} + \{\rho u v \}_{j} - \{\rho u v\}_{j+1} }{S_R - S_L} \right]
\end{align}

\begin{align}
 \{\rho u v\}_{j-1/2} =  \{ \rho u v \}_{j-1/2}^{HLL} + S_L \left[ \{\rho u \}_{j-1} \left( \frac{S_L-\{v\}_{j-1}}{S_L - \{S_M\}_{j-1/2}} \right) - 
 \frac{S_R \{\rho u\}_{j} -S_L\{\rho u\}_{j-1} + \{\rho u v \}_{j-1} - \{\rho u v\}_{j} }{S_R - S_L} \right]
\end{align}
with $\{S_M\}_{j+1/2}$ and $\{S_M\}_{j-1/2}$ defined as in Eqs.(\ref{eqn:middlewaveatj+1/2}) and (\ref{eqn:middlewaveatj-1/2}). The dissipation term in the flux difference would be,

\begin{align}
 \Delta D^{\{\rho u v\}_{j+1/2} - \{\rho u v\}_{j-1/2}}_{HLLC} = \Delta D^{\{\rho u v\}_{j+1/2} - \{\rho u v\}_{j-1/2}}_{HLL} + S_L \Biggl[ \{\rho u \}_j \left( \frac{S_L-\{v\}_j}{S_L - \{S_M\}_{j+1/2}} \right)  -  \{\rho u \}_{j-1} \left( \frac{S_L-\{v\}_{j-1}}{S_L - \{S_M\}_{j-1/2}} \right) \nonumber \\  
  + \frac{S_R \left( -\{\Delta \rho u \}_{j+1/2}\right) + S_L \left( \{\Delta \rho u \}_{j-1/2} \right) - \{\Delta \rho u v \}_{j-1/2} + \{\Delta \rho u v\}_{j+1/2} }{S_R -S_L}  \Biggr]
  \label{eqn:dissipationofmomentumfluxdifferences_hllc}
\end{align}
where the term $\Delta D^{\{\rho u v\}_{j+1/2} - \{\rho u v\}_{j-1/2}}_{HLL}$ denotes the HLLE dissipation component that is inherent in the HLLC scheme. Once again 
we recollect that the order of magnitude of this term has been already estimated in Sec.(\ref{sec:dissipationanalysisofhllescheme_momentumflux}) and turns out to be $\mathcal{O}(\delta)$. 
The contribution from the remaining antidiffusive term is referred to as $\Delta A^{\{\rho u v\}_{j+1/2} - \{\rho u v\}_{j-1/2}}_{HLLC}$ and is given as,

\begin{align}
  \Delta A^{\{\rho u v\}_{j+1/2} - \{\rho u v\}_{j-1/2}}_{HLLC} =  (\tilde{v}-\tilde{a}) \Biggl[ \cancelto{\mathcal{O}(\delta)} {\{\Delta \rho u \}_{j-1/2}} \ \ \ \ \ \ \ \ \ \ \ \cancelto{\mathcal{O}(1)}{ \left( \frac{(\tilde{v}-\tilde{a})}{(\tilde{v}-\tilde{a}) - \mathcal{O}(\delta)} \right)}  \nonumber \\  
  + \frac{-(\tilde{v}+\tilde{a}) \cancelto{\mathcal{O}(\delta)}{\{\Delta \rho u \}_{j+1/2}} +(\tilde{v}-\tilde{a})\cancelto{\mathcal{O}(\delta)}{\{\Delta \rho u \}_{j-1/2}}  - \cancelto{\mathcal{O}(\delta)}{\{\Delta \rho u v \}_{j-1/2}} + \cancelto{\mathcal{O}(\delta)}{\{\Delta \rho u v\}_{j+1/2}} }{2\tilde{a}}  \Biggr]
  \label{eqn:dissipationofmomentumfluxdifferences_hllc_simplified1}
\end{align}
which effectively means,
\begin{align}
 \Delta A^{\{\rho u v\}_{j+1/2} - \{\rho u v\}_{j-1/2}}_{HLLC} \sim \mathcal{O} (\delta)
 \label{eqn:orderofantidissipationterms_hllcmomentumflux}
\end{align}
We observe again that in the presence of perturbations of $\mathcal{O}(\delta)$ in flow quantities, the antidiffusive terms are activated along with the inherent HLL-type
dissipative terms. These terms are also of the same order as that of the perturbations.

\subsection{Possible cause for shock instability in HLLC scheme}

The analysis presented above in conjuction with perturbation analysis presented in Sec.(\ref{sec:quirkanalysis_hllandhllc}) provides an explanation for the cause of shock instability 
in the HLLC scheme. Suppose during a simulation, random numerical perturbations of $\mathcal{O} (\delta)$ exist in flow variables along a strong normal shock wave.
From Eq.(\ref{eqn:quirk_analysis_hllc}) we know that the HLLC scheme has an inherent damping mechanism specifically for pressure perturbations that may exist. 
However, as seen from Eqs.(\ref{eqn:orderofantidissipationterms_hllcmassflux}) and (\ref{eqn:orderofantidissipationterms_hllcmomentumflux}) these perturbations inadvertently
activate the antidiffusive terms in the mass flux component and interface-normal momentum flux component on the interfaces transverse to the shock front. These activated terms are also of $\mathcal{O} (\delta)$.
Due to their activation, the perturbation evolution equations for $\hat{\rho}$ and $\hat{u}$ would be, 

      \begin{align}
      \nonumber
      \hat{\rho}^{n+1}&= \underbrace{\hat{\rho}^n (1-2\lambda)}_{HLL \ component} +  \  \underbrace{ (2\lambda(\hat{\rho}^n - \frac{\hat{p}^n}{\gamma}))}_{Antidiffusive \ component} \\\nonumber
      \hat{u}^{n+1} &= \underbrace{\hat{u}^n (1-2\lambda)}_{HLL \ component} + \  \underbrace{ (2\lambda \hat{u}^n)}_{Antidiffusive \ component}\\
      \label{eqn:quirk_analysis_rearranged_hllc}
      \end{align}
In these evolution equation,  we have clearly distinguished the contribution from the inherent HLL-type diffusive terms and the antidiffusive terms. Note that the 
antidiffusive terms introduce additional density 
and pressure perturbations into the density evolution equation (obtained from mass flux equation) and extra x-velocity perturbation into the x-velocity perturbation evolution
equation(obtained from x-momentum equation). These additional terms counteract the dissipative action
of the HLL terms and causes growth in the $\hat{\rho}$ and $\hat{u}$ perturbations. In fact by their very construction, the antidiffusive terms of the HLLC scheme are designed to reduce the dissipation 
associated with its inherent HLL two wave approximation and provide accuracy on linear wavefields. Recollect that the interface-normal momentum flux component $\rho u v$ in y-direction
is chiefly responsible for the advection of the quantity $\rho u$ in this direction.
Thus, undamped perturbations in $\hat{\rho}$ and $\hat{u}$ can independently
cause undesirable variations in this quantity along the shock front \cite{shen2014}. Further due to enforcement of Rankine-Hugoniot jump conditions,
these unphysical variations along the front could result in non-conservation of the quantity $\rho u$ along the shock wave. These errors could propagate into other equations due to the 
nonlinear coupling nature of the Euler equations. This forces the shock structure to adjust through a typical 'bulge' or in worst cases a complete 
breakdown of it. On certain problems (like the standing shock instability problem) it may be enough to damp only the x-velocity perturbations by controlling the antidiffusive 
terms as observed in the result shown in Fig.(\ref{fig:X(1F2F3F4F)_Y(1F2F3N4F)_standingshock_finergrid}) while on some other problem (like the moving shock instability problem) it may
be necessary to damp both the density and x-velocity perturbations to ensure consistency of quantity $\rho u$ (see result in 
Fig.(\ref{fig:X(1F2F3F4F)_Y(1N2F3N4F)_movingshock})). This explains why the configuration $X\{F1,F2,F3,F4\}-Y\{N1,F2,F3,F4\}$ that damps on $\hat{\rho}$ does not guarentee shock stability while the
configuration $X\{F1,F2,F3,F4\}-Y\{N1,F2,N3,F4\}$ that damps both $\hat{\rho}$ and $\hat{u}$ remains shock stable over wider range of Mach numbers as 
seen in Fig.(\ref{fig:effectofhllontransversefluxes_hllc_hll_hybrid}) and is a desirable configuration for a hybrid flux.  
In this context, note that withdrawing antidiffusive terms from energy discretization, achieved by the configuration $X\{F1,F2,F3,F4\}-Y\{F1,F2,F3,N4\}$, 
may not be much effective in ensuring stability. This may be because the pressure perturbations, whose linear evolution is governed by this flux component, 
is inherently damped by the HLLC scheme. However as seen above, what is more important is the rate at which these pressure 
perturbations feeds into the density perturbations (see Eq.(\ref{eqn:quirk_analysis_hllc})) and withdrawing of antidiffusive terms from this flux component
does not have any effect on this process. Based on these observations, we formulate a shock stable HLLC scheme in the next section.

\section{A shock stable HLLC scheme}
\label{sec:formulation}

\begin{figure}[H]
	    \centering
	    \includegraphics[scale=0.25]{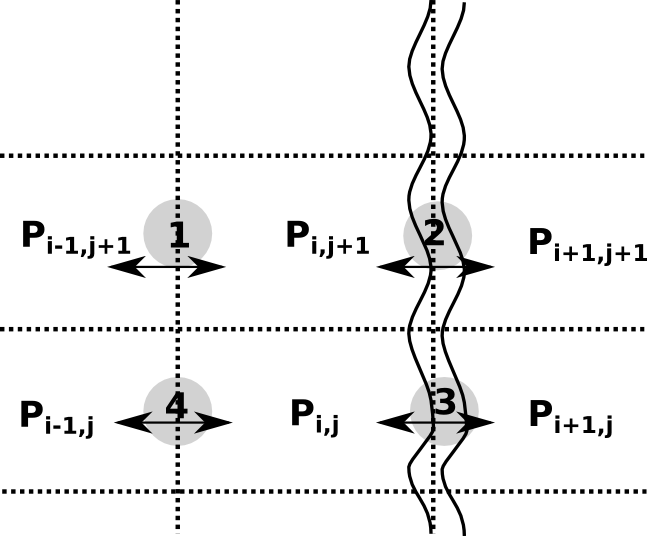}
	    \caption{Typical stencil used for evaluation of pressure ratio based switching coefficient $\omega$.}
	    \label{fig:stencilforomega}
	    \end{figure}

The above analyses recognizes that in case of the HLLC scheme, random numerical perturbations in the vicinity of a 
computed shock causes a weakening of its inherent HLL-type dissipation due to the counteraction of its antidiffusive term $S_{L/R}(\mathbf{U}^{*HLLC}_{L/R} - \mathbf{U}^{*HLL})$. 
In general, this loss of dissipation, particularly in the mass and interface-normal momentum flux components on interfaces 
that are not aligned to the shock front could initiate shock instabilities in the HLLC scheme.
Based on this observation, we realize that an adequate control of these antidiffusive terms near shock waves, 
especially on these vulnerable interfaces, 
could prove beneficial in suppressing the instability. 
Specifically, a strategy 
that can reduce the order of magnitude of these terms as compared to that of the inherent HLL-type diffusion terms is sought. 
In Eq.(\ref{eqn:hllc-modifiedformwithomega}) we introduced the idea of pre-multiplying the antidiffusive terms of the HLLC scheme with a factor 
$\omega$ that can help us recover the full HLLE or HLLC schemes based on its value 
($\omega=0$ or $\omega=1$ respectively). To achieve a smoother withdrawal of the antidiffusion term at interface ($i,j+1/2$) in the vicinity of the shock, we define a pressure ratio based $\omega$ 
(described here for a structured Cartesian mesh) as \cite{zhang2017}, 

\begin{align}
 \omega = min_k (f_k),\ \ \ k=1...4
 \label{eqn:definitionofomega}
\end{align}
where $f_k$'s are pressure ratio based functions evaluated on a predefined stencil around the ($i,j+1/2$) interface shown in Fig.(\ref{fig:stencilforomega}). 
At any interface $k$, $f_k$ is defined as,  
\begin{align}
 f_k = min\left ( \frac{p_R}{p_L},\frac{p_L}{p_R} \right )_k^\alpha
 \label{eqn:definitionofpressureratiofucntion}
\end{align}
Here $p_R$ and $p_L$ denotes the right and left cell center pressures across any $k^{th}$ interface. The parameter $\alpha>1$ denotes a tunable parameter that affects the 
magnitude
of $f_k$ across the interface. In Sec.(\ref{sec:matrixanalysis}) a strategy to obtain a value of $\alpha$ that enables the modified HLLC scheme to 
work over a wide range of freestream Mach numbers is discussed. We emphasize that on a regular two-dimensional Cartesian mesh, the switching coefficient $\omega$ 
need to be applied only on the mass and interface-normal momentum flux components of cell interfaces that are orthogonal to the shock front while the full HLLC scheme 
may be used to safely discretize the remaining flux components in both directions. However on a generic mesh where the interfaces do not align with global Cartesian basis 
or where the shock front
is not aligned with the interfaces, it may be beneficial 
to apply the coefficient $\omega$ to mass and interface-normal momentum flux components in both directions to ensure shock stability. Note that until otherwise stated, 
the HLLC-ADC scheme used in the rest of the paper uses the switching parameter $\omega$ only on the mass and interface-normal momentum flux components on the interfaces
that are not aligned to the shock front. In the presence of the 
switching coefficient $\omega$, the evolution equations of the primitive variables subjected to a saw-tooth perturbation will turn out to be,

\begin{align}
\nonumber
\hat{\rho}^{n+1}&= \underbrace{\hat{\rho}^n (1-2\lambda)}_{HLL \ component} +  \ \omega \underbrace{ (2\lambda(\hat{\rho}^n - \frac{\hat{p}^n}{\gamma}))}_{Antidiffusive \ component} \\\nonumber
\hat{u}^{n+1} &= \underbrace{\hat{u}^n (1-2\lambda)}_{HLL \ component} + \  \omega \underbrace{ (2\lambda \hat{u}^n)}_{Antidiffusive \ component}\\
\hat{p}^{n+1}&= \hat{p}^n(1-2\lambda)
\label{eqn:quirk_analysis_rearranged_hllc_hll_hybrid}
\end{align}

\begin{figure}[H]
	    \centering
	    \setcounter{subfigure}{0}
	    \subfloat[$\boldsymbol{\left(\hat{\rho} = 0.01, \hat{u} = 0, \hat{p} = 0\right)} $ ]{\label{fig:HLLC_HLL_hybrid_density_perturbation}\includegraphics[scale=0.25]{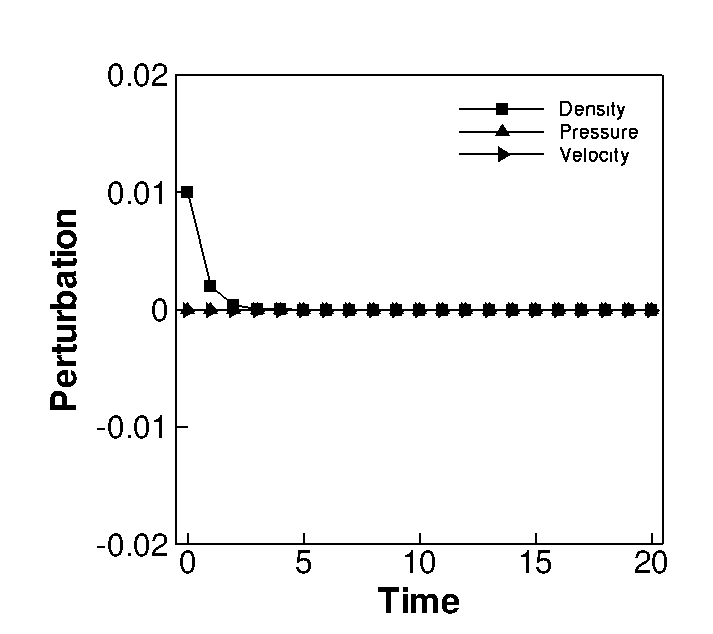}}
            \subfloat[$\boldsymbol{\left( \hat{\rho} = 0, \hat{u} = 0.01, \hat{p} = 0 \right) }$]{\label{fig:HLLC_HLL_hybrid_velocity_perturbation}\includegraphics[scale=0.25]{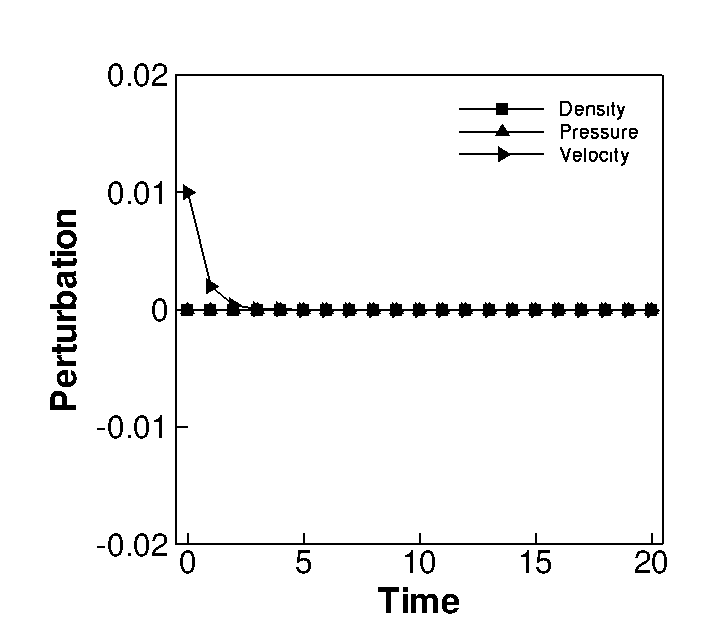}}\\
            \subfloat[$\boldsymbol{ \left(\hat{\rho} = 0, \hat{u} = 0, \hat{p} = 0.01\right) }$]{\label{fig:HLLC_HLL_hybrid_pressure_perturbation}\includegraphics[scale=0.25]{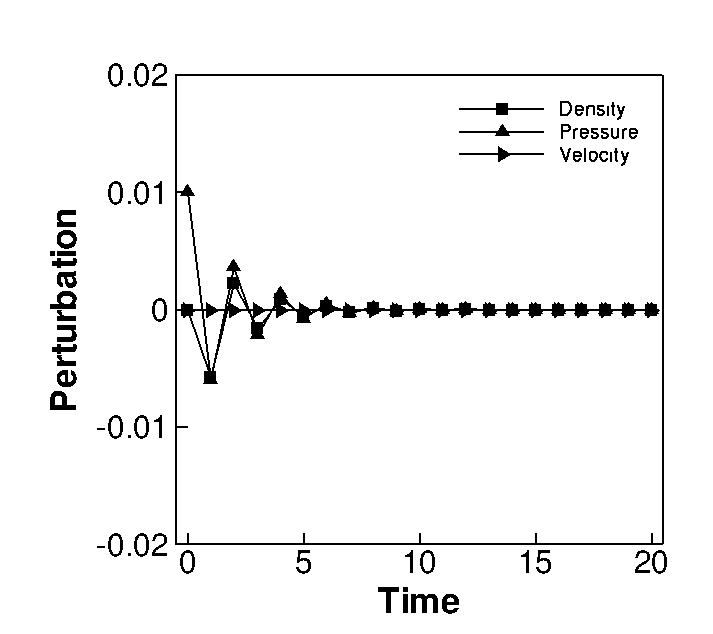}} 
	    \caption{Comparison of evolution of density, x-velocity and pressure perturbations in the proposed HLLC-ADC scheme.}
	    \label{fig:hllc_hll_hybrid_perturbationstudies}
	    \end{figure}

From Eq.(\ref{eqn:quirk_analysis_rearranged_hllc_hll_hybrid}) it is seen that the switching coefficient $\omega$ has affected only the antidiffusive terms of 
the evolution equations concerning $\hat{\rho}$ and $\hat{u}$. 
As expected the pressure perturbation evolution equation, which remains 
free of $\omega$, is same as the HLLC and the HLLE schemes 
and is damping in nature by itself. The amplification factors of the perturbations 
$\hat{\rho},\hat{u},\hat{p}$ are respectively $\left( 1- 2\lambda(1-\omega),1-2\lambda(1-\omega),1-2\lambda \right)$.
Consider a situation where $\omega<1$ but $\omega\neq0$. Then a von-Neumann like stability criterion for $\lambda$ can be derived from 
Eq.(\ref{eqn:quirk_analysis_rearranged_hllc_hll_hybrid}) as 
  \begin{align}
  \label{eqn:limitonlambda_hllc_hll_hybrid}
  0 \leq \lambda \leq \frac{1}{1-\omega}
  \end{align}
For an $0\leq\omega<1$ and an appropriate choice of $\lambda$ under the above criterion, it can be readily seen that switching coefficient $\omega$ acts as
an additional damping term that acts exclusively
in suppressing the perturbations introduced by the antidiffusive terms. Hence the pressure perturbations that feeds into the density perturbations 
are damped recursively by $\omega$. The additional x-velocity perturbations are also damped in a similar fashion. In effect, $\omega$ ensures that any $\hat{\rho}$
and $\hat{u}$ that may exist during the course of computation are damped. This ensures elimination of unphysical mass flux 
variations and guarantees shock stability. Thus using a simple switching coefficient $\omega$ on the antidiffusive terms, we have been able to overcome 
the \textit{marginally} stable behavior of the HLLC scheme to make it a \textit{strictly} stable scheme.
Further, in the presence of an infinitely strong shock, where $\omega\sim0$, we recover the full HLLE behaviour with the 
antidiffusive terms largely suppressed. We refer to a HLLC scheme with an antidiffusion control using $\omega$ as described above simply as HLLC-ADC 
(\textbf{A}nti \textbf{D}iffusion \textbf{C}ontrol) henceforth. It is interesting to note that introduction of $\omega$ does not introduce additional constraints on the 
CFL criterion. Hence the HLLC-ADC scheme obeys the usual first order and higher order stability restrictions.
The behaviour of these equations can be 
seen in Fig.(\ref{fig:hllc_hll_hybrid_perturbationstudies}) where any initial perturbation in density, x-velocity and pressure is damped with time. 
The experiments were done with $\lambda=0.8$ and $\omega=0.5$.

A switching parameter $\omega$ described by Eqs.(\ref{eqn:definitionofomega})
and (\ref{eqn:definitionofpressureratiofucntion}) presents several advantages. 
Firstly, Sanders et al \cite{sanders1998} have shown through linear analysis that for upwind schemes in multidimensional computations, 
a multidimensional dissipation is more effective in ensuring shock stability. The above definition of $\omega$ incorporates the required multidimensional flavor because
the value of $\omega$ on transverse interfaces are a function of the solution values on the longitudinal interfaces and vice versa. Secondly, we note that 
a pressure based shock sensor works better than the characteristics based sensor proposed in \cite{sanders1998,pandolfi2001} in terms of ability to clearly distinguish between 
shock waves and contact/shear waves. Lastly, due to its differentiability, it renders the whole HLLC-ADC scheme for straightforward extension to its implict variants.
We remark that although $\omega$ is defined above for 
regular Cartesian mesh, we remark that it can be easily extended to unstructured grids using the strategy discussed in \cite{phongthanapanich2015}.
In the next section, use the matrix based stability analysis to estimate the parameter $\alpha$ that is used to define $\omega$ and complete our description of 
the HLLC-ADC scheme.

\section{Estimation of $\alpha$ using the matrix based stability analysis}
\label{sec:matrixanalysis}
 
\begin{figure}[H]
	    \centering
	    \includegraphics[scale=0.3]{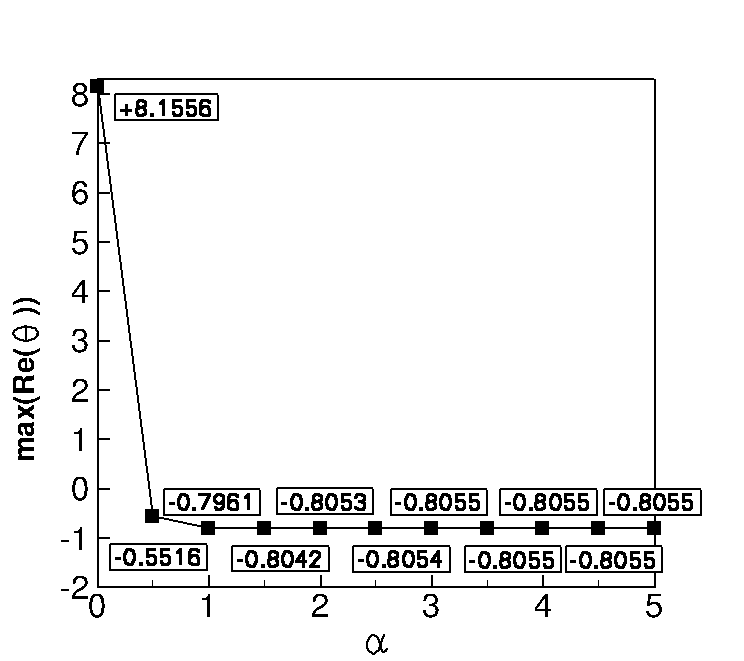}
	   \caption{Plot showing the effect of $\alpha$ on $max(Re(\theta))$ for a $M=7$ isolated two dimensional thin steady shock.}
	   \label{fig:effectofalpha_hllc_hll_hybrid}
	    \end{figure}

As described in Sec.(\ref{sec:formulation}) the selective control of the necessary antidiffusive terms of the HLLC scheme is achieved through a pressure ratio based shock sensor
$\omega$ which in turn depends on a tunable parameter $\alpha \in \mathbf{R}$.  However the appropriate choice of value for $\alpha$ remains undetermined as yet and has to be 
obtained. A reasonable estimate for $\alpha$ can be obtained by studying the effect of varying $\alpha$ on the stability of the shock profile found in
the steady simulation of an isolated two dimensional thin shock discussed previously. We resort to technique of matrix based 
stability analysis of this problem introduced in Sec.(\ref{sec:matrixanalysis_hllandhllc}) when computed by the HLLC-ADC scheme on a $11$ by $11$ grid. 
Fig.(\ref{fig:effectofalpha_hllc_hll_hybrid}) shows the variation of the 
max Real part of Eigenvalues $\theta$ (indicated in the boxes) as a function of varying $\alpha$
values for this problem.

\noindent Note that $\alpha=0.0$ corresponds to the unmodified HLLC scheme which is known to be unstable on this problem. 
The positive $max(Re(\theta))$ value of +8.1556 
associated with $\alpha=0.0$ confirms that the numerical errors have a propensity to grow and eventually spoil the solution.
Further, the analysis predicts that any value of $\alpha>0.0$ will quickly stabilize the solution. The stability of the solution increases progressively with increasing 
$\alpha$ until an asymptotic stability limit is achieved at around $\alpha =3$. The Eigenvalues converge asymptotically to a value of -0.8055 beyond this $\alpha$. 
Any increase in 
$\alpha$ beyond a value of 3.0 is predicted to not have much effect on the overall solution's stability. 

\begin{figure}[H]
	    \centering
	    \subfloat[$\alpha=3.0$]{\label{fig:hllcswm_alpha_3_aftershockdensity}\includegraphics[scale=0.2]{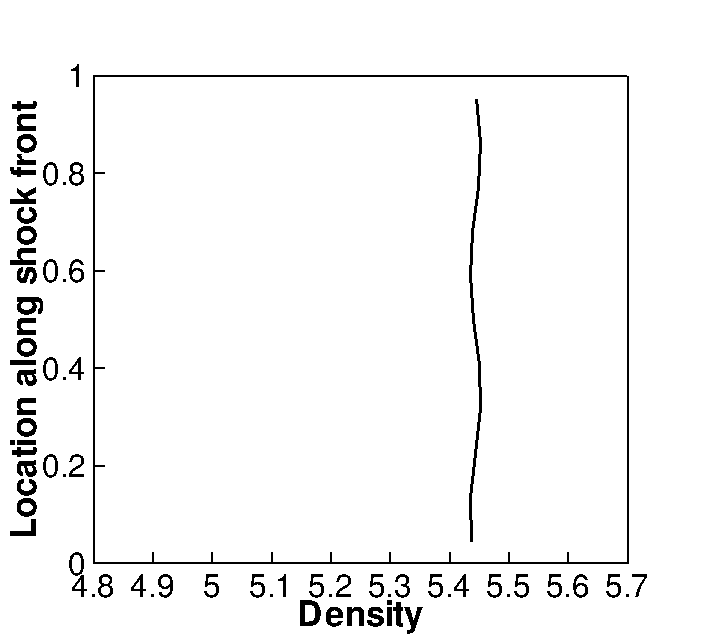}}
	    \subfloat[$\alpha=4.0$]{\label{fig:hllcswm_alpha_4_aftershockdensity}\includegraphics[scale=0.2]{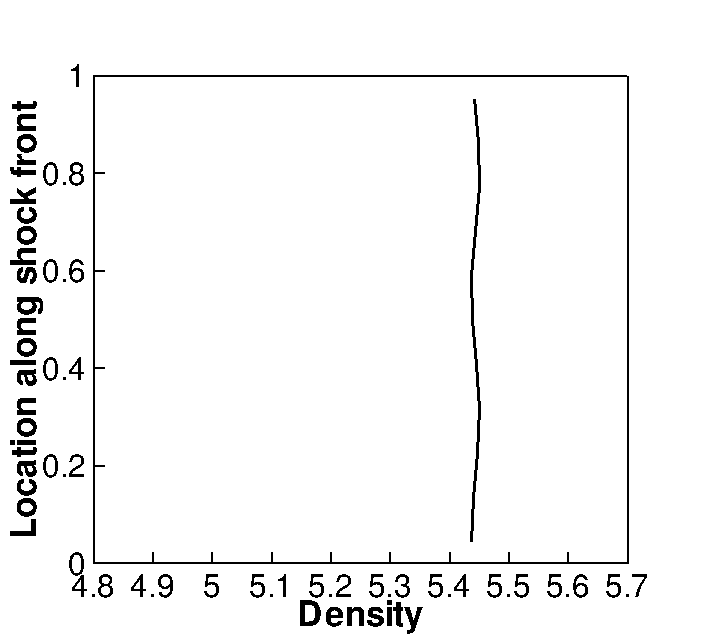}}
	    \subfloat[$\alpha=5.0$]{\label{fig:hllcswm_alpha_5_aftershockdensity}\includegraphics[scale=0.2]{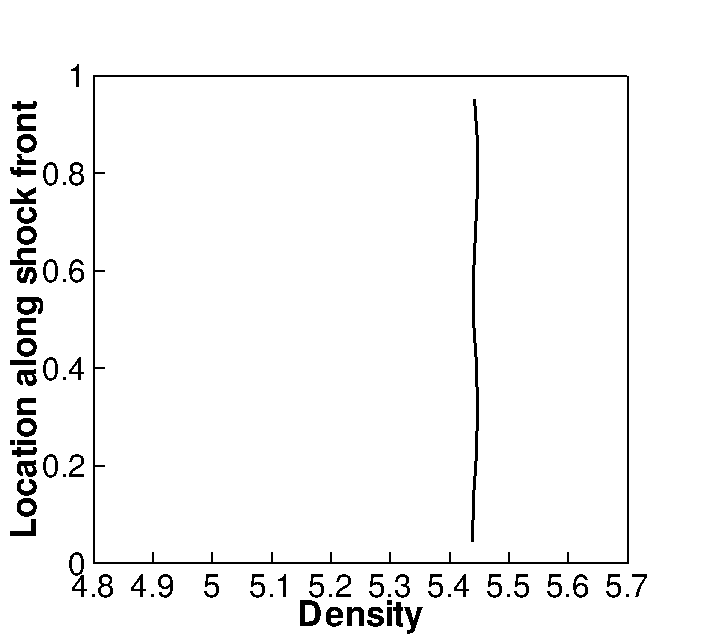}}
	    \caption{Variation of density along the computed shock front extracted from a coloum of cells behind shock (7th coloum). The solution was obtained using the proposed HLLC-ADC scheme 
	    for varying $\alpha$ values (only results for selected $\alpha$'s are shown). All solutions correspond to t=20 units. The deviation of value of density variable from its exact steady value of 5.44 indicates presence of instabilities.}
	    \label{fig:densityaftershock_hllc_hll_hybrid}
	    \end{figure}

\noindent As mentioned previously, these predictions can be numerically verified by directly observing the physical quantites during an actual numerical simulation.
Fig.(\ref{fig:densityaftershock_hllc_hll_hybrid}) shows plots of cell centered density values versus the y locations extracted right behind the 
shock, ie. from the 7th coloum of cells in the computational domain for selected values of $\alpha$.

	    \begin{figure}[H]
	    \centering
	    \subfloat[$\alpha=0.0$]{\label{fig:hllc_aftershockdensity}\includegraphics[scale=0.2]{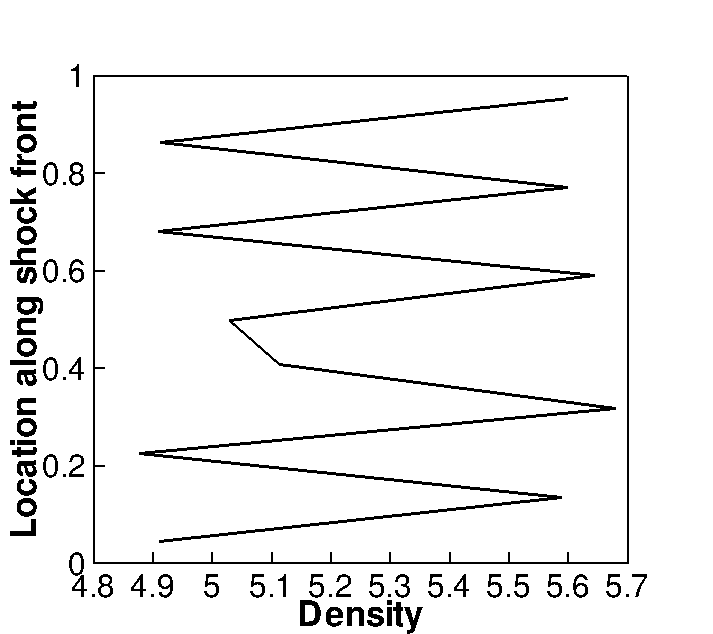}}
	    \subfloat[$\alpha=1.0$]{\label{fig:hllcswm_alpha_1_aftershockdensity}\includegraphics[scale=0.2]{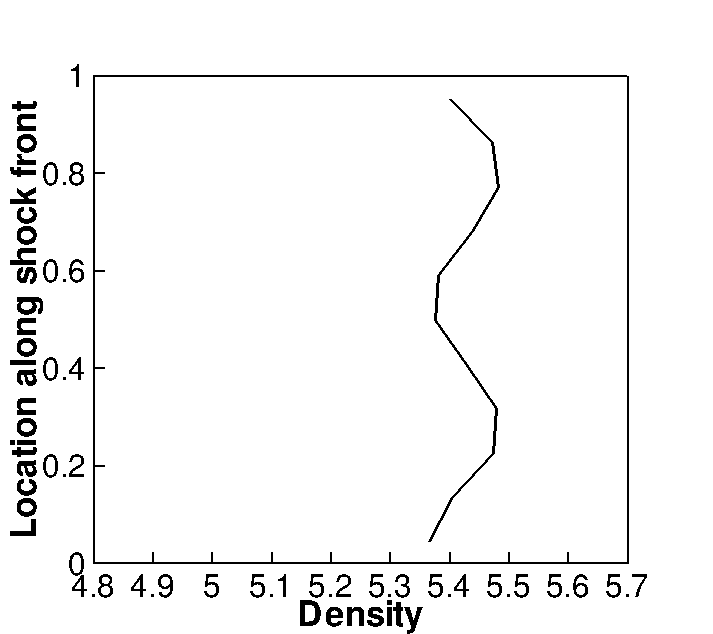}}
	    \subfloat[$\alpha=2.0$]{\label{fig:hllcswm_alpha_2_aftershockdensity}\includegraphics[scale=0.2]{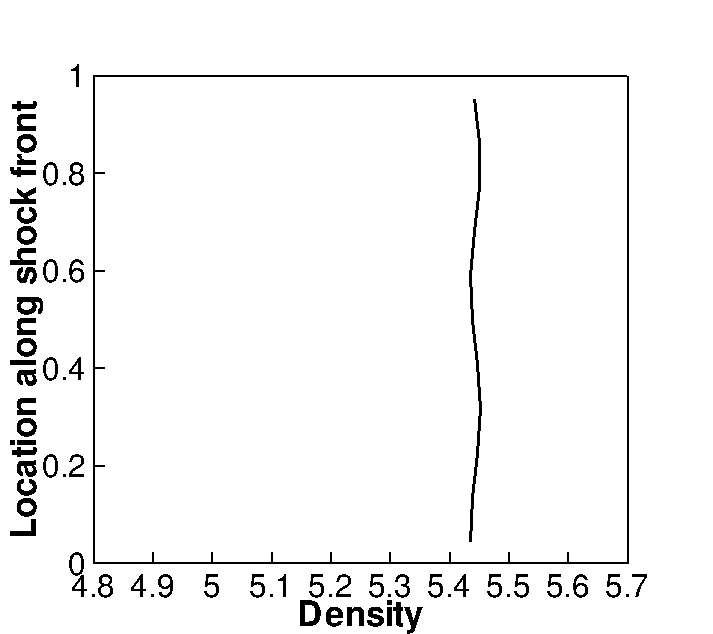}}
	    \end{figure}

\begin{figure}[H]
	    \centering
	    \subfloat[]{\label{fig:effectofalphaforfourMachnumbers_hllc_hll_hybrid}\includegraphics[scale=0.3]{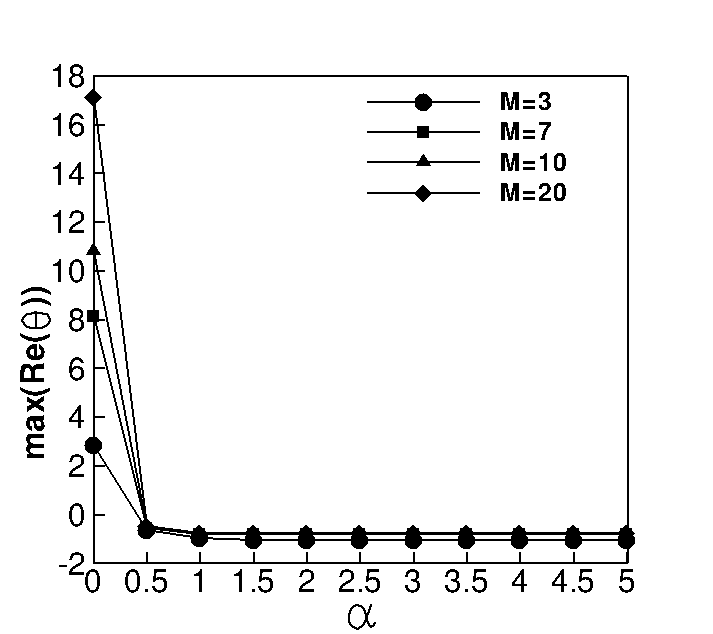}}
            \subfloat[]{\label{fig:effectofalphaforfourMachnumbers_zoomed_hllc_hll_hybrid}\includegraphics[scale=0.3]{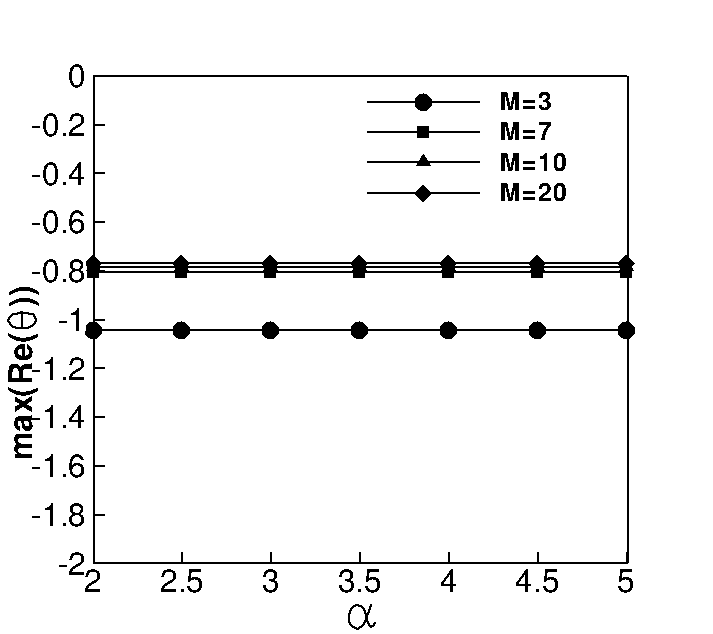}}
	   \caption{(a) Effect of $\alpha$ on $max(Re(\theta))$ for varying inflow Mach numbers 
	   M$=3,7,10$ and $20$ for the two dimensional isolated steady thin shock problem. (b) A zoomed-in version that 
	   shows the asymptotic linear stability limits achieved at each Mach number.}
	    \end{figure}

\noindent Fig.(\ref{fig:hllc_aftershockdensity}) clearly shows the saw-tooth profiled density variation that is typical of a shock unstable solution. In general, 
the numerical solutions agree with the predictions from linear analysis that increasing $\alpha$ values results in increasing shock stability. 
Note that at the linear 
asymptotic stability limit of $\alpha=3.0$ predicted by the analysis, the perturbations in the density profile have reduced substantially. 
Hence it can be safely assumed that for the given problem, $\alpha=3.0$ will guarantee a shock stable 
solution. Dumbser et al have reported in \cite{dumbser2004} that inflow Mach number can have a predominant effect on the proliferation of these instabilities. 
Hence, a choice of $\alpha$ that is capable of dealing with instabilities over a wide range of Mach numbers is always preferred. To consider the effect of Mach numbers, 
the above analysis is repeated 
for three additional Mach numbers; M=3, M=10 and M=20. Fig.(\ref{fig:effectofalphaforfourMachnumbers_hllc_hll_hybrid}) shows the 
variation of $max(Re(\theta))$ with respect to $\alpha$ for these additional Mach numbers.	 
 
\begin{figure}[H]
	    \centering
	    \includegraphics[scale=0.3]{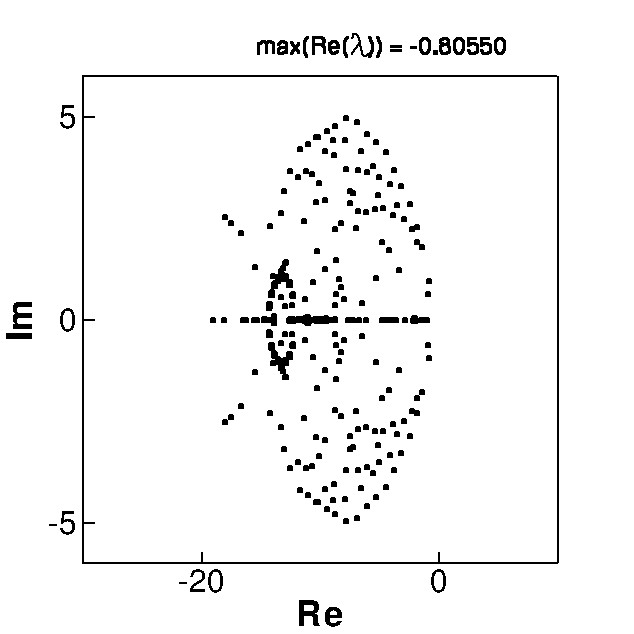}
	   \caption{Eigenvalue spectrum for the proposed HLLC-ADC scheme corresponding to the M=7 steady shock problem. Here $\alpha$ is taken to be 3.0
	    for configuring the HLLC-ADC scheme.}
	   \label{fig:hllc_hll_hybrid_matrixanalysis}
	    \end{figure}

\noindent It is seen from Fig.(\ref{fig:effectofalphaforfourMachnumbers_hllc_hll_hybrid}) that for $\alpha=0.0$ which corresponds to HLLC scheme, an increase in 
Mach number leads to a 
drastic decrease in theoretical shock stability. This is an expected behaviour from a typical shock instability prone scheme like the HLLC scheme \cite{dumbser2004}. 
It is interesting to note that for all Mach numbers considered here, $\alpha=0.5$ is sufficient to achieve theoretical linear stability. 
This means that any slight reduction 
of the contribution from the antidiffusion term promotes a tendency towards a shock stable solution. Further, with increasing values of $\alpha$, 
it is once again observed that 
the $max(Re(\theta))$ corresponding to each Mach number tends to asymptotically converge. 
Fig.(\ref{fig:effectofalphaforfourMachnumbers_zoomed_hllc_hll_hybrid}) shows the asymptotic stability limit for 
each Mach number more clearly. The asymptotic value of $max(Re(\theta))$ for $M=3,10,20$ are 
respectively -1.04261, -0.78238 and -0.76874. By comparing these with the asymptotic value corresponding to $M=7$, it can be said that in general the linear stability of 
this problem is inversely proportional to the Mach number. By studying the aftershock density contours for various Mach numbers, similar to the case of $M=7$,
it was found that $\alpha=3.0$ is a sufficient 
value to ensure stability on this problem. Additionally, the Eigenvalue spectrum corresponding to $M=7$ case for 
the proposed HLLC-ADC scheme with $\alpha=3.0$ is given in Fig.(\ref{fig:hllc_hll_hybrid_matrixanalysis}). Comparing this eigenvalue distribution with the one corresponding to the original
HLLC scheme provided in Fig.(\ref{fig:hllc_matrixanalysis}) the effectiveness of the present strategy is easily observable. The $max(Re(\theta))$ has 
dropped one order of magnitude from the unstable growth rate of +8.15562 to the stable growth rate of -0.80550. 
It is important to note that the shock stability predicted by the matrix analysis for the proposed HLLC-ADC 
scheme is independant of the time discretization technique adopted or the CFL number employed because only the spatial discretization has been used to construct the 
stability matrix. 

\section{Numerical results}
\label{sec:numericaltests}
In this section, robustness of the proposed HLLC-ADC scheme is demonstrated through a series of strict shock instability test problems on which the HLLC scheme is
known to fail. Solution comptued by the HLLE scheme is also provided for comparison in all cases. 
It is well known that 
spatial order of accuracy has varying effects on the intensity of instability manifestation on various problems \cite{gressier2000}.
Hence we selectively compute certain test cases to first order accuracy while some others to second order accuracy based on which exhibits more instability
to clearly demostrate the effectiveness of the cure
proposed. The 
second order spatial accuracy in primitive variables 
is achieved by limiting the gradients obtained using
Green Gauss method \cite{balzek2005} with Barth Jersperson limiter \cite{barth1989} while the second order time accuracy is achieved using 
strong stability preserving variant of Runge Kutta method \cite{Gottlieb2001}. All boundary conditions are set using ghost cells.
The value of $\alpha =3.0$ is used in all the calculations with the HLLC-ADC method.

\subsection{Odd-even decoupling problem}
\begin{figure}[H]
	    \begin{center}
	    \setcounter{subfigure}{0}
	    \subfloat[HLLC]{\label{fig:hllc_movingshock}\includegraphics[scale=0.27]{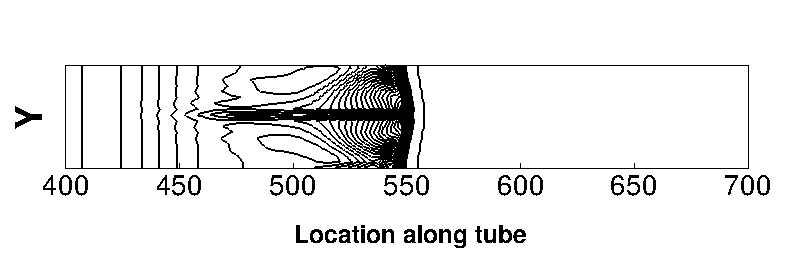}}
	    \subfloat[HLLE]{\label{fig:hlle_movingshock}\includegraphics[scale=0.27]{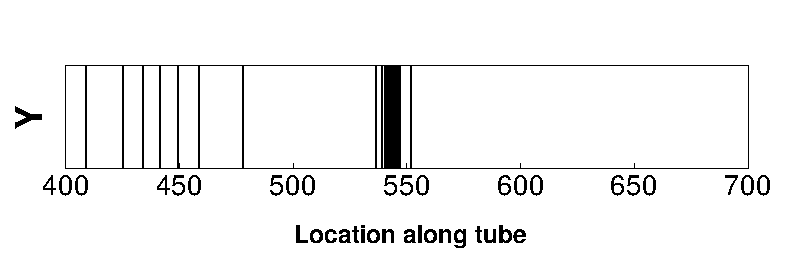}}\\
	    \subfloat[HLLC-ADC]{\label{fig:hllc_hll_hybrid_movingshock}\includegraphics[scale=0.27]{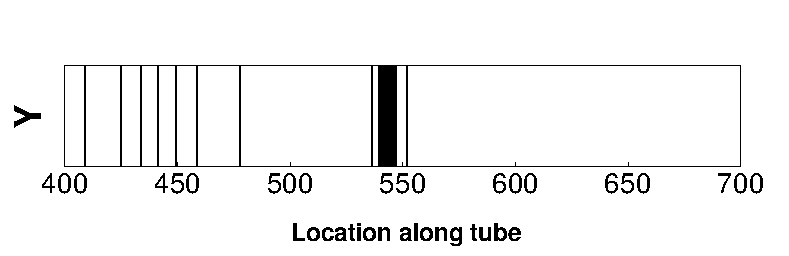}}
	    \end{center}
	    \caption{Density contours for M=6 Quirk's odd-even decoupling problem.}
	   \label{fig:movingshockresults_hybrid}
	    \end{figure}

Quirk \cite{quirk1994} first reported the occurrence of shock instability in a moving shock propagating down a computational tube. 
A normal shock of strength M=6 is made to propagate down a computational tube consisting of 800 by 20 cells.
The tube is initialized as a stationary fluid with primitive values $(\rho,u,v,p)=(1.4,1.0,0.0,1.0)$.
The instability is triggered by perturbing the centerline grid of the domain to an order of 1E-6. First order solution is sought. The CFL for the computations were
taken to be 0.5 and simulations were run till shock reached a location x=550. The results showing fifty density
contours spanning value from 1.4 to 7.34 is shown in Fig.(\ref{fig:movingshockresults_hybrid}).

\noindent Fig.(\ref{fig:hllc_movingshock}) clearly shows the deteriorated condition of the shock profile calculated by the HLLC scheme. 
Fig.(\ref{fig:hllc_hll_hybrid_movingshock}) demonstrates that HLLC-ADC scheme is capable of providing the necessary dissipation and 
keeping the solution unaffected by the instabilities that occurs in the moving shock. This solution is identical to the solution computed by HLLE scheme shown in 
Fig.(\ref{fig:hlle_movingshock}).

\subsection{Inclined stationary shock instability problem}

\begin{figure}[H]
	    \centering
	    \setcounter{subfigure}{0}
	    \subfloat[HLLC]{\label{fig:hllc_inclinedshock}\includegraphics[scale=0.2]{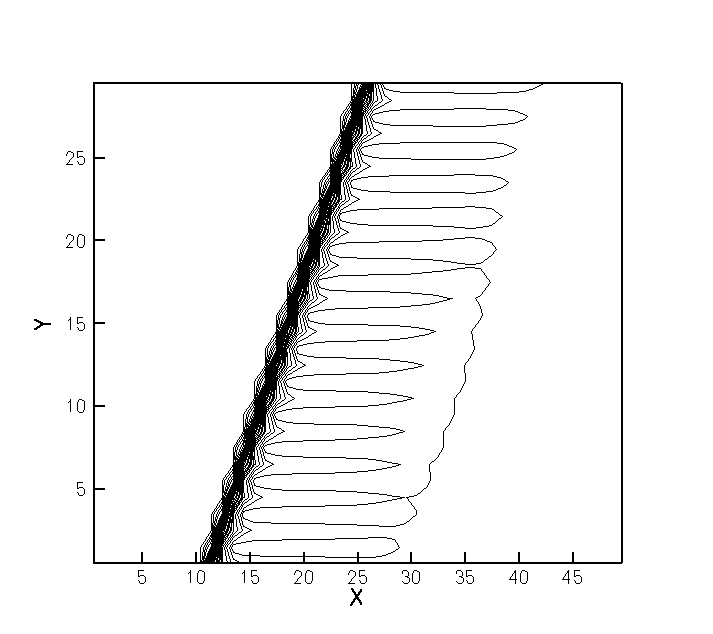}}
            \subfloat[HLLE]{\label{fig:hlle_inclinedshock}\includegraphics[scale=0.19]{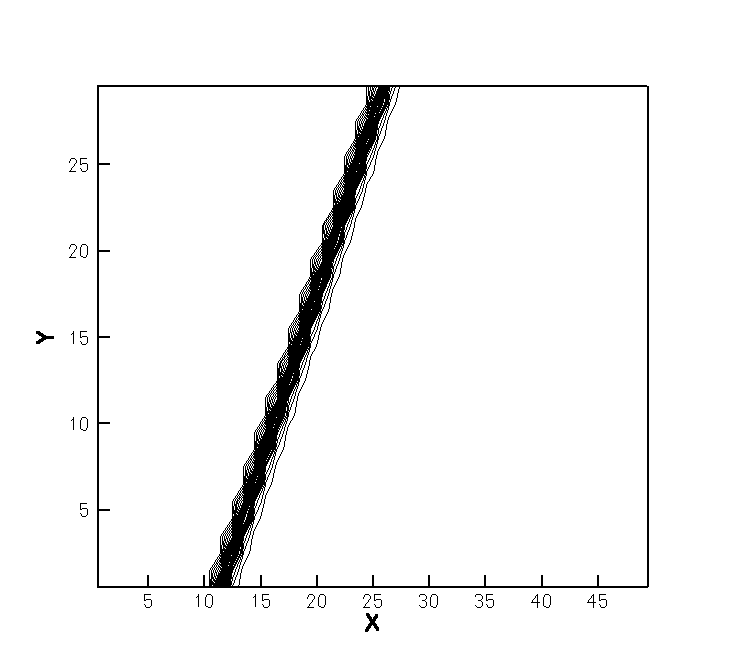}}
	    \subfloat[HLLC-ADC]{\label{fig:hllc_hll_hybrid_inclinedshock}\includegraphics[scale=0.19]{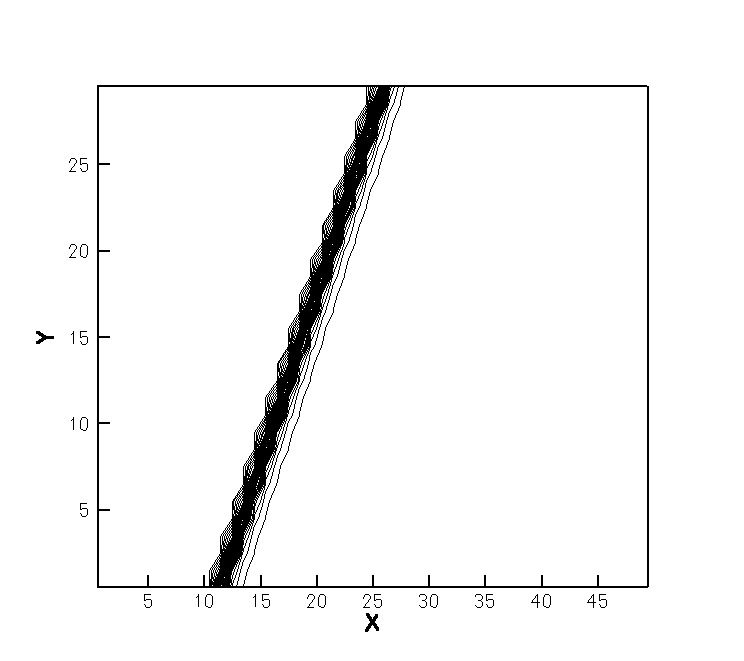}}
	   \caption{Density contours for M=7 stationary inclined shock problem.}
	   \label{fig:inclinedshockresults_hybrid}
	    \end{figure}

This problem was reported in \cite{ohwada2013}. In the present setup a stationary shock is initialized at an angle of $63.43^o$ with respect to the
global x-direction. The computation mesh consists of 50$\times$30 cells uniformly spanning a domain of size 50.0$\times$30.0.
The initial shock wave is set up along the line y=2(x-12). The conditions before the shock wave are provided as
$(\rho,u,v,p)_L=(1.0,447.26,-223.50,3644.31)$ and it describes a supersonic flow normal to the shock front. The after shock conditions are given as
$(\rho,u,v,p)_R=(5.444,82.15,-41.05,207725.94)$. The left and right boundaries are maintained as supersonic inlet and subsonic outlet respectively.
At the top boundary, cells numbered from $i=$ 1 to 15 are maintained as supersonic inlet while all the remaining cells are set periodic to the corresponding 
first row of bottom internal cells. At the bottom boundary, cells numbered from $i=$ 36 to 50 are set as subsonic outlet while the remaining cells are set periodic to the 
corresponding cells at the first row of top internal cells. The CFL of the computation was chosen as 0.5 and first order solution was sought after 1000 iterations. 
Fig.(\ref{fig:inclinedshockresults_hybrid}) shows the result of this experiment for the proposed HLLC-ADC scheme along with reference solutions for HLLC and HLLE. 
Thirty density contours 
uniformly spanning 1.0 to 5.4 are shown in the Fig.(\ref{fig:inclinedshockresults_hybrid}).

\begin{figure}[H]
	    \centering
	    \setcounter{subfigure}{0}
	    \subfloat[HLLC]{\label{fig:hllc_inclinedshock_diagonaldensity}\includegraphics[scale=0.2]{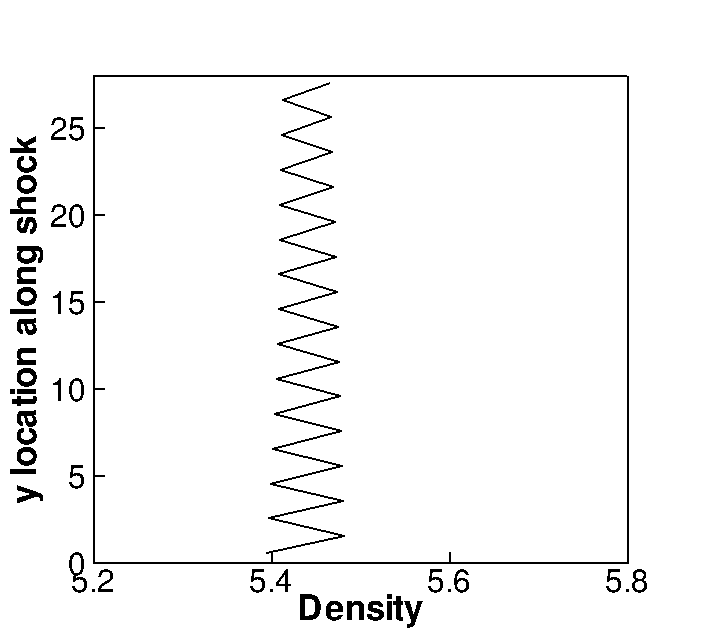}}
            \subfloat[HLLE]{\label{fig:hlle_inclinedshock_diagonal_density}\includegraphics[scale=0.2]{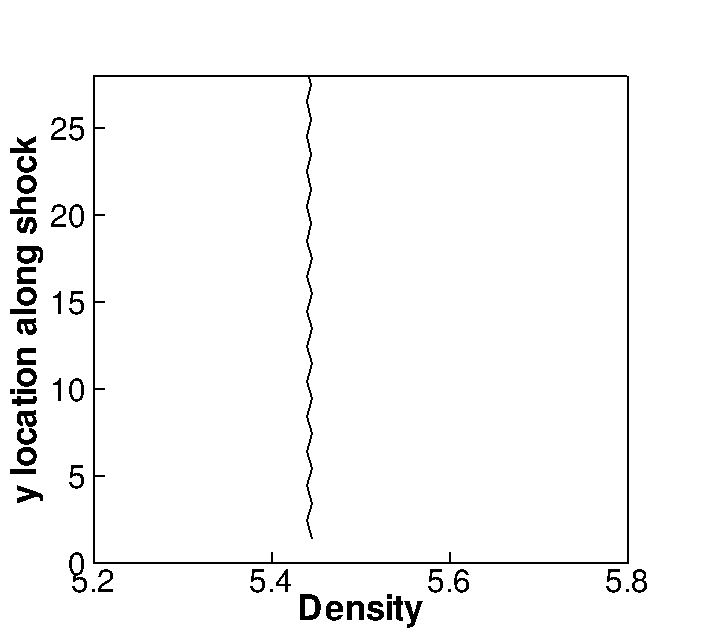}}
	    \subfloat[HLLC-ADC]{\label{fig:hllc_hll_hybrid_inclinedshock_diagonaldensity}\includegraphics[scale=0.2]{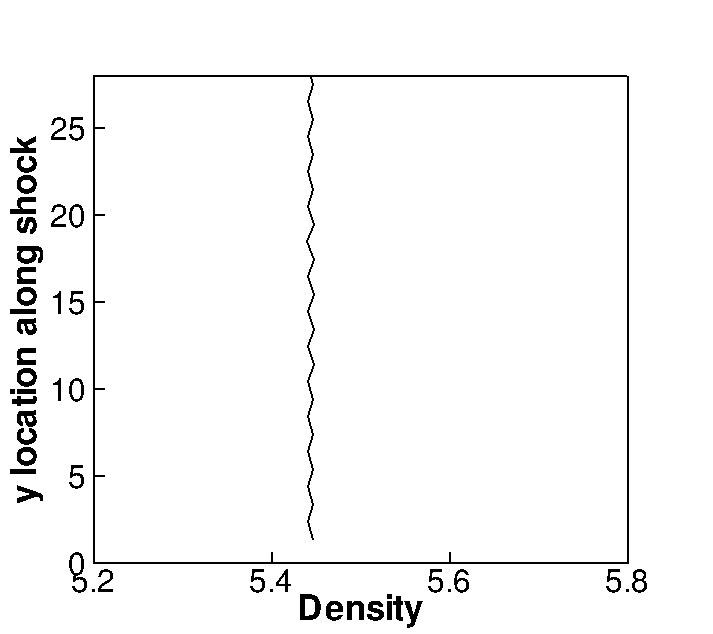}}
	   \caption{Plot showing density values extracted from the immediate subsonic region along the shock profile for the M=7 stationary inclined shock.}
	   \label{fig:inclinedshock_diagonal_density_hybrid}
	    \end{figure}

\noindent Fig.(\ref{fig:hllc_inclinedshock}) shows the presence of instability in the original HLLC scheme. Fig.(\ref{fig:hllc_inclinedshock_diagonaldensity}) 
shows the density variation along the shock front extracted just behind the discontinuity. This density variation has a typical saw tooth profile that is characteristic of
most shock unstable solution. Fig.(\ref{fig:hllc_hll_hybrid_inclinedshock}) shows the result computed
by HLLC-ADC scheme wherein a clean shock profile is observed. This result closely matches the shock profile computed by HLLE scheme 
shown in Fig.(\ref{fig:hlle_inclinedshock}) which also lacks any oscillations.
The density plots in Figs.(\ref{fig:hllc_hll_hybrid_inclinedshock_diagonaldensity}) and (\ref{fig:hlle_inclinedshock_diagonal_density}) corroborate
this observation.

\subsection{Supersonic flow over forward facing step}
\begin{figure}[H]
	    \centering
	    \setcounter{subfigure}{0}
	    \subfloat[HLLC]{\label{fig:hllc_ffs}\includegraphics[scale=0.3]{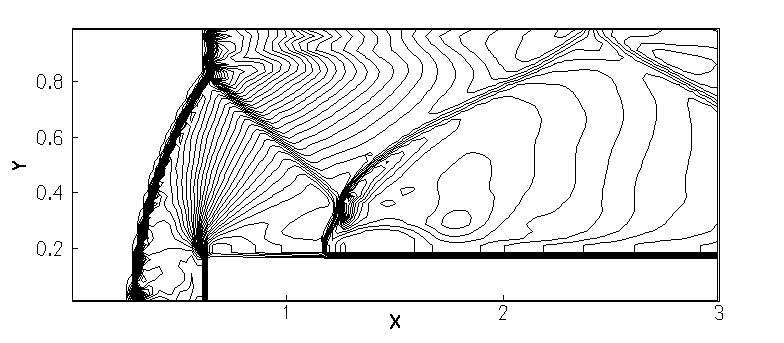}}
            \subfloat[HLLE]{\label{fig:hlle_ffs}\includegraphics[scale=0.3]{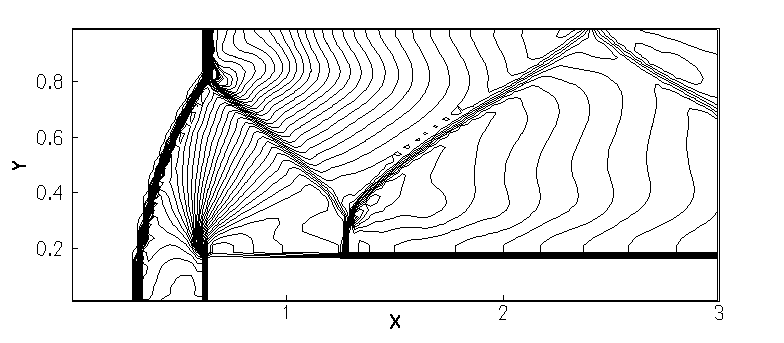}}\\
	    \subfloat[HLLC-ADC]{\label{fig:hllc_hll_hybrid_ffs}\includegraphics[scale=0.3]{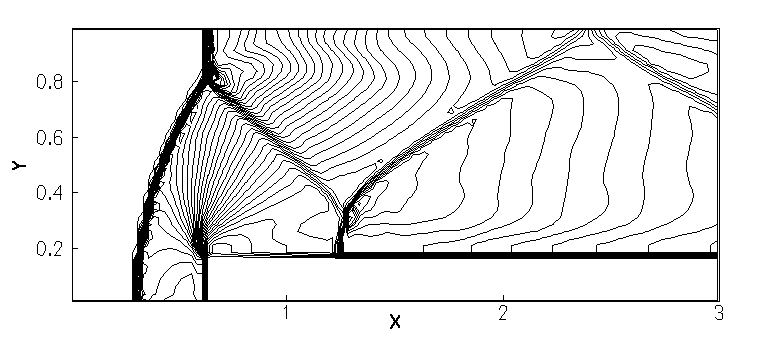}}
	   \caption{Density contours for M=3 flow over a forward facing step problem.}
	   \label{fig:ffs__hybrid}
	    \end{figure}

This problem was studied extensively 
by Woodward and Colella \cite{woodward1984}. The problem consists of a M=3 supersonic flow that passes over a step of 0.2 units high located 
at a distance of 0.6 units from the inlet. The computational domain is of size $[0,3]\times[0,1]$. The domain is meshed with $120\times40$ 
structured Cartesian cells. The whole domain is initialized with the value of $(\rho,u,v,p)= (1.4,3,0,1)$. The inlet boundary is maintained as 
freestream while the outlet boundary is set to zero gradient. The top and the bottom walls are
set as inviscid walls. The problem is computed to second order accuracy. The simulation is run for t=4 with CFL as 0.5. 
Fig.(\ref{fig:ffs__hybrid}) shows 40 density contours spanning 0.2 to 7.0. 
	    
Fig.(\ref{fig:hllc_ffs}) shows the solution computed by the HLLC scheme. The oscillations in the regions surrounding the
normal shock stems of the primary shock near the bottom and top boundaries
are clearly 
visible in this solution.  Further, the HLLC scheme also computes a severe kinked stem in the reflected shock. In comparison, 
solution computed by the HLLC-ADC scheme shown in Fig.(\ref{fig:hllc_hll_hybrid_ffs}) demonstrate a clean shock profile with no 
instabilities visible. The infamous kink has also dissapeared. This solution is most comparable to the HLLE solution shown in Fig.(\ref{fig:hlle_ffs}).

\subsection{Diffraction of a moving normal shock over a 90$^{0}$ corner}
\begin{figure}[H]
	    \centering
	    \setcounter{subfigure}{0}
	    \subfloat[HLLC]{\label{fig:hllc_supersoniccorner}\includegraphics[scale=0.3]{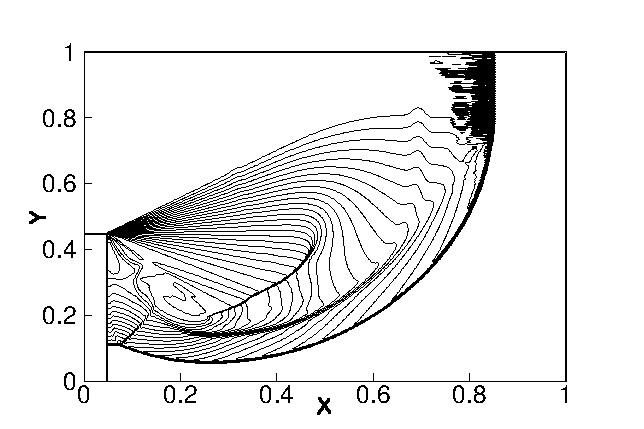}}
            \subfloat[HLLE]{\label{fig:hlle_supersoniccorner}\includegraphics[scale=0.3]{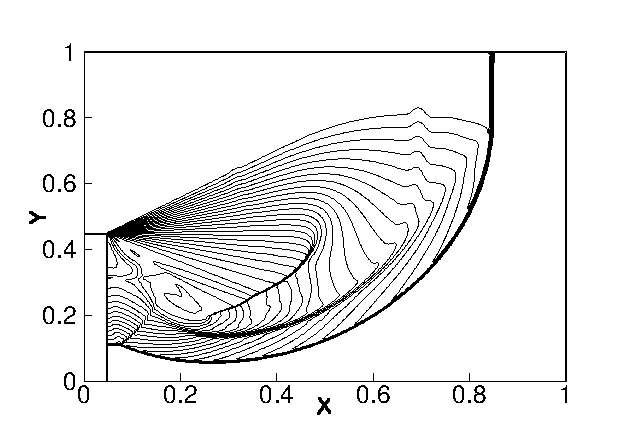}}\\
            \subfloat[HLLC-ADC]{\label{fig:hllc_hlle_hybrid_supersoniccorner}\includegraphics[scale=0.3]{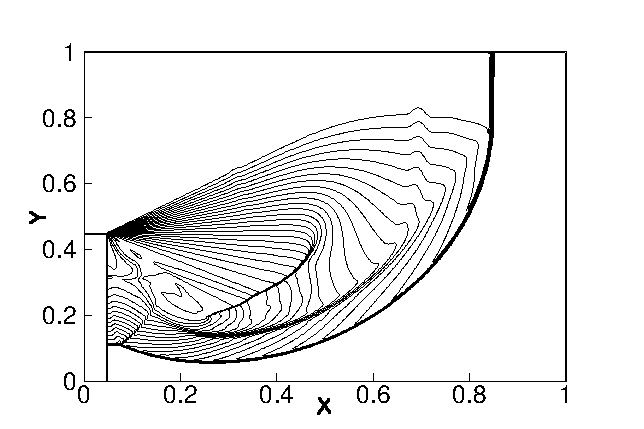}}
	   \caption{Density contours for M=5.09 normal shock diffraction around a $90^o$ corner.}
	   \label{fig:supersoniccornerresults_hybrid}
	    \end{figure}

Quirk \cite{quirk1994} showed that many popular Riemann solvers produce unacceptable shock profiles when computing a Mach 5.09 normal shock 
expanding around a $90^o$ corner. The problem is set up on a unit dimensional domain with 400 $\times$ 400 regular Cartesian cells. 
A corner is located at $x=0.05, y=0.45$ on top of which a normal shock is located at $x=0.05$. The cells lying outside post shock region
are initialized as a stationary fluid with properties $\rho=1.4, u=0.0, v=0.0$ and $p=1.0$. The inlet boundary is maintained as post shock 
conditions while outlet boundary is set to zero gradient. Top boundary is adaptively reconfigured to allow for the shock motion. Bottom boundary 
behind the corner uses extrapolated values from within the domain. The corner 
surface is maintained as reflective wall. The problem is computed to second order accuracy. Simulation is run for $t=0.1561$ units with CFL of 
0.4. Thirty density contours equally spanning values of 0.1 to 7 is
shown in Figure \ref{fig:supersoniccornerresults_hybrid}.
	    
Fig.(\ref{fig:hllc_supersoniccorner}) shows result computed by the HLLC scheme. A part of the normal shock in the right corner is completely distorted in this case.
In comparison, the HLLE scheme produces a clean shock profile as seen in Fig.(\ref{fig:hlle_supersoniccorner}). Fig.(\ref{fig:hllc_hlle_hybrid_supersoniccorner})
shows the corresponding solution computed by the HLLC-ADC scheme. It is evident that the proposed scheme is able to satisfactorily 
compute the whole solution and does not produce any trace of instability.

\subsection{Double Mach Reflection problem}
\begin{figure}[H]
	    \centering
	    \setcounter{subfigure}{0}
	    \subfloat[HLLC]{\label{fig:hllc_dmr}\includegraphics[scale=0.27]{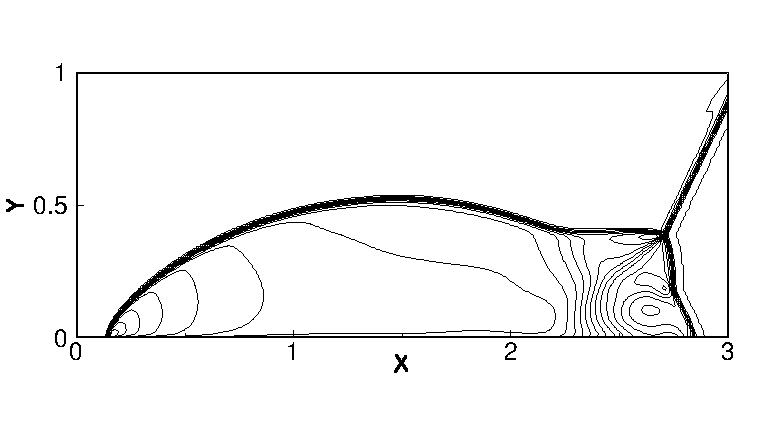}}
	    \subfloat[HLLE]{\label{fig:hlle_dmr}\includegraphics[scale=0.29]{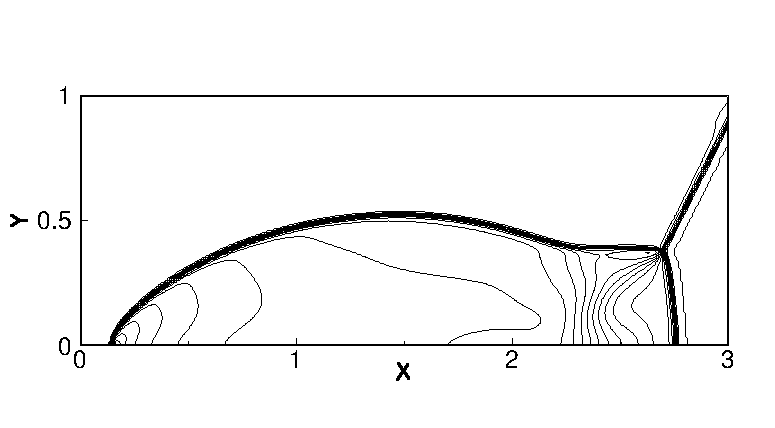}}\\
            \subfloat[HLLC-ADC]{\label{fig:hllc_hlle_hybrid_dmr}\includegraphics[scale=0.27]{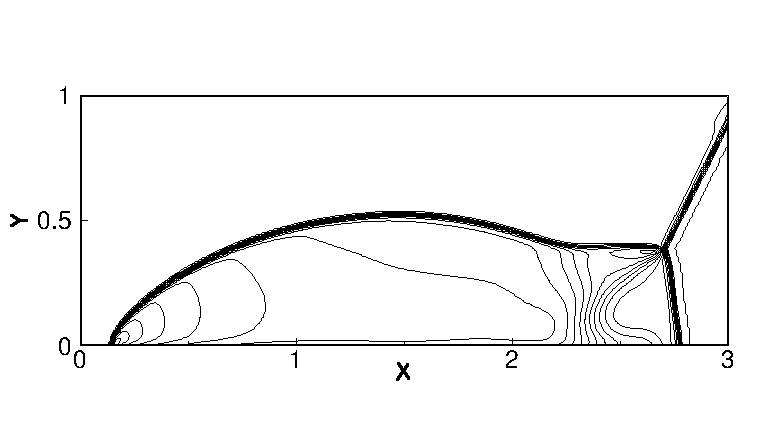}}\\
	   \caption{Density contours for double Mach reflection problem.}
	   \label{fig:dmrresults_hybrid}
	    \end{figure}

To demonstrate the robustness of the proposed scheme, the standard test problem called Double Mach reflection problem is used. Various popular 
Riemann solvers are known to produce severely kinked principle Mach stem causing existence of a triple point on this problem \cite{quirk1994, woodward1984}.
For the present test, a domain of $4.0\times1.0$ is chosen and is constituted of $480\times120$ structured Cartesian cells. An 
oblique shock corresponding to $M=10$, making a $60^o$
angle with the bottom wall at $x=0.16667$ is made to propagate through the domain. Cells ahead of the shock are initialized with values 
$(\rho,u,v,p) = (1.4,0,0,1)$ while those after the shock are set to appropriate post shock conditions. Inlet boundary is maintained at post shock 
conditions while zero gradient condition is used at outlet boundary. Top boundary conditions are adjusted to allow for propagation of shock front. 
At the bottom, post shock conditions are maintained till $x=0.16667$ after which invisid wall conditions are used. The simulation is run till 
$t=0.02$ with CFL of 0.8. The problem is computed to first order accuracy. Fig.(\ref{fig:dmrresults_hybrid})
shows results of the experiment with twenty five density contours equally spanning values 
from 1.4 to 21.
	    
In Fig.(\ref{fig:hllc_dmr}), the presence of kinked Mach stem and the subsequent triple point produced by the HLLC scheme is clearly visible. 
The solution computed by the proposed HLLC-ADC scheme shown in Fig.(\ref{fig:hllc_hlle_hybrid_dmr}) is devoid of these unphysical 
features and matches well with solution by the HLLE scheme given in Fig.(\ref{fig:hlle_dmr}).

\subsection{Hypersonic flow over a blunt body}
\begin{figure}[H]
	    \centering
	    \setcounter{subfigure}{0}
	    \subfloat[HLLC]{\label{fig:hllc_bluntbody}\includegraphics[scale=0.3]{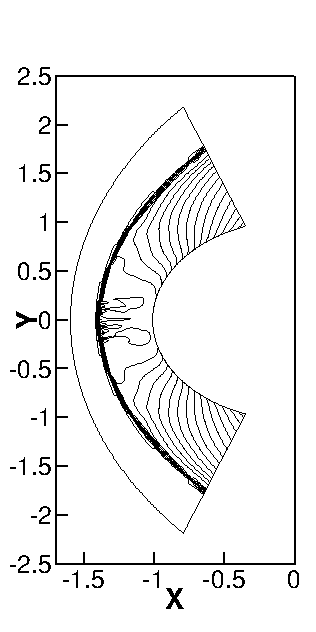}}
            \subfloat[HLLE]{\label{fig:hlle_bluntbody}\includegraphics[scale=0.3]{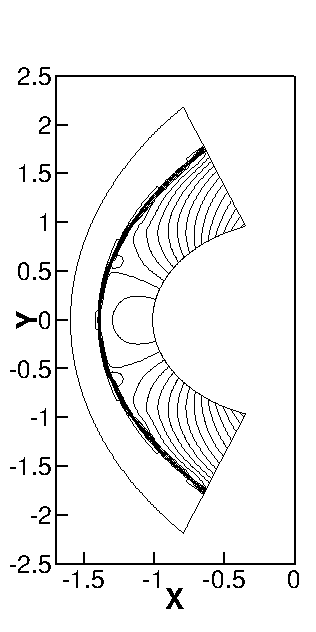}}
            \subfloat[HLLC-ADC]{\label{fig:hllc_hlle_hybrid_bluntbody}\includegraphics[scale=0.3]{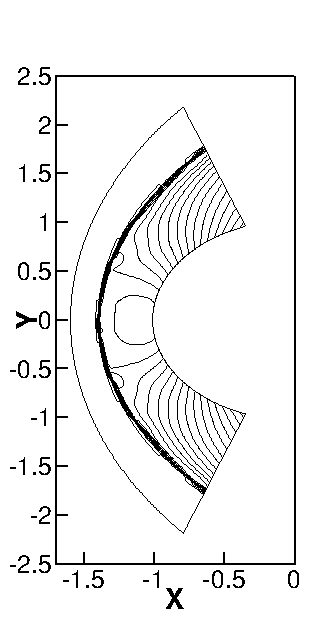}}
	   \caption{Density contours for M=20 supersonic flow over a cylindrical body.}
	   \label{fig:carbuncleresults_hllc_hll_hybrid}
	    \end{figure}

Another routine test problem that is used to investigate the susceptibility of a numerical scheme to Carbuncle phenomenon is the steady state 
numerical solution of a hypersonic flow over a cylindrical body. 
The problem is set up by placing a cylindrical body of radius 1 unit in a M=20 flow with free stream conditions given as 
$(\rho,u,v,p) = (1.4,20.0,0.0,1.0)$.
The computational mesh is prepared using method described in \cite{huang2011} wherein 320$\times$40 body fitted structured quadrilateral cells 
are used in circumferential and radial directions respectively. The inlet boundary is maintained as supersonic inlet while at the solid wall, impermeability is prescribed
with density and pressure are extrapolated 
from the internal cell. Simple extrapolation
is employed at top and bottom boundaries. The computation of this problem is carried out to first order accuracy. The CFL for the computations were taken to be 0.5 
and simulations were run for 
80,000 iterations. The results showing twenty density contours equally spanning value from 1.4 to 8.5 is shown in Fig.(\ref{fig:carbuncleresults_hllc_hll_hybrid}). 

\begin{figure}[H]
	    \centering
	    \setcounter{subfigure}{0}
	    \subfloat[Comparison of centerline pressures]{\label{fig:centerline_comparison_overall}\includegraphics[scale=0.3]{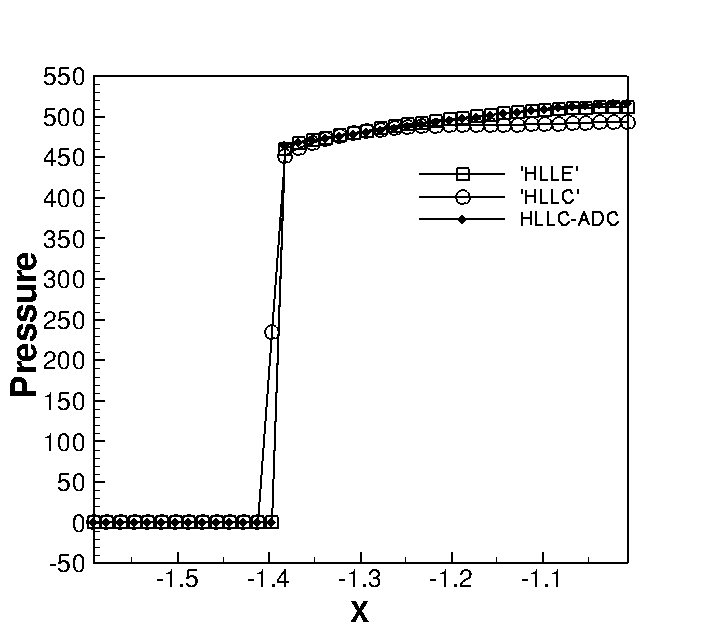}}
            \subfloat[Location A]{\label{fig:centerline_comparison_A}\includegraphics[scale=0.3]{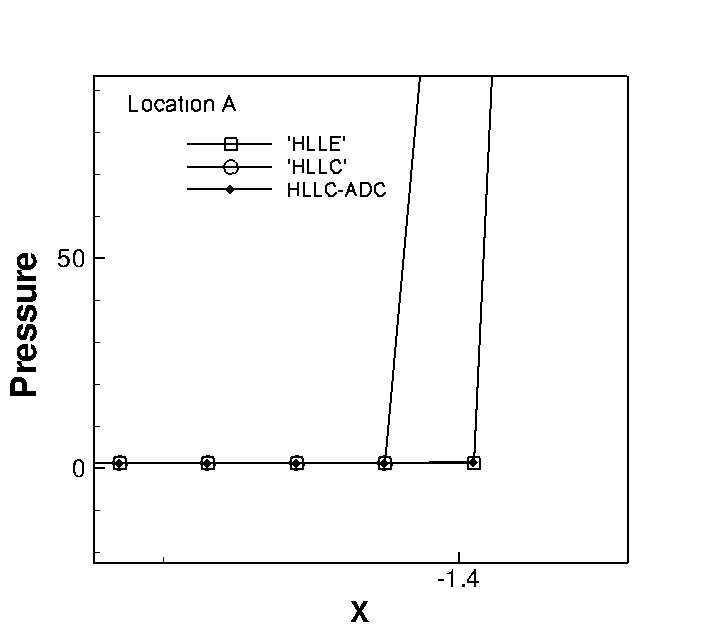}}\\
            \subfloat[Location B]{\label{fig:centerline_comparison_B}\includegraphics[scale=0.3]{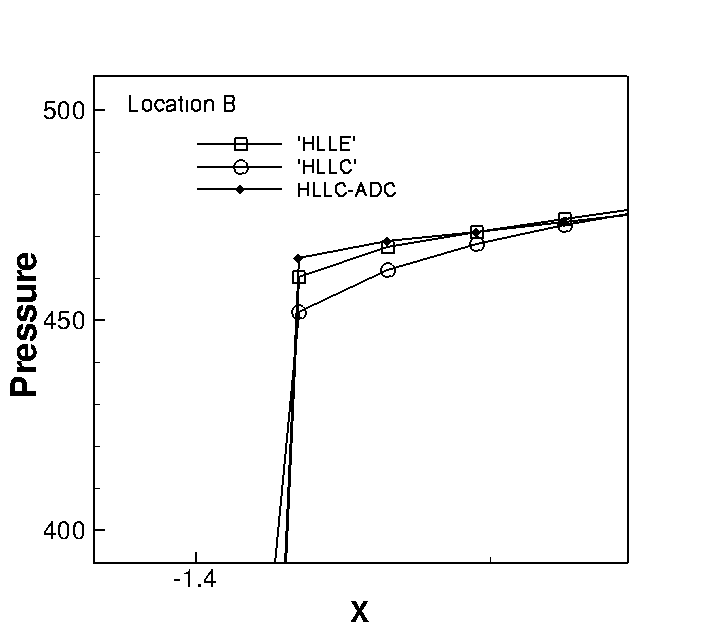}}
            \subfloat[Location C]{\label{fig:centerline_comparison_C}\includegraphics[scale=0.3]{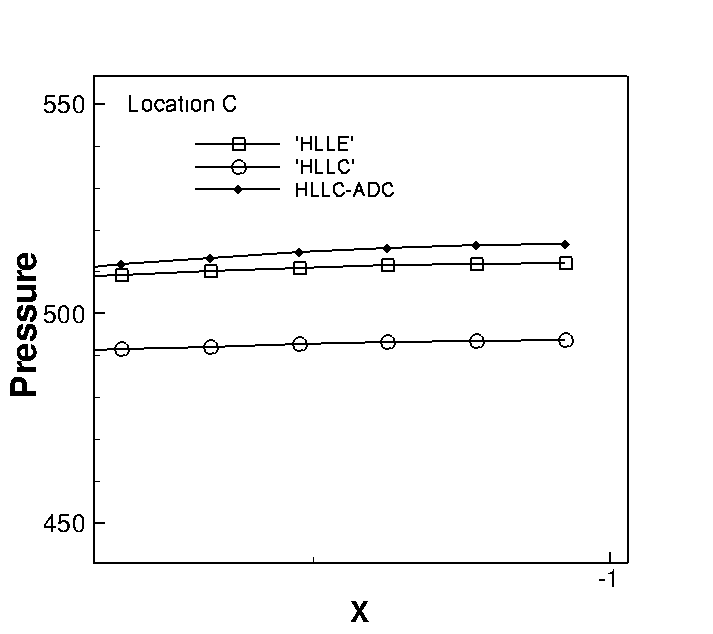}}\\
	   \caption{Comparison of centerline pressures between the HLLC, the HLLE and the proposed HLLC-ADC schemes.}
	   \label{fig:centerline_comparison_hllc_hll_hybrid}
	    \end{figure}

The perturbations in the isodensity contours near the normal shock region of the computed bow shock in Fig.(\ref{fig:hllc_bluntbody}) reveals the propensity
of the HLLC scheme to produce a Carbuncle solution. The solution computed by HLLE scheme is given in Fig.(\ref{fig:hlle_bluntbody}) and depicts smooth isodensity contours. 
Fig.(\ref{fig:hllc_hlle_hybrid_bluntbody}) shows the corresponding result for the proposed HLLC-ADC scheme which is as smooth as the one computed by the HLLE scheme. 
The quality of these solutions can be probed further
by inspecting their centerline pressure data. Fig.(\ref{fig:centerline_comparison_overall}) shows a comparison of the centerline pressures (at j=160) 
between solutions computed by these schemes. Specific locations on the plots are labeled as  \textquoteleft A \textquoteright,
\textquoteleft B \textquoteright, \textquoteleft C \textquoteright and are shown magnified in Figs.(\ref{fig:centerline_comparison_A}),(\ref{fig:centerline_comparison_B})
and (\ref{fig:centerline_comparison_C}).

Refering to Fig.(\ref{fig:centerline_comparison_A}), at location \textquoteleft A \textquoteright, the instabilities has caused unphysical thickening of the 
shock in the solution computed by the HLLC scheme while both the HLLE and the HLLC-ADC schemes produce a much crisper shock. Fig.(\ref{fig:centerline_comparison_B}) 
exposes further how the HLLC scheme computes a much diffused shock while the shock computed by the stable HLLC-ADC scheme seems to be the most accurate amongst all 
the three schemes. Location 
\textquoteleft C \textquoteright which is close to the stagnation point clearly shows that the HLLC-ADC scheme produces a much accurate stagnation pressure value 
(which is closer to the expected inviscid value of 515.60) as compared to both the HLLC and the HLLE schemes.
From basic gasdynamics it is known that the stagnation pressure 
values across a normal shock can be related to the entropy changes incurred by the fluid as it passes through it. Clearly then spurious entropy changes 
are introduced into the solution by the HLLC scheme because of the Carbuncle instabilities while the proposed HLLC-ADC scheme remains completely free of them.

\subsection{Two dimensional supersonic shear flow}
\label{sec:inviscidshearflow}
\begin{figure}[H]
	    \centering
	    \setcounter{subfigure}{0}
	    \subfloat[HLLC]{\label{fig:hllc_shear}\includegraphics[scale=0.21]{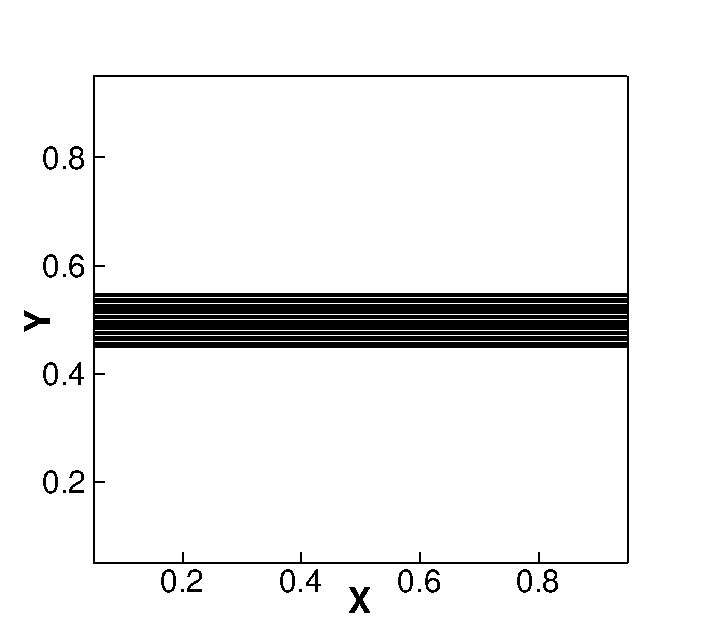}}
            \subfloat[HLLE]{\label{fig:hll_shear}\includegraphics[scale=0.21]{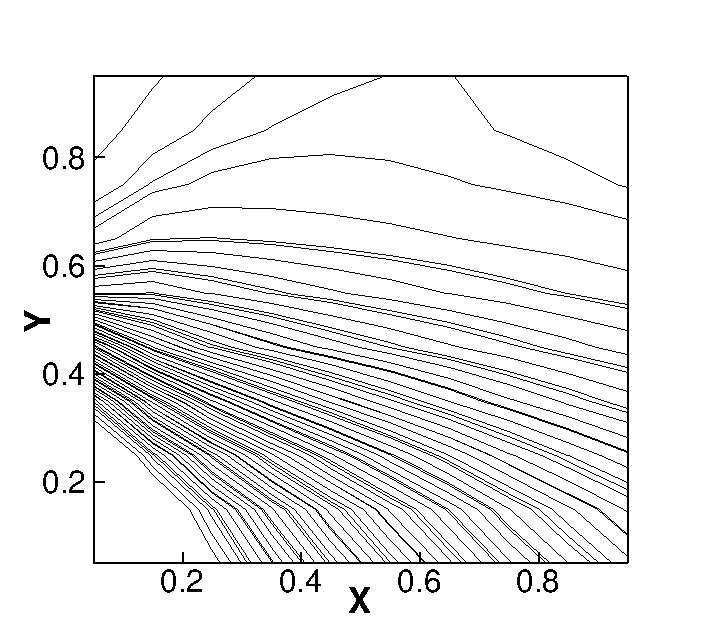}}\\
            \subfloat[HLLC-ADC]{\label{fig:hllc_hll_shear}\includegraphics[scale=0.21]{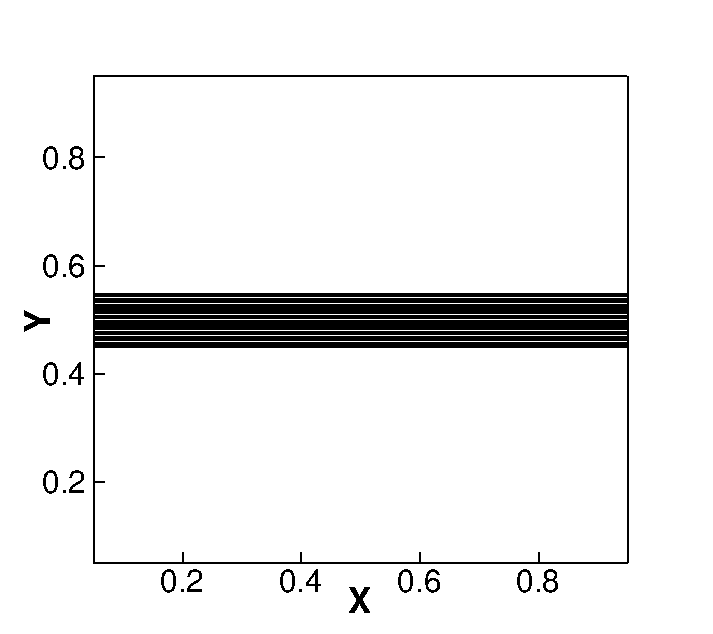}}
            \subfloat[Comparison]{\label{fig:comparison_shear}\includegraphics[scale=0.21]{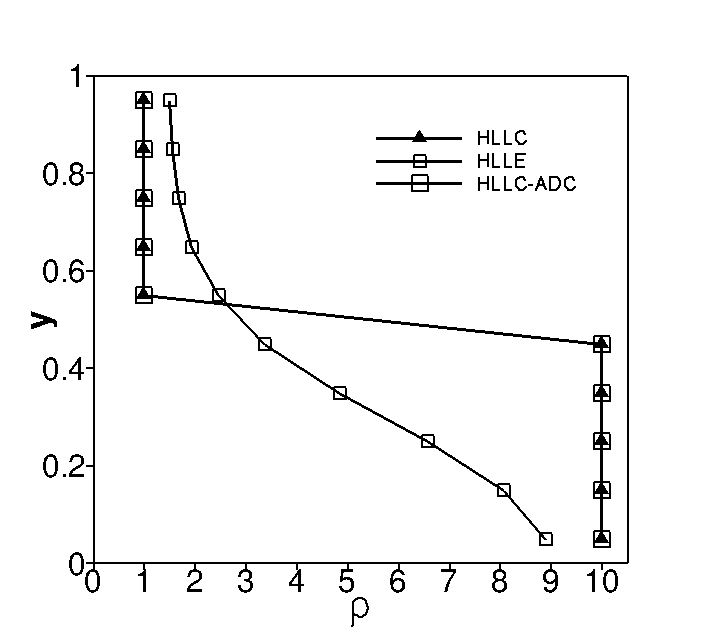}}
	   \caption{Figures (a), (b) and (c) show the density contours for the two dimensional supersonic shear flow computed by HLLC, HLLE and proposed HLLC-ADC scheme 
	   while Figure (d) shows a comparison of density variation along y-direction at the center of the domain.}
	   \label{fig:shearresults_hllc_hll}
	    \end{figure}

The problem describes two fluids with different densities sliding at different speeds over each other and investigates the inviscid contact capturing ability of a given scheme
\cite{wada1997}.
The conditions for the top and bottom fluids are chosen as $(\rho, p,M)_{top}$ = (1, 1, 2) and $(\rho, p,M)_{bottom}$ = (10, 1, 1.1) respectively.
A coarse grid consisting of 10$\times$10 cells on a domain of 1.0$\times$1.0 is used. CFL for the simulations are taken to be 1.0 and simulations were run for 1000 iterations. 
All simulations are plain first order accurate.
The results showing fifty equispaced density contours spanning 1.0 to 10.0 is shown in Fig.(\ref{fig:shearresults_hllc_hll}).  

\noindent It can be seen from Fig.(\ref{fig:shearresults_hllc_hll}) that behaviour of the proposed HLLC-ADC schemes closely matches that of the HLLC scheme 
in being able to preserve an inviscid contact. On the other hand, the HLLE scheme produces a very diffused contact surface.
Fig.(\ref{fig:comparison_shear}) shows a comparison of density variation along y-location at the center of the domain which confirms the observation made above.
From the plot it can be inferred 
that the pressure based sensor used to control the antidiffusive terms in the HLLC-ADC scheme does not affect the
ability to resolve inviscid contact discontinuities.

\subsection{Laminar flow over flat plate}
\label{sec:flatplate}

\begin{figure}[H]
	    \centering
	    \includegraphics[scale=0.4]{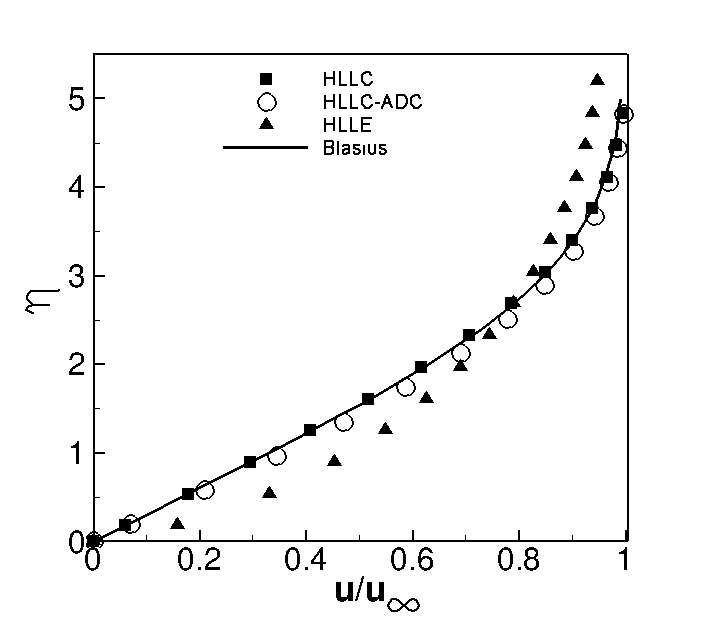}
	   \caption{Result for laminar flow over a flat plate.}
	   \label{fig:laminarflatplate}
	    \end{figure}

Ability of the HLLC-ADC scheme in being able to resolve viscous shear dominated flows will be investigated here using the test case of
a $M = 0.1$ laminar flow of air at pressure of 41368.5 Pa and temperature of 388.88 K over a flat plate of length L=0.3048 m. The total length of the domain 
is 0.381 m in x direction and 0.1 m in y direction. The domain is divided into 31$\times$33 Cartesian cells. While uniform meshing is 
done in the x direction, a non-uniform grid spacing is preferred in the y direction with atleast 15 cells within the boundary layer.
Viscous fluxes were discretized using the Coirier diamond path method discussed in \cite{coirier1994}. CFL was taken to be 0.7. 
The flow was considered to have achieved steady state when the horizontal velocity residuals dropped to the order of $1E^{-7}$. 
The normalized longitudinal velocity profiles $(\frac{u}{u_{\infty}})$ are plotted against the Blasius parameter $\eta=y\sqrt{u_{\infty}/\mu L}$ in Fig.(\ref{fig:laminarflatplate}).  
Fig.(\ref{fig:laminarflatplate}) shows that the HLLC-ADC scheme is almost as accurate as the HLLC scheme in resolving the gradients. 
The slight amount of dissipation that exists in the solution could probably be due to the factor $\omega$ getting activated due to pressure change
in the direction of the flow.

\section{Conclusions}
\label{sec:conclusions}

In this paper we presented a simple and effective strategy to save the Harten\ -Lax\ -van Leer with Contact (HLLC) approximate Riemann solver from numerical shock 
instability. Firstly, we suggested an alternative form of the standard HLLC scheme that clearly distinguishes its inherent diffusive HLL term from its antidiffusive 
term which is responsible for its accuracy on contact and shear wave. 
Through a matrix based stability analysis and associated numerical simulations, we identified that 
numerical discretization of both mass and interface-normal momentum fluxes, on interfaces that are not aligned to the shock front, indepedently contribute 
to the shock instability behaviour of a scheme.
Further, through a linear scale analysis of these critical flux components in the HLLC scheme, we realized that 
instability may be triggered due to weakening of its diffusive HLL terms, that is responsible for damping perturbations in flow quantities, through 
the action of its antidiffusive component. The undamped perturbations specifically in density and primary-flow-velocity variables, grows unbounded, causing unphysical
variations in mass flux across a shock wave culminating in a distorted shock profile. 
To avert the development of such spurious solutions, we employ a simple differentiable pressure ratio based multidimensional shock sensor to prevent the erroneous 
activation of the antidiffusive terms of these critical flux components in the vicinity of a shock wave. 
The resulting scheme called the HLLC-ADC scheme
was found to effectively damp out perturbations and guarantee theoretical shock stability over a wide range of flow Mach numbers. We corroborated this fact 
using a suite of test cases that demonstrated that  
the HLLC-ADC scheme is quite capable of dealing with most common manifestations of shock instability without compromising on the contact and shear preserving abilities 
of its underlying HLLC scheme.

\bibliographystyle{unsrtnat}

\bibliography{References}

\end{document}